\documentclass[onecolumn,secnumarabic,amssymb,nobibnotes, aps, prb,10pt]{revtex4-2}

\usepackage{mathrsfs,amsmath,bm,graphicx,comment,esint,old-arrows}
\usepackage{dcolumn}  
\renewcommand{\arraystretch}{2}
\usepackage{verbatim}
\usepackage{upgreek}
\usepackage{xcolor}
\usepackage[normalem]{ulem}
\usepackage{dutchcal}
\newcommand{\stkout}[1]{\ifmmode\text{\sout{\ensuremath{#1}}}\else\sout{#1}\fi}
\begin{document}

\title{Optical multipolar torque in structured electromagnetic fields: on `helicity gradient' torque, quadrupolar torque and the spin of field gradient}

\author{Lei Wei}
\email{lei.wei@kcl.ac.uk}
\affiliation{Department of Physics, King's College London, Strand, London, WC2R 2LS, United Kingdom}
\author{Francisco J. Rodr\'{i}guez-Fortu\~{n}o}
\email{francisco.rodriguez\_fortuno@kcl.ac.uk}
\affiliation{Department of Physics, King's College London, Strand, London, WC2R 2LS, United Kingdom}

\date{\today}

\begin{abstract}
Structured light mechanically interacts with matter via optical forces and torques. The optical torque is traditionally calculated via the flux of total angular momentum (AM) into a volume enclosing an object. In [Phys. Rev. A 92, 043843 (2015)] a powerful method was suggested to calculate optical torque separately from the flux of the spin and the orbital parts of optical AM, rather than the total, providing useful physical insight. However, the method predicted a new type of dipolar torque dependent on the gradient of the helicity density of the optical beam, inconsistent with prior torque calculations. In this work we intend to clarify this discrepancy and clear up the confusion. We re-derive, from first principles and with detailed derivations, both the traditional dipolar total torque using total AM flux, and the spin and orbital torque components based on the corresponding AM contributions, ensuring that their sum agrees with the total torque. We also test our derived analytical expressions against numerical integration, with exact agreement. We find that `helicity gradient' torque terms indeed exist in the spin and orbital components separately, but we present corrected prefactors, such that upon adding them, they cancel out, and the `helicity gradient' term vanishes from the total dipolar torque, reconciling literature results. We also derive the analytical expression of the quadrupolar torque, showing that it is proportional to the spin of the EM field \textit{gradient}, rather than the local EM field spin, as sometimes wrongly assumed in the literature. We provide examples of counter-intuitive situations where the spin of the EM field gradient behaves very differently to the local EM spin. Naively using the local EM field spin leads to wrong predictions of the torque on large particles with strong contributions of quadrupole and higher-order multipoles, especially in a structured incident field.
\end{abstract}

\maketitle
\section{Introduction}
The mechanical interaction between a structured optical beam and structured photonic matter is a very important subject to study, in both fundamental and applied research. Such interaction is often complex, and simple analytical models like a multipole theory of optical force and torque in a general (inhomogeneous) electromagnetic field can greatly help our understanding of the physics involved. 

Since the realisation of orbital angular momentum in a paraxial laser beam (related to a helical phase) in 1992, there has been much confusion and debate on the separation of angular momentum into its spin and orbital parts  \cite{SMBarnettJO2016,KYBliokhNJP2014}. In the strict sense, the spin and orbital parts of the angular momentum are not separately meaningful physical quantities though both of them have the unit of angular momentum. However, it is still possible to separate the total angular momentum into spin and orbital parts in a laboratory frame of reference such that both satisfy the proper continuity relations and are separately conserved quantities \cite{RPCameronNJP2012, KYBliokhNJP2014, SMBarnettJO2016}. Both being conserved quantities, their net flow into a volume can be associated with a torque, it is Ref. \cite{MNVesperinasPRA2015} that first proposed the interesting concept of deriving the optical torque from the separate spin and orbital parts of angular momentum. The analytical expression given in Ref. \cite{MNVesperinasPRA2015j2} of optical torque acting on a dipolar particle takes into account complex spatial structures of the incident electromagnetic field. Such a treatment, if done properly, should give more physical insights on the separation of spin and orbital angular momentum involved in the interaction of structured light and objects \cite{YELeeNanophoton2014,AAWuPRA2020}. 

One of the most interesting results of Ref. \cite{MNVesperinasPRA2015j2} is the theoretical prediction of a torque dependent on the gradient of helicity density. Studying early literature on optical torque since Ashkin's invention of optical tweezers \cite{AAshkinPRL1970, AAshkinOL1986}, a few pioneering works actually studied the optical torque on spherical particles exerted by an inhomogenous electromagnetic field based on generalised Mie theory \cite{PLMarstonPRA1984,SChangJOSAB1985,JPBartonJAP1989, SChangOC1998, ACanaguierDurandPRA2014}. In some of these early works \cite{SChangJOSAB1985, JPBartonJAP1989, SChangOC1998}, the helicity density gradient is generally nonzero in the incident optical beam. However, the dipolar components of optical torque in these early results \cite{SChangJOSAB1985, JPBartonJAP1989, SChangOC1998} do not seem to contain the `gradient' torque term as predicted in Ref. \cite{MNVesperinasPRA2015j2}, thus pointing to a contradiction.

In this work we intend to clarify this discrepancy and clear up the confusion. We re-derive, from first principles and with detailed derivations, both the traditional dipolar total torque using total AM flux, and the spin and orbital torque components based on the corresponding AM contributions, ensuring that their sum agrees with the total torque. We also test our derived analytical expressions against numerical integration, with exact agreement. We find that `gradient' torque terms indeed exist in the spin and orbital components separately, but we present corrected prefactors, such that upon adding them, they cancel out, and the `gradient' term vanishes from the total torque, reconciling literature results. The concept proposed in \cite{MNVesperinasPRA2015j2} is still a very powerful tool to study the separate SAM and OAM contributions of optical torque, providing many advantages in understanding optical manipulations in the interacting structured light and photonic nanostructures. 

In the second part of the work, we derive the analytical expression of the optical torque acting on an isotropic electromagnetic quadrupole in a general (inhomogeneous) electromagnetic field. With the analytical result, we show that the quadrupolar torque on an isotropic Mie particle is still proportional to the absorption cross section, confirming the transfer of the angular momentum to mechanical action through absorption. However, the relevant physical property of the incident beam that must be used to indicate the orientation of the optical torque is not the spin of the electromagnetic field, but the spin of the EM field gradient. Using the two-wave interference as an example, we show that there are significant differences between the spin of the electromagnetic field and the spin of the electromagnetic field gradient. As a result, simply using the local EM field spin can give rise to wrong predictions on the orientation of optical torque on large particles with strong quadrupole resonance. We show that an extraordinary transverse spin of the magnetic field gradient appears in a field formed by purely TM-polarised two wave interference, even though the magnetic field spin is zero. Furthermore, some of the non-intuitive ‘negative’ torque that arises in the two-wave interference is often due to an incorrect interpretation of the physical properties related to the optical torque acting on a quadrupole and other higher-order multipoles.

\section{Optical torque and Angular Momentum}
It is well known that electromagnetic fields can carry linear and angular momentum, and that angular momentum is a conserved vector quantity. Consequently, if an electromagnetic field shows a net flow of electromagnetic angular momentum constantly flowing into a volume containing a material object, we can conclude that the 'missing' angular momentum is being transferred to the object via a mechanical torque.

The flux density of the electromagnetic angular momentum at every point in space is given by a tensor denoted as $\langle\overleftrightarrow{\mathbf{M}}\rangle$. This is a tensor because it contains information about the flow of electromagnetic angular momentum (itself a vector quantity) along each spatial direction. The time-averaged mechanical torque vector $\mathbf{\Gamma}$ can therefore be calculated as the total flux integral of the time-averaged angular momentum flux density $\langle\overleftrightarrow{\mathbf{M}}\rangle$ over a closed surface surrounding the object, as:
\begin{equation}
\begin{aligned}
\mathbf{\Gamma}=\varoiint \langle\overleftrightarrow{\mathbf{M}}\rangle \cdot d\mathbb{S},
\end{aligned}
\end{equation}

In turn, the angular momentum flux is calculated as $\langle\overleftrightarrow{\mathbf{M}}\rangle=\mathbf{r}\times\langle\overleftrightarrow{\mathbf{T}}\rangle$ where $\langle\overleftrightarrow{\mathbf{T}}\rangle$ represents the time-averaged flux density of electromagnetic linear momentum and is referred to as Maxwell's Stress Tensor $\langle\overleftrightarrow{\mathbf{T}}\rangle$ of the total field:
\begin{equation}
\begin{aligned}
\langle\overleftrightarrow{\mathbf{T}}\rangle=\frac{1}{2}\Re\bigg\{&\varepsilon_0 \mathbf{E}_{\mathrm{tot}}\otimes\mathbf{E}_{\mathrm{tot}}^*+ \mu_0 \mathbf{H}_{\mathrm{tot}}\otimes\mathbf{H}_{\mathrm{tot}}^*-\frac{\varepsilon_0 |\mathbf{E}_{\mathrm{tot}}|^2+\mu_0 |\mathbf{H}_{\mathrm{tot}}|^2}{2}\overleftrightarrow{\mathbf{I}}\bigg\},
\end{aligned}
\end{equation}
where $\mathbf{r}=(\mathbf{r}'-\mathbf{r}_0)$, $\mathbf{r}_0$ is the location of the object, $\mathbf{r}'$ denotes a point on the surface $\mathbb{S}$, $\mathbf{E}_{\mathrm{tot}}=\mathbf{E}_{\mathrm{inc}}+\mathbf{E}_{\mathrm{sca}}$ and $\mathbf{H}_{\mathrm{tot}}=\mathbf{H}_{\mathrm{inc}}+\mathbf{H}_{\mathrm{sca}}$ are the total electromagnetic fields. 

In this report, we study the interaction between tiny objects and a time harmonic general (inhomogeneous) incident electromagnetic field. A time dependence of $e^{-i\omega t}$ is assumed.
Since the integration of the angular momentum flux is done relative to the centre of the object, the mechanical action on the object related to the optical torque corresponds to a rotation about the object's own centre.

The time-averaged spin angular momentum flux $\langle\overleftrightarrow{\mathbf{M}}^{s}\rangle$ and orbital angular momentum flux $\langle\overleftrightarrow{\mathbf{M}}^{o}\rangle$, which together make up the total angular momentum flux density $\langle\overleftrightarrow{\mathbf{M}}\rangle=\langle\overleftrightarrow{\mathbf{M}}^{s}\rangle+\langle\overleftrightarrow{\mathbf{M}}^{o}\rangle$, can be separately written as \cite{KYBliokhNJP2014,MNVesperinasPRA2015j2}, 
\begin{equation}
\begin{aligned}
\langle\overleftrightarrow{\mathbf{M}}^{s}\rangle=\frac{1}{2\omega}\Im\Big\{&\mathbf{E}_{\mathrm{tot}}\otimes\mathbf{H}_{\mathrm{tot}}^*+\mathbf{H}_{\mathrm{tot}}^*\otimes\mathbf{E}_{\mathrm{tot}}-\left(\mathbf{E}_{\mathrm{tot}}\cdot\mathbf{H}_{\mathrm{tot}}^*\right)\overleftrightarrow{\mathbf{I}}\Big\},
\end{aligned}
\end{equation}
\begin{equation}
\begin{aligned}
\langle\overleftrightarrow{\mathbf{M}}^{o}\rangle=\frac{1}{4\omega}\Im\Big\{&\mathbf{E}_{\mathrm{tot}}^*\otimes\mathbf{H}_{\mathrm{tot}}+\mathbf{H}_{\mathrm{tot}}\otimes\mathbf{E}_{\mathrm{tot}}^*+\left[(\mathbf{r}\times\nabla)\otimes\mathbf{E}_{\mathrm{tot}}^*\right]\times\mathbf{H}_{\mathrm{tot}}+\left[(\mathbf{r}\times\nabla)\otimes\mathbf{H}_{\mathrm{tot}}\right]\times\mathbf{E}_{\mathrm{tot}}^*\Big\},
\end{aligned}
\end{equation}

The optical torque $\mathbf{\Gamma}^s$ attributed to the spin angular momentum flux can be calculated by 
\begin{equation}
\begin{aligned}\label{eq:spinSTtorque}
\mathbf{\Gamma}^s=&\varoiint \langle\overleftrightarrow{\mathbf{M}}^{s}\rangle\cdot \mathrm{d}\mathbb{S},
\end{aligned}
\end{equation}
while the optical torque $\mathbf{\Gamma}^o$ attributed to the orbital angular momentum flux can be calculated by
\begin{equation}
\begin{aligned}\label{eq:orbitalSTtorque}
\mathbf{\Gamma}^o=&\varoiint \langle\overleftrightarrow{\mathbf{M}}^{o}\rangle\cdot \mathrm{d}\mathbb{S},
\end{aligned}
\end{equation}
such that they represent two physically distinct parts of the total torque $\mathbf{\Gamma}=\mathbf{\Gamma}^s+\mathbf{\Gamma}^o$. Like the optical torque using the total angular momentum flux, the torque $\mathbf{\Gamma}^o$ arising from the orbital angular momentum flux depends on a certain reference point as $\langle\overleftrightarrow{\mathbf{M}}^{o}\rangle$ has components that are dependent on the position vector $\mathbf{r}$. As before, this reference point is often considered at the object's own centre when one only considers the optical torque resulting in self-rotation of the object. However, the optical torque $\mathbf{\Gamma}^s$ attributed to the spin angular momentum flux does not have this dependence and therefore this torque can be calculated for any choice of coordinate origin. 

\section{Prediction of a `gradient' torque}
We first follow the formulism in Ref. \cite{MNVesperinasOL2015} and write down the analytical expression (in SI units) of optical dipolar torque $\mathbf{\Gamma}_{\mathrm{N}}$ derived from the total angular momentum flux,
\begin{equation}
\begin{aligned}
\mathbf{\Gamma}_{\mathrm{N}}=&\frac{1}{2}\Re\left\{\mathbf{p}^*\times\mathbf{E}_{\mathrm{inc}}\right\}+\frac{1}{2}\Re\left\{\mathbf{m}^*\times\mu_0\mathbf{H}_{\mathrm{inc}}\right\}-\frac{k^3}{12\pi\varepsilon_0}\Im\{\mathbf{p}^*\times\mathbf{p}\}-\frac{k^3\mu_0}{12\pi}\Im\{\mathbf{m}^*\times\mathbf{m}\}\\
&+\frac{3}{4\omega}\Im\left\{\frac{1}{\varepsilon_0}(\mathbf{p}\cdot\nabla)\mathbf{H}_{\mathrm{inc}}^*-(\mathbf{m}\cdot\nabla)\mathbf{E}_{\mathrm{inc}}^*\right\}
\end{aligned}
\end{equation}

We then follow the formulism in Ref. \cite{MNVesperinasPRA2015j2} and separate the total optical torque $\mathbf{\Gamma}_{\mathrm{N}}$ into a SAM related torque $\mathbf{\Gamma}_{\mathrm{N}}^{s}$ and OAM related torque $\mathbf{\Gamma}_{\mathrm{N}}^{o}$.  $\mathbf{\Gamma}_{\mathrm{N}}^{s}$ and $\mathbf{\Gamma}_{\mathrm{N}}^{o}$ are derived following Eq. \ref{eq:spinSTtorque} and Eq. \ref{eq:orbitalSTtorque}, from the spin angular momentum flux and the orbital angular momentum flux that satisfy separate conservation laws, 
\begin{equation}
\begin{aligned}
\mathbf{\Gamma}_{\mathrm{N}}^{s}=&\frac{1}{2}\Re\left\{\mathbf{p}^*\times\mathbf{E}_{\mathrm{inc}}\right\}+\frac{1}{2}\Re\left\{\mathbf{m}^*\times\mu_0\mathbf{H}_{\mathrm{inc}}\right\}-\frac{k^3}{24\pi\varepsilon_0}\Im\{\mathbf{p}^*\times\mathbf{p}\}-\frac{k^3\mu_0}{24\pi}\Im\{\mathbf{m}^*\times\mathbf{m}\}\\
&+\frac{1}{2\omega}\Im\left\{\frac{1}{\varepsilon_0}(\mathbf{p}\cdot\nabla)\mathbf{H}_{\mathrm{inc}}^*-(\mathbf{m}\cdot\nabla)\mathbf{E}_{\mathrm{inc}}^*\right\},
\end{aligned}
\end{equation}
\begin{equation}
\begin{aligned}
\mathbf{\Gamma}_{\mathrm{N}}^{o}=&-\frac{k^3}{24\pi\varepsilon_0}\Im\{\mathbf{p}^*\times\mathbf{p}\}-\frac{k^3\mu_0}{24\pi}\Im\{\mathbf{m}^*\times\mathbf{m}\}+\frac{1}{4\omega}\Im\left[\frac{1}{\varepsilon_0}(\mathbf{p}\cdot\nabla)\mathbf{H}_{\mathrm{inc}}^*-(\mathbf{m}\cdot\nabla)\mathbf{E}_{\mathrm{inc}}^*\right]
\end{aligned}
\end{equation}

For induced dipoles in an isotropic Mie particle as described in Appendix \ref{appendixA}, the total optical torque $\mathbf{\Gamma}_{\mathrm{N}}$ can further be separated into an intrinsic part $\mathbf{\Gamma}_{\mathrm{N}}^{int}$ that is closely related to the local spin density and an extrinsic part $\mathbf{\Gamma}_{\mathrm{N}}^{ext}$ closely related to the dipole moments and the local field gradient. 
\begin{equation}
\begin{aligned}
\mathbf{\Gamma}_{\mathrm{N}}^{int}=&\frac{1}{2}\Re\{\mathbf{p}^*\times\mathbf{E}_{\mathrm{inc}}\}-\frac{k^3}{12\pi\varepsilon_0}\Im\{\mathbf{p}^*\times\mathbf{p}\}+\frac{1}{2}\Re\{\mathbf{m}^*\times\mu_0\mathbf{H}_{\mathrm{inc}}\}-\frac{k^3\mu_0}{12\pi}\Im\left\{\mathbf{m}^*\times\mathbf{m}\right\}\\
=&\frac{6\pi}{k^3}\left[\Re(a_1)-|a_1|^2\right]\mathbf{s}^{\mathrm{e}} + \frac{6\pi}{k^3}[\Re(b_1)-|b_1|^2]\mathbf{s}^{\mathrm{m}},\\
\mathbf{s}^{\mathrm{e}}=&\frac{1}{2}\varepsilon_0\Im\left\{\mathbf{E}_{\mathrm{inc}}^*\times\mathbf{E}_{\mathrm{inc}}\right\},\\
\mathbf{s}^{\mathrm{m}}=&\frac{1}{2}\mu_0\Im\left\{\mathbf{H}_{\mathrm{inc}}^*\times\mathbf{H}_{\mathrm{inc}}\right\},\\
\end{aligned}
\end{equation}
where $a_1$ is the Mie coefficient for an induced isotropic electric dipole and $b_1$ is the Mie coefficient for an induced isotropic magnetic dipole, whose relations with the induced dipolar polarisabilities are given in Appendix \ref{appendixA}.

The extrinsic torque $\mathbf{\Gamma}_{\mathrm{N}}^{ext}$, as given in Ref. \cite{MNVesperinasPRA2015j2}, can be re-formulated using vector calculus identities,  
\begin{equation}
\begin{aligned}\label{eq:TNext}
\mathbf{\Gamma}_{\mathrm{N}}^{ext}=&\frac{3}{4\omega}\Im\left\{\frac{1}{\varepsilon_0}(\mathbf{p}\cdot\nabla)\mathbf{H}_{\mathrm{inc}}^*-(\mathbf{m}\cdot\nabla)\mathbf{E}_{\mathrm{inc}}^*\right\}\\
=&\frac{9\pi}{2k^4c_0}\Re\Big\{a_1(\mathbf{E}_{\mathrm{inc}}\cdot\nabla)\mathbf{H}_{\mathrm{inc}}^*-b_1(\mathbf{H}_{\mathrm{inc}}\cdot\nabla)\mathbf{E}_{\mathrm{inc}}^*\Big\}\\
=&-\frac{9\pi}{2k^4c_0}\frac{\Re\{a_1\}+\Re\{b_1\}}{2}\nabla\times\Re\{\mathbf{E}_{\mathrm{inc}}\times\mathbf{H}_{\mathrm{inc}}^*\}\\
&-\frac{9\pi}{2k^4c_0}\frac{\Im\{a_1\}+\Im\{b_1\}}{2}\nabla \Im\{\mathbf{E}_{\mathrm{inc}}\cdot\mathbf{H}_{\mathrm{inc}}^*\}\\
&+\frac{9\pi}{2k^4c_0}\frac{\Re\{a_1\}-\Re\{b_1\}}{2}\bigg\{\nabla\Re\{\mathbf{E}_{\mathrm{inc}}\cdot\mathbf{H}_{\mathrm{inc}}^*\}+\mu_0\Im\{\mathbf{H}_{\mathrm{inc}}^*\times\mathbf{H}_{\mathrm{inc}}\}-\varepsilon_0\Im\{\mathbf{E}_{\mathrm{inc}}^*\times\mathbf{E}_{\mathrm{inc}}\}\bigg\}\\
&+\frac{9\pi}{2k^4c_0}\frac{\Im\{a_1\}-\Im\{b_1\}}{2}\nabla\times\Im\{\mathbf{E}_{\mathrm{inc}}\times\mathbf{H}_{\mathrm{inc}}^*\},
\end{aligned}
\end{equation}

The total dipolar optical torque discussed in Ref. \cite{MNVesperinasPRA2015j2} should correspond to the mechanical action of a particle rotating about its own centre. However, the extrinsic part in the total dipolar torque expression implies that under certain conditions, i. e. $[\Im\{a_1\}+\Im\{b_1\}]\neq 0$, the torque is dependent on the gradient of the helicity density of the incident beam $\nabla \Im\{\mathbf{E}_{\mathrm{inc}}\cdot\mathbf{H}_{\mathrm{inc}}^*\}$. Based on this, Ref. \cite{MNVesperinasPRA2015j2} predicted the existence of the `gradient' torque. The gradient torque is said to exist as long as the incident beam has a non-zero gradient of helicity density $\nabla \Im\{\mathbf{E}_{\mathrm{inc}}\cdot\mathbf{H}_{\mathrm{inc}}^*\}$, and the nanoparticle has a non-zero imaginary part of the dipolar Mie coefficients $\Im\{a_1\}$ and $\Im\{b_1\}$. However, unlike a spinning torque that introduces a rotation around the particle's own centre, or a revolution torque that introduces a rotation around a fixed reference point, it is difficult to interpret what mechanical action the gradient torque introduces on the dipolar particle.

In some of the early works \cite{SChangJOSAB1985, JPBartonJAP1989, SChangOC1998} on optical torque, the helicity density gradient is generally nonzero in the incident optical beam. Therefore, according to the prediction of Ref. \cite{MNVesperinasPRA2015j2}, there should exist a gradient torque in these early results. However, the dipolar components of the optical torque in \cite{SChangJOSAB1985, JPBartonJAP1989, SChangOC1998} only show dependence on the absorption cross section of the dipole coefficients, i.e. $\Re\{a_1\}-|a_1|^2$ and $\Re\{b_1\}-|b_1|^2$. None of them has a dependence on $\Im\{a_1\}$ and $\Im\{b_1\}$ which are linked to the existence of a gradient torque. These early results seem to contradict the gradient torque prediction in Ref. \cite{MNVesperinasPRA2015j2}. Ref. \cite{YJiangArxiv2015} also presented a multipolar theory of optical torque on an isotropic Mie particle in a general time-harmonic electromagnetic field. The analytical expression of dipolar torque in \cite{YJiangArxiv2015} is exactly the same as the intrinsic torque $\mathbf{\Gamma}_{\mathrm{N}}^{int}$ in Ref. \cite{MNVesperinasPRA2015j2}. Just as the intrinsic torque $\mathbf{\Gamma}_{\mathrm{N}}^{int}$, the dipolar torque derived in \cite{YJiangArxiv2015} depends only on the absorption cross section and the local spin density, agreeing with previous results \cite{SChangJOSAB1985, JPBartonJAP1989, SChangOC1998}. However, the analytical expression in \cite{YJiangArxiv2015} does not include extrinsic torque $\mathbf{\Gamma}_{\mathrm{N}}^{ext}$, and thus no torque that depends on the gradient of helicity density, again pointing to a contradiction. 

In this work, we try to clarify this discrepancy and confusion on gradient torque. We rederive from first principles, with detailed calculations in the Appendices, the analytical expressions of the dipolar torque both from the total angular momentum flux but also from the separate spin and orbital AM fluxes, ensuring that their sum exactly matches with the result obtained from the total AM flux. We also test our analytical expressions for the total torque against numerical integration of the angular flux density, finding exact agreement. Our results on total dipolar torque agree with \cite{YJiangArxiv2015}, and contain only the intrinsic torque $\mathbf{\Gamma}_{\mathrm{N}}^{int}$ but not the extrinsic torque $\mathbf{\Gamma}_{\mathrm{N}}^{ext}$ presented in \cite{MNVesperinasPRA2015j2}. However, in our derivation of optical torques from the separate spin and orbital AM fluxes, we find that both the spin AM part and the orbital AM part of the dipolar torque contain components that are proportional to the extrinsic torque $\mathbf{\Gamma}_{\mathrm{N}}^{ext}$. To be specific, we obtained the same analytical expression for SAM related torque as $\mathbf{\Gamma}_{\mathrm{N}}^{s}$ in \cite{MNVesperinasPRA2015j2}. However, we obtained a different coefficient for the extrinsic part of the OAM related torque compared to the expression $\mathbf{\Gamma}_{\mathrm{N}}^{o}$ given in \cite{MNVesperinasPRA2015j2}. In our derivation, the extrinsic parts in the SAM and OAM related torques exactly cancel each other out, and thus do not show in the total dipolar torque, which allows their sum to match the torque calculated via the more traditional conservation of total AM.

\section{Analytical expression of the dipolar optical torque}\label{section:Ttotl}
In this section, we study the interaction between a time harmonic electromagnetic wave (described by its electric and magnetic fields $\mathbf{E}_{\mathrm{inc}}$ and $\mathbf{H}_{\mathrm{inc}}$) and a tiny particle that can be described by induced electromagnetic dipole moments. By definition, the torque can be calculated by integrating the total angular momentum flux $\langle\overleftrightarrow{\mathbf{M}}\rangle=\mathbf{r}\times\langle\overleftrightarrow{\mathbf{T}}\rangle$ over a closed surface centered at the origin, 
\begin{equation}
\begin{aligned}
\mathbf{\Gamma}=\varoiint \langle\overleftrightarrow{\mathbf{M}}\rangle \cdot d\mathbb{S},
\end{aligned}
\end{equation}

Without loss of generality, we consider this enclosed surface to be spherical. The time-averaged torque on the Mie particle can then be expressed as, 
\begin{equation}
\begin{aligned}\label{eq:TorqueAM}
\mathbf{\Gamma}=&\varoiint \langle\overleftrightarrow{\mathbf{M}}\rangle \cdot d\mathbb{S}\\
=&\int_0^{2\pi}\int_{0}^{\pi} \mathbf{r}\times\Big(\langle\overleftrightarrow{\mathbf{T}}\rangle\cdot\hat{\mathbf{n}}\Big) r^2\sin\theta \mathrm{d}\theta \mathrm{d}\phi,\\
=&\Re \int_0^{2\pi}\int_{0}^{\pi} (r\hat{\mathbf{n}})\times\Big\{\frac{\varepsilon_0}{2}\mathbf{E}_{\mathrm{tot}}\left(\mathbf{E}_{\mathrm{tot}}^*\cdot\hat{\mathbf{n}}\right)+\frac{\mu_0}{2}\mathbf{H}_{\mathrm{tot}}\left(\mathbf{H}_{\mathrm{tot}}^*\cdot\hat{\mathbf{n}}\right)\Big\}  r^2\sin\theta \mathrm{d}\theta \mathrm{d}\phi,
\end{aligned}
\end{equation}
where $r$ is the radius of the spherical surface and $\hat{\mathbf{n}}$ is the outward radial unit vector normal to the surface. 

Given the induced electric dipole moment $\mathbf{p}$, the corresponding radiation field $\mathbf{E}_{\mathrm{p}}$ and $\mathbf{H}_{\mathrm{p}}$ can be analytically calculated, as expressed in appendix \ref{appendixB}. Knowing the total electromangetic field $\mathbf{E}_{\mathrm{tot}}=\mathbf{E}_{\mathrm{inc}}+\mathbf{E}_{\mathrm{p}}$ and $\mathbf{H}_{\mathrm{tot}}=\mathbf{H}_{\mathrm{inc}}+\mathbf{H}_{\mathrm{p}}$, the analytical expression of optical torque on an induced dipole by a general time-hamonic electromagnetic field can be derived, as detailed in Appendix \ref{appendixC} based on the angular momentum flux of the total field.

The dipolar torque $\mathbf{\Gamma}_{\mathrm{p}}$, attributed to the interaction between induced electric dipole and incident field, can be decomposed into different parts, 
\begin{equation}
\begin{aligned}\label{eq:edTorque_components}
\mathbf{\Gamma}_{\mathrm{p}}=\mathbf{\Gamma}_{\mathrm{inc}}+\mathbf{\Gamma}_{\mathrm{p,mix}}+\mathbf{\Gamma}_{\mathrm{p,recoil}}.
\end{aligned}
\end{equation}

From the derivation in Appendix \ref{appendixC}, the torque component purely dependent on the incident field is
\begin{equation}
\begin{aligned}
\mathbf{\Gamma}_{\mathrm{inc}}&=0,
\end{aligned}
\end{equation}
the extinction torque component $\mathbf{\Gamma}_{\mathrm{p, mix}}$, dependent on the interference between incident and radiation fields, is analytically given 
\begin{equation}
\begin{aligned}
\mathbf{\Gamma}_{\mathrm{p, mix}}&=\frac{1}{2}\Re\{\mathbf{p}^*\times\mathbf{E}_{\mathrm{inc}}\},
\end{aligned}
\end{equation}
and the recoil torque $\mathbf{\Gamma}_{\mathrm{p, recoil}}$, as a result of self-interaction of the induced electric dipole, is
\begin{equation}
\begin{aligned}
\mathbf{\Gamma}_{\mathrm{p, recoil}}&=-\frac{k^3}{12\pi\varepsilon_0}\Im\{\mathbf{p}^*\times\mathbf{p}\}. \\
\end{aligned}
\end{equation}

The total optical torque for an induced electric dipole in a general optical field can therefore be written as, 
\begin{equation}
\begin{aligned}
\mathbf{\Gamma}_{\mathrm{p}}=&\frac{1}{2}\Re\{\mathbf{p}^*\times\mathbf{E}_{\mathrm{inc}}\}-\frac{k^3}{12\pi\varepsilon_0}\Im\{\mathbf{p}^*\times\mathbf{p}\},
\end{aligned}
\end{equation}
In the special case of an isotropic electric dipole as described in Appendix \ref{appendixA}, one can easily relate the induced electric dipole $\mathbf{p}$ to the incident electric field $\mathbf{E}_{\mathrm{inc}}$, via the Mie polarizabilities, and therefore the optical torque can be written as a function of polarizabilities and incident fields only, as: 
\begin{equation}
\begin{aligned}
\mathbf{\Gamma}_{\mathrm{p}}=&\frac{6\pi}{k^3}[\Re(a_1)-|a_1|^2]\mathbf{s}^{\mathrm{e}},\\
\mathbf{s}^{\mathrm{e}}=&\frac{\varepsilon_0}{2}\Im\left\{\mathbf{E}_{\mathrm{inc}}^*\times\mathbf{E}_{\mathrm{inc}}\right\},
\end{aligned}
\end{equation}
which indicates that the optical torque for an isotropic electric dipole is proportional to the absorption cross section and the local spin density of the incident electric field. 

The analytical expression of the optical torque for a magnetic dipole can be derived in a similar manner,
\begin{equation}
\begin{aligned}
\mathbf{\Gamma}_{\mathrm{m}}=\frac{1}{2}\Re\{\mathbf{m}^*\times\mu_0\mathbf{H}_{\mathrm{inc}}\}-\frac{k^3\mu_0}{12\pi}\Im\{\mathbf{m}^*\times\mathbf{m}\},
\end{aligned}
\end{equation}
and in the case of an isotropic magnetic dipole,
\begin{equation}
\begin{aligned}\label{eq:magneticfieldgradientspin}
\mathbf{\Gamma}_{\mathrm{m}}=&\frac{6\pi}{k^3}[\Re(b_1)-|b_1|^2]\mathbf{s}^{\mathrm{m}},\\
\mathbf{s}^{\mathrm{m}}=&\frac{\mu_0}{2}\Im\left\{\mathbf{H}_{\mathrm{inc}}^*\times\mathbf{H}_{\mathrm{inc}}\right\}.
\end{aligned}
\end{equation}

As we can see from the above expressions, the total optical torque on isotropic electric and magnetic dipoles derived from the conservation of total AM don't have the extrinsic torque terms as in Ref. \cite{MNVesperinasOL2015,MNVesperinasPRA2015j2}. The transfer from spin to torque is purely through absorption, and the only relevant property of the incident field is its local spin angular momentum even in the case of a complex incident optical field. 

\section{Dipolar torque and conservation laws for spin and orbital parts of the angular momentum}\label{section:Ttso}
In this section, we follow the optical torque calculation devised from the separate conservation laws for the spin and orbital parts of angular momentum, first outlined in \cite{MNVesperinasPRA2015,KYBliokhNJP2014}, and we will see that after a careful derivation, their sum will equal the total torque derived in the previous section. 

The dipolar optical torque attributed to the spin angular momentum flux $\mathbf{\Gamma}^s$ and the torque attributed to the orbital angular momentum flux $\mathbf{\Gamma}^o$ can be calculated by integrating the corresponding angular momentum flux as shown Eq. \ref{eq:spinSTtorque} and Eq. \ref{eq:orbitalSTtorque}. Based on this method, the spin and orbital torques of a general electromagnetic field acting on an induced electric dipole can be analytically derived as detailed in Appendix \ref{appendixD},
\begin{equation}
\begin{aligned}
&\mathbf{\Gamma}^s_{\mathrm{p,mix}}=\frac{1}{2}\Re\left\{\mathbf{p}^*\times\mathbf{E}_{\mathrm{inc}}\right\}+\frac{1}{2\omega\varepsilon_0}\Im\left[(\mathbf{p}\cdot\nabla)\mathbf{H}_{\mathrm{inc}}^*\right],\\
&\mathbf{\Gamma}^o_{\mathrm{p,mix}}=-\frac{1}{2\omega\varepsilon_0}\Im\left[(\mathbf{p}\cdot\nabla)\mathbf{H}_{\mathrm{inc}}^*\right],\\
&\mathbf{\Gamma}^s_{\mathrm{p,recoil}}=-\frac{k^3}{24\pi\varepsilon_0}\Im\left(\mathbf{p}^*\times\mathbf{p}\right),\\
&\mathbf{\Gamma}^o_{\mathrm{p,recoil}}=-\frac{k^3}{24\pi\varepsilon_0}\Im\left(\mathbf{p}^*\times\mathbf{p}\right),
\end{aligned}
\end{equation}

Similarly, the `spin' and `orbital' optical torque of a general electromagnetic field acting on an induced magnetic dipole can be analytically derived as,
\begin{equation}
\begin{aligned}
&\mathbf{\Gamma}^s_{\mathrm{m,mix}}=\frac{\mu_0}{2}\Re\left\{\mathbf{m}^*\times\mathbf{H}_{\mathrm{inc}}\right\}-\frac{1}{2\omega}\Im\left\{(\mathbf{m}\cdot\nabla)\mathbf{E}_{\mathrm{inc}}^*\right\},\\
&\mathbf{\Gamma}^o_{\mathrm{m,mix}}=+\frac{1}{2\omega}\Im\left\{(\mathbf{m}\cdot\nabla)\mathbf{E}_{\mathrm{inc}}^*\right\},\\
&\mathbf{\Gamma}^s_{\mathrm{m,recoil}}=-\frac{k^3\mu_0}{24\pi}\Im\left\{\mathbf{m}^*\times\mathbf{m}\right\},\\
&\mathbf{\Gamma}^o_{\mathrm{m,recoil}}=-\frac{k^3\mu_0}{24\pi}\Im\left\{\mathbf{m}^*\times\mathbf{m}\right\},
\end{aligned}
\end{equation}

By adding up both the spin and orbital angular momentum flux contributions, the optical torque on an electromagnetic dipole, described by an electric dipole moment $\mathbf{p}$ and a magnetic dipole moment $\mathbf{m}$, is given as
\begin{equation}
\begin{aligned}
&\mathbf{\Gamma}_{\mathrm{p,mix}}=\mathbf{\Gamma}^s_{\mathrm{p,mix}}+\mathbf{\Gamma}^o_{\mathrm{p,mix}}=\frac{1}{2}\Re\left\{\mathbf{p}^*\times\mathbf{E}_{\mathrm{inc}}\right\},\\
&\mathbf{\Gamma}_{\mathrm{p,recoil}}=\mathbf{\Gamma}^s_{\mathrm{p,recoil}}+\mathbf{\Gamma}^o_{\mathrm{p,recoil}}=-\frac{k^3}{12\pi\varepsilon_0}\Im\left\{\mathbf{p}^*\times\mathbf{p}\right\}, \\
&\mathbf{\Gamma}_{\mathrm{m,mix}}=\mathbf{\Gamma}^s_{\mathrm{m,mix}}+\mathbf{\Gamma}^o_{\mathrm{m,mix}}=\frac{\mu_0}{2}\Re\left\{\mathbf{m}^*\times\mathbf{H}_{\mathrm{inc}}\right\},\\
&\mathbf{\Gamma}_{\mathrm{m,recoil}}=\mathbf{\Gamma}^s_{\mathrm{m,recoil}}+\mathbf{\Gamma}^o_{\mathrm{m,recoil}}=-\frac{k^3\mu_0}{12\pi}\Im\left\{\mathbf{m}^*\times\mathbf{m}\right\},
\end{aligned}
\end{equation}
The resulting total dipolar torque
\begin{equation}
\begin{aligned}
\mathbf{\Gamma}_{\mathrm{d}}=&\mathbf{\Gamma}_{\mathrm{p,mix}}+\mathbf{\Gamma}_{\mathrm{p,recoil}}+\mathbf{\Gamma}_{\mathrm{m,mix}}+\mathbf{\Gamma}_{\mathrm{m,recoil}},\\
=&\frac{1}{2}\Re\left\{\mathbf{p}^*\times\mathbf{E}_{\mathrm{inc}}\right\}-\frac{k^3}{12\pi\varepsilon_0}\Im\left\{\mathbf{p}^*\times\mathbf{p}\right\}+\frac{\mu_0}{2}\Re\left\{\mathbf{m}^*\times\mathbf{H}_{\mathrm{inc}}\right\}-\frac{k^3\mu_0}{12\pi}\Im\left\{\mathbf{m}^*\times\mathbf{m}\right\},
\end{aligned}
\end{equation}
agrees with the analytical expressions of dipolar torques devised from total angular momentum flux. The `orbital' optical torque on a source of combined electric and magnetic dipole has the same structure as the extrinsic torque $\mathbf{\Gamma}_{\mathrm{N}}^{\mathrm{ext}}$ in Ref. \cite{MNVesperinasPRA2015j2}, apart from a different coefficient. However, as it can be seen, the difference means that this `extrinsic' type of torque, which includes the `gradient torque', is exactly cancelled when adding the spin and orbital angular momentum contributions together.

\section{On the (non-) existence of `gradient' torque}
As discussed previously, Ref. \cite{MNVesperinasPRA2015j2} predicts the existence of a gradient torque, i.e. an optical torque arises from a non-zero gradient of helicity density and is depedent on the imaginary part of the dipolar Mie coefficients $\Im(a_1)$ and $\Im(b_1)$. The existence of a gradient torque challenges our understanding of the nature of optical torque. Conventionally, we know that optical torque introduces a mechanical action on an object being either rotation around its own centre or revolution around a fixed reference point. However, it is difficult to interpret what mechanical action the gradient torque introduces on the object. We try to examine this with an electromagnetic field designed such that there are no torques except for the gradient torque, if it exists at all.

We consider the special case of an electromagnetic field built up by the coherent interference of multiple $N$ circularly polarised plane waves with constant radial wavevector $k_{\rho}=k$ and evenly distributed over the full $2\pi$ azimuthal directions in the $z=0$ plane. Each plane wave has an electric field distribution $\mathbf{E}_{v}(\phi_v)=(\hat{\mathbf{e}}_p+i\hat{\mathbf{e}}_s)\frac{E_0}{N}\mathrm{exp}\big\{i(k\hat{\mathbf{e}}_{v})\cdot(\rho\hat{\mathbf{e}}_{\rho})\big\}$, where $\hat{\mathbf{e}}_s (\phi_v)=-\hat{\mathbf{e}}_{\phi_v}=\sin\phi_{v}\hat{\mathbf{e}}_{x}-\cos\phi_{v}\hat{\mathbf{e}}_{y}$ and $\hat{\mathbf{e}}_p (\phi_v)=-\hat{\mathbf{e}}_{z}$ are the unit vectors for Transverse Magnetic (TM or $p$-) and Transverse Electric (TE or $s$-) polarisation, and $\hat{\mathbf{e}}_{v}=-\cos\phi_{v}\hat{\mathbf{e}}_{x}-\sin\phi_{v}\hat{\mathbf{e}}_{y}$. In the limit of infinitely many beams as $N\to\infty$, the electromagnetic field built up can be analytically calculated as, 
\begin{equation}
\begin{aligned}
\mathbf{E}_{\mathrm{inc}}(\rho,\phi,z)=&\frac{1}{2\pi}\int_{0}^{2\pi}\big[\hat{\mathbf{e}}_p(\phi_v)+i\hat{\mathbf{e}}_s(\phi_v)\big]E_0\mathrm{exp}\big\{ik\rho(\hat{\mathbf{e}}_{v}\cdot\hat{\mathbf{e}}_{\rho})\big\}\mathrm{d}\phi_v\\
=&\mathbf{\hat{e}}_{x}E_0 J_{1}(k\rho)\sin\phi-\mathbf{\hat{e}}_{y}E_0 J_{1}(k\rho)\cos\phi-\mathbf{\hat{e}}_{z}E_0J_{0}(k\rho),\\
\mathbf{H}_{\mathrm{inc}}(\rho,\phi,z)=&-\mathbf{\hat{e}}_{x}i\frac{E_0}{Z_0} J_{1}(k\rho)\sin\phi+\mathbf{\hat{e}}_{y}i\frac{E_0}{Z_0} J_{1}(k\rho)\cos\phi+\mathbf{\hat{e}}_{z}i\frac{E_0}{Z_0} J_{0}(k\rho),
\end{aligned}
\end{equation}
where $Z_0=\sqrt{\mu_0/\varepsilon_0}$ is the impedance of free space. 

This electromagnetic field has the following properties
\begin{equation}
\begin{aligned}
&\mathbf{E}_{\mathrm{inc}}\times\mathbf{H}_{\mathrm{inc}}^*=0,\\
&\Im\{\mathbf{E}_{\mathrm{inc}}^*\times\mathbf{E}_{\mathrm{inc}}\}=0,\,\, \Im\{\mathbf{H}_{\mathrm{inc}}^*\times\mathbf{H}_{\mathrm{inc}}\}=0,\\
&\mathbf{E}_{\mathrm{inc}}\cdot\mathbf{H}_{\mathrm{inc}}^*=i\frac{E_0^2}{Z_0}\left[J_1^2(k\rho)+J_0^2(k\rho)\right].
\end{aligned}
\end{equation}

Based on the analytical results in Ref. \cite{MNVesperinasPRA2015j2} as given in Eq. \ref{eq:TNext}, placing an isotropic magnetic dipole in the designed optical field would introduce an optical torque arising only from the gradient torque:
\begin{equation}
\begin{aligned}
\mathbf{\Gamma}_{\mathrm{N}}=&\mathbf{\Gamma}_{\mathrm{N}}^{s}+\mathbf{\Gamma}_{\mathrm{N}}^{o}\\
=&\mathbf{\Gamma}_{\mathrm{N}}^{ext}\\
=&-\frac{3}{4\omega}\Im\big[(\mathbf{m}\cdot\nabla)\mathbf{E}_{\mathrm{inc}}^*\big]\\
=&-\frac{9\pi}{4k^4c_0}\Im(b_1)\nabla \Im\{\mathbf{E}_{\mathrm{inc}}\cdot\mathbf{H}_{\mathrm{inc}}^*\}.
\end{aligned}
\end{equation}

\begin{figure}[!htp]
\centering
\includegraphics[width=0.7\textwidth]{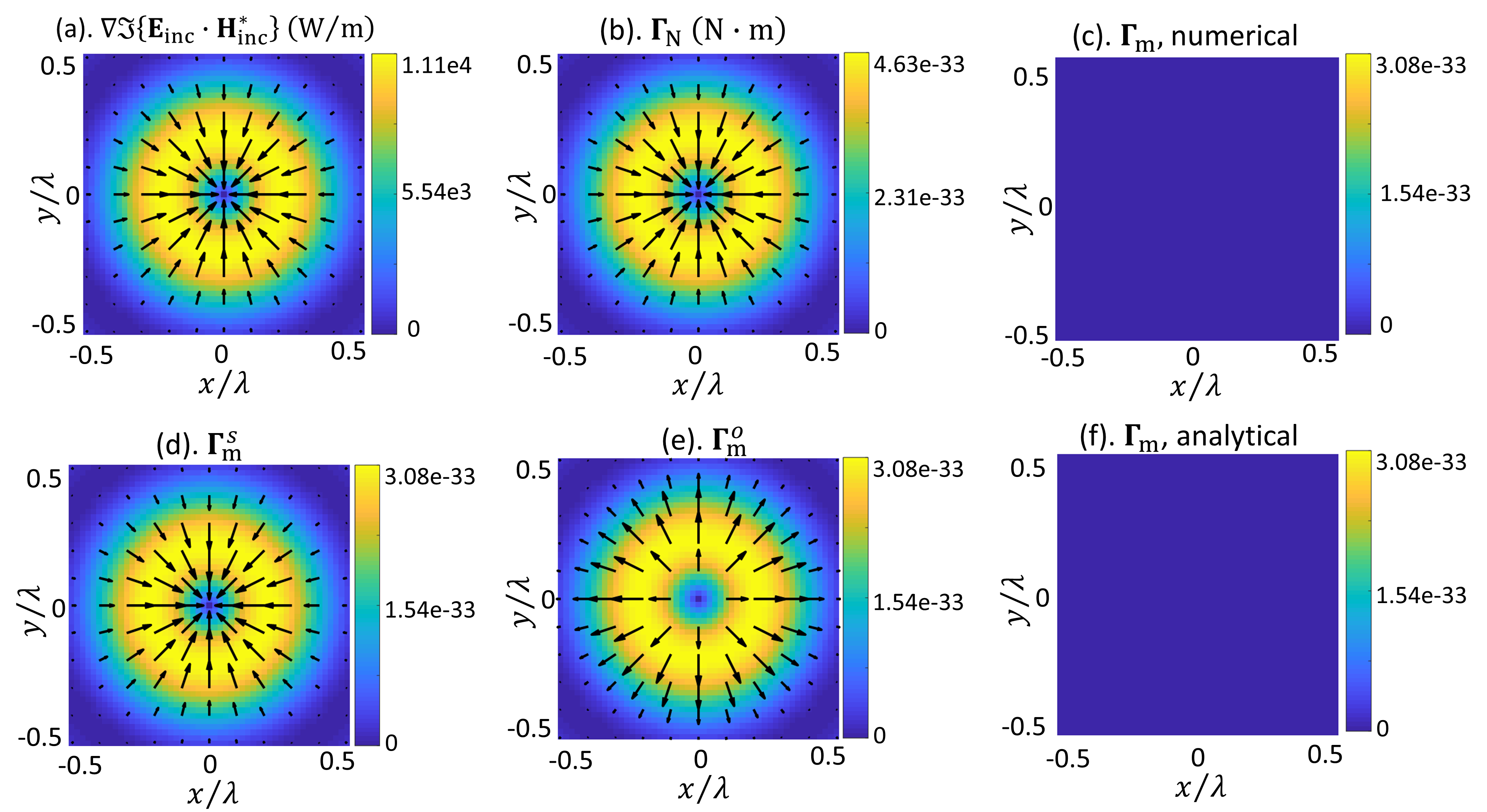}
\caption{(a) Gradient of helicity density of the designed optical field; (b) Total optical torque (unit $\mathrm{N}\cdot\mathrm{m}$) on a dipolar particle with Mie coefficient $b_1=0.969-0.173i$ in the designed optical field ($E_0=1$ V/m), based on the analytical results in Ref. \cite{MNVesperinasOL2015, MNVesperinasPRA2015j2}; (c) Optical torque on the same particle calculated based on the numerical integration of total angular momentum flux; Analytical results of the optical torque attributed to (d) the spin AM flux, (e) the orbital AM flux, and (f) the total AM flux.}
\label{fig:Torque_Nieto_MST}
\end{figure}

Fig. \ref{fig:Torque_Nieto_MST}(b) shows the optical torque distribution based on this analytical result where a dipolar particle with $b_1=0.969-0.173i$ is placed in the $z=0$ plane of the designed beam. Since $\Im\{\mathbf{H}_{\mathrm{inc}}^*\times\mathbf{H}_{\mathrm{inc}}\}=0$, the instrinsic parts of both the mixed and recoil dipolar torque are zero. The optical torque shown in Fig. \ref{fig:Torque_Nieto_MST}(b), calculated according to the expressions on Ref. \cite{MNVesperinasPRA2015j2}, has only an extrinsic part and depends on $\Im(b_1)$ and on the gradient of helicity density $\Im\{\mathbf{E}_{\mathrm{inc}}\cdot\mathbf{H}_{\mathrm{inc}}^*\}$ as shown in Fig. \ref{fig:Torque_Nieto_MST}(a). As a result, the analytical results of Ref. \cite{MNVesperinasPRA2015j2} give rise to a radially oriented torque across the incident beam in the $z=0$ plane. 

In order to test the existence of the gradient torque, we first calculate the optical torque numerically by integrating the total angular momentum flux $\langle\overleftrightarrow{\mathbf{M}}\rangle=\mathbf{r}\times\langle\overleftrightarrow{\mathbf{T}}\rangle$ over an enclosed surface surrounding the particle. The numerically calculated optical torque for the same dipole in the designed beam is shown in Fig. \ref{fig:Torque_Nieto_MST}(c), whose values are close to zero (down to numerical errors) across the beam. On the other hand, Fig. \ref{fig:Torque_Nieto_MST}(d)-(f) show the torque distributions attributed to the spin, orbital and total AM fluxes based on our analytical results given in sections \ref{section:Ttso} and \ref{section:Ttotl}. The numerically calculated torque based on total angular momentum flux in Fig. \ref{fig:Torque_Nieto_MST}(c) does not show the radially oriented `gradient' torque across the incident beam as predicted by by Ref. \cite{MNVesperinasPRA2015j2}. Instead, it agrees well with the `intrinsic' torque  in Fig. \ref{fig:Torque_Nieto_MST}(f) which is null and dependent on local spin density $\frac{\mu_0}{2}\Im\{\mathbf{H}^*\times\mathbf{H}\}=0$. This gradient torque structure does show up in the spin and orbital parts of the torque as in Fig. \ref{fig:Torque_Nieto_MST}(d)-(f). However, as discussed previously, the spin and orbital parts of the torque exactly cancel each other, leading to a null total optical torque.

\section{Analytical expression of the isotropic quadrupolar optical torque}
Given the electric quadrupole moment $\overleftrightarrow{\mathbf{Q}}^{\mathrm{e}}$, the corresponding radiation field $\mathbf{E}_{\mathrm{Qe}}$ and $\mathbf{H}_{\mathrm{Qe}}$ can be analytically calculated, as expressed in appendix \ref{appendixB}. Knowing the total electromagnetic field $\mathbf{E}_{\mathrm{tot}}=\mathbf{E}_{\mathrm{inc}}+\mathbf{E}_{\mathrm{Qe}}$ and $\mathbf{H}_{\mathrm{tot}}=\mathbf{H}_{\mathrm{inc}}+\mathbf{H}_{\mathrm{Qe}}$, the analytical expression of optical torque on an induced quadrupole by a general time-hamonic electromagnetic field can be derived, as detailed in Appendix \ref{appendixE} based on the angular momentum flux of the total field.

The quadrupolar torque $\mathbf{\Gamma}_{\mathrm{Qe}}$ and $\mathbf{\Gamma}_{\mathrm{Qm}}$, attributed to the interaction between an induced electric and magnetic quadrupole in an isotropic Mie particle and a general electromagnetic field, can be analytically expressed as, 
\begin{equation}
\begin{aligned}
\mathbf{\Gamma}_{\mathrm{Qe}}=&\frac{120\pi}{k^5}\left[\Re\{a_2\}-|a_2|^2\right]\mathbf{s}^{\mathrm{Qe}},\\
\mathbf{\Gamma}_{\mathrm{Qm}}=&\frac{120\pi}{k^5}\left[\Re\{b_2\}-|b_2|^2\right]\mathbf{s}^{\mathrm{Qm}},
\end{aligned}
\end{equation}
where,
\begin{equation}
\begin{aligned}\label{eq:fieldgradientspin}
&\mathbf{s}^{\mathrm{Qe}}=\sum_{u=x,y,z}\frac{\varepsilon_0}{6}\Im\big\{\mathbf{D}^{\mathrm{e}*}_{u}\times\mathbf{D}^{\mathrm{e}}_{u}\big\},\\
&\mathbf{s}^{\mathrm{Qm}}=\sum_{u=x,y,z}\frac{\mu_0}{6}\Im\left\{\mathbf{D}^{m*}_u\times\mathbf{D}^{\mathrm{m}}_u\right\},\\
&\mathbf{D}^{\mathrm{e}}_{u}=\sum_{v=x,y,z}\mathbf{\hat{e}}_{v}\left[\overleftrightarrow{\mathbcal{D}^\mathrm{e}}\right]_{uv},\,\,\,\mathbf{D}^{\mathrm{m}}_u=\sum_{v=x,y,z}\mathbf{\hat{e}}_{v}\left[\overleftrightarrow{\mathbcal{D}^\mathrm{m}}\right]_{uv},\\
&\left[\overleftrightarrow{\mathbcal{D}^\mathrm{e}}\right]_{uv}=\frac{\partial_{u}E_{v}+\partial_{v}E_{u}}{2},\,\,\,\left[\overleftrightarrow{\mathbcal{D}^{\mathrm{m}}}\right]_{uv}=\frac{\partial_{u}H_{v}+\partial_{v}H_{u}}{2}. 
\end{aligned}
\end{equation}

In previous literature, the local EM field spin densities $\mathbf{s}^{\mathrm{e}}$ and $\mathbf{s}^{\mathrm{m}}$ are often used to indicate the orientation of the optical torque \cite{JChenSC2014, MNVesperinasOL2015, MNVesperinasPRA2015j2, GTkachenkoOptica2020}. Just like the dipolar case, the optical torque acting on a quadrupole are proportional to the corresponding quadrupolar absorption cross sections $\Re\{a_2\}-|a_2|^2$ and $\Re\{b_2\}-|b_2|^2$. However, the relevant physical properties of the incident beam is not the spin densities of the EM field $\mathbf{s}^{\mathrm{e}}$ and $\mathbf{s}^{\mathrm{m}}$, but rather the spin densities of the EM field gradient $\mathbf{s}^{\mathrm{Qe}}$ and $\mathbf{s}^{\mathrm{Qm}}$. As a result, simply using the local EM field spin densities can give rise to wrong predictions on the orientation and magnitude of optical torque.

The EM field spin vectors $\big\{\mathbf{s}^{\mathrm{e}}, \mathbf{s}^{\mathrm{m}}\big\}$ and the EM field gradient spin vectors $\big\{\mathbf{s}^{\mathrm{Qe}}, \mathbf{s}^{\mathrm{Qm}}\big\}$ can, perhaps unintuitively, show very different behaviours, including having opposite orientations. This can be shown through examples. In the remaining of this section, we will illustrate the differences in the very simple case of two-wave interference, but differences will appear in any general structured EM field.

The EM field, built up by two free-propagating plane waves along the wavevectors $\mathbf{k}_1=k_x\mathbf{\hat{e}}_{x}+k_z\mathbf{\hat{e}}_{z}$ and $\mathbf{k}_2=-k_x\mathbf{\hat{e}}_{x}+k_z\mathbf{\hat{e}}_{z}$, can be described in the transverse magnetic ($p-$) and transverse electric ($s-$) polarisation basis, 
\begin{equation}
\begin{aligned}
&\mathbf{E}_1=\left(E_1^p\mathbf{\hat{e}}_{1}^p+E_1^s\mathbf{\hat{e}}_1^s\right)\mathrm{exp}(ik_xx+ik_zz),\\
&\mathbf{E}_2=\left(E_2^p\mathbf{\hat{e}}_{2}^p+E_2^s\mathbf{\hat{e}}_2^s\right)\mathrm{exp}(i\varphi_0-ik_xx+ik_zz),\\
&\mathbf{H}_1=\frac{\mathbf{k}_1}{Z_0k}\times\mathbf{E}_1,\,\,\, \mathbf{H}_2=\frac{\mathbf{k}_2}{Z_0k}\times\mathbf{E}_2,\\
&\mathrm{where},\\
&\mathbf{\hat{e}}_{1}^s=\mathbf{\hat{e}}_{y},\,\,\,\mathbf{\hat{e}}_{2}^s=-\mathbf{\hat{e}}_{y},\,\,\,\mathbf{\hat{e}}_1^p=\mathbf{\hat{e}}_1^s\times\frac{\mathbf{k}_1}{k},\,\,\, \mathbf{\hat{e}}_2^p=\mathbf{\hat{e}}_2^s\times\frac{\mathbf{k}_2}{k},
\end{aligned}
\end{equation}

Here we focus on the field properties that are related to the optical torque in dipoles and quadrupoles. The field spin $\mathbf{s}^{\mathrm{m}}$ (as given in Eq. \ref{eq:magneticfieldgradientspin}) and the field gradient spin $\mathbf{s}^{\mathrm{Qm}}$ (as given in Eq. \ref{eq:fieldgradientspin}) are evaluated on the $z=0$ plane:
\begin{equation}
\begin{aligned}\label{eq:twowavedQs}
\mathbf{s}^{\mathrm{m}}_x=&\frac{k_x\varepsilon_0}{k}\Im\big\{E_1^{p*}E_1^s-E_{2}^{p*}E_{2}^{s}\big\}+\frac{k_x\varepsilon_0}{k}\Im\Big\{\big(E_1^{p*}E_{2}^{s}+E_{1}^{s*}E_{2}^{p}\big)\mathrm{exp}(i\varphi_0-2ik_xx)\Big\},\\
\mathbf{s}^{\mathrm{m}}_y=&\frac{2k_xk_z\varepsilon_0}{k^2}\Im\big\{E_{1}^{s*}E_{2}^{s}\mathrm{exp}(i\varphi_0-2ik_xx)\big\},\\
\mathbf{s}^{\mathrm{m}}_z=&\frac{k_z\varepsilon_0}{k}\Im\big\{E_{1}^{p*}E_{1}^{s}+E_{2}^{p*}E_{2}^{s}\big\}+\frac{k_z\varepsilon_0}{k}\Im\big\{(E_{1}^{s*}E_{2}^{p}-E_{1}^{p*}E_{2}^{s})\mathrm{exp}(i\varphi_0-2ik_xx)\big\},\\
\mathbf{s}^{\mathrm{Qm}}_x=&\frac{k_xk\varepsilon_0}{12}\Im\big\{E_{1}^{p*}E_{1}^{s}-E_{2}^{p*}E_{2}^{s}\big\}+\frac{k_x(3k_z^2-k_x^2)\varepsilon_0}{12k}\Im\big\{(E_{1}^{p*}E_{2}^{s}+E_{1}^{s*}E_{2}^{p})\mathrm{exp}(i\varphi_0-2ik_xx)\big\},\\
\mathbf{s}^{\mathrm{Qm}}_y=&\frac{2k_xk_z(k_z^2-k_x^2)\varepsilon_0}{3k^2}\Im\big\{E_{1}^{s*}E_{2}^{s}\mathrm{exp}(i\varphi_0-2ik_xx)\big\}+\frac{k_xk_z\varepsilon_0}{6}\Im\big\{E_{1}^{p*}E_{2}^{p}\mathrm{exp}(i\varphi_0-2ik_xx)\big\},\\
\mathbf{s}^{\mathrm{Qm}}_z=&\frac{k_zk\varepsilon_0}{12}\Im\big\{E_{1}^{p*}E_{1}^{s}+E_{2}^{p*}E_{2}^{s}\big\}+\frac{k_z(k_z^2-3k_x^2)\varepsilon_0}{12k}\Im\big\{(E_{1}^{s*}E_{2}^{p}-E_{1}^{p*}E_{2}^{s})\mathrm{exp}(i\varphi_0-2ik_xx)\big\}.
\end{aligned}
\end{equation}

\begin{figure}[!htp]
\centering
\includegraphics[width=0.8\textwidth]{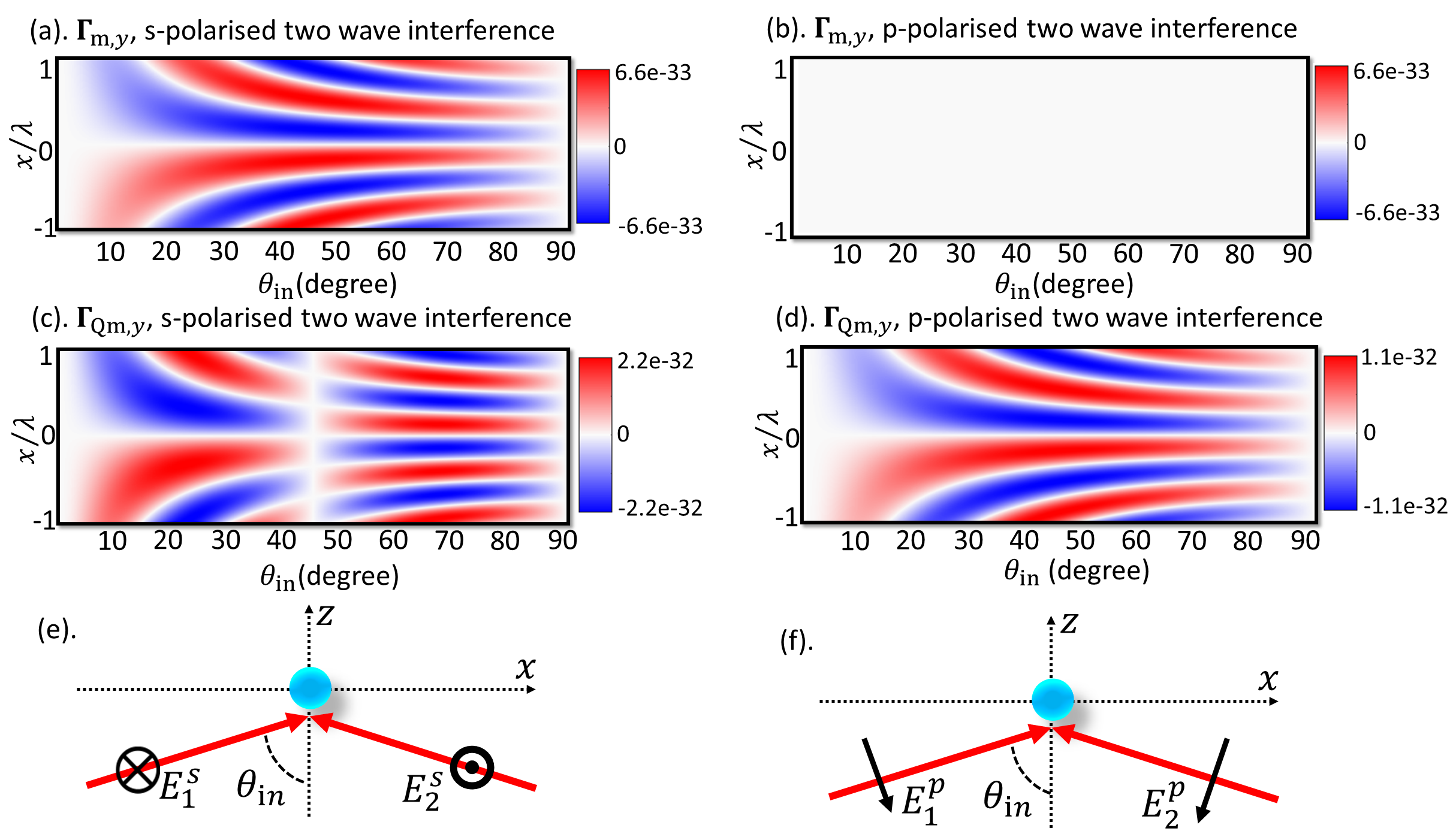}
\caption{Optical torques on an induced magnetic dipole and magnetic quadrupole (with equal Mie coeffcients $b_1=b_2=0.802-0.047i$) in an electromagnetic field formed by two-wave interference, at varying $x$ and incident angles $\theta_{\mathrm{in}}$. The $y$ component of the optical torque $\Gamma_{\mathrm{m},y}$ ($\mathrm{N}\cdot\mathrm{m}$) on a pure magnetic dipole in the EM field formed by (a) two $s$-polarised plane waves ($E_{1}^{p}=E_{2}^{p}=1$ V/m) and (b) two $p$-polarised plane waves ($E_{1}^{s}=E_{2}^{s}=1$ V/m); The $y$ component of the optical torque on a pure magnetic quadrupole $\Gamma_{\mathrm{Qm},y}$ in the EM field formed by (c) two $s$-polarised plane waves  and (d) two $p$-polarised plane waves; Illustrations of two-wave interference by (e) purely $s$-polarised plane waves and (f) purely $p$-polarised plane waves.}
\label{fig:ps_2wave_dQT}
\end{figure}

\begin{figure}[!htp]
\centering
\includegraphics[width=0.8\textwidth]{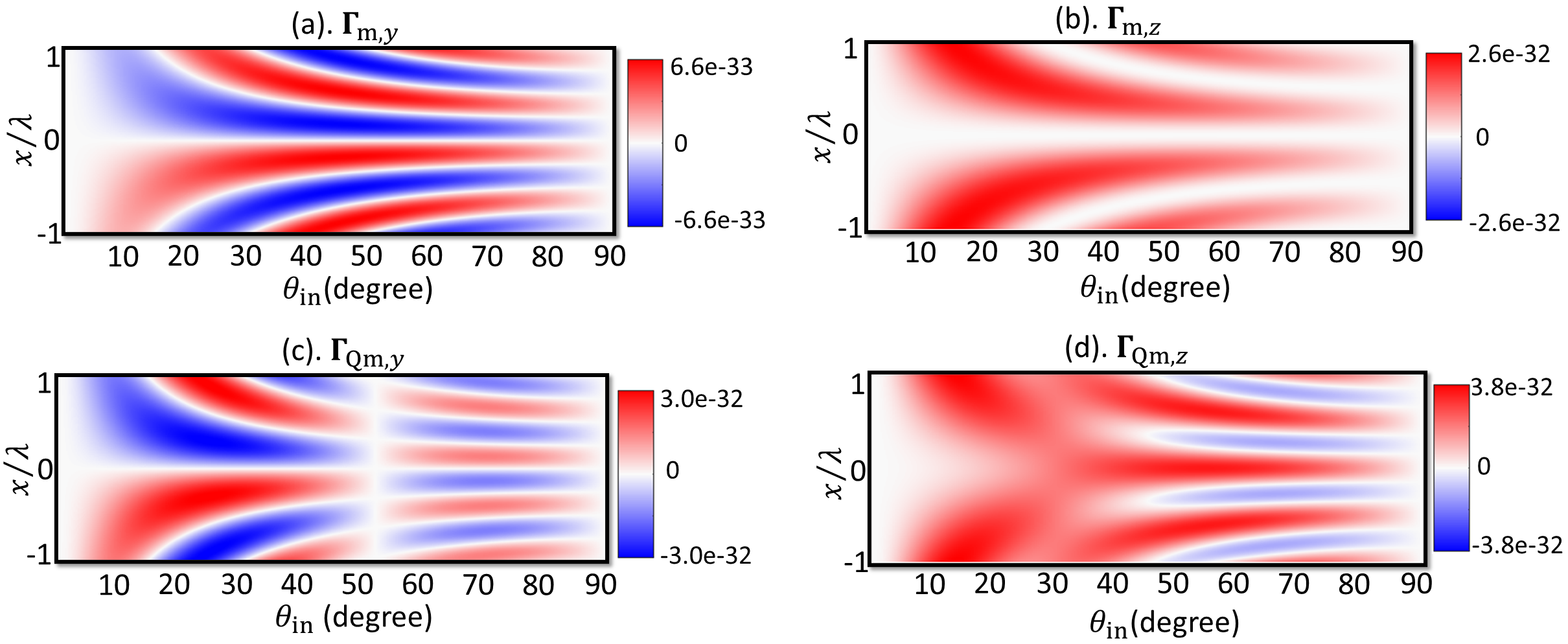}
\caption{Non-zero components of optical torque on an induced magnetic dipole and magnetic quadrupole (with equal Mie coeffcients $b_1=b_2=0.802-0.047i$), at varying $x$ and incident angles $\theta_{\mathrm{in}}$, in an electromagnetic field formed by two circularly polarised plane waves ($E_{1}^{p}=E_{2}^{p}=1$ V/m, $E_{1}^{s}=E_{2}^{s}=i$ V/m). (a) $y$ component of the optical torque $\Gamma_{\mathrm{m},y}$ on a pure magnetic dipole; (b) $z$ component of the optical torque $\Gamma_{\mathrm{m},z}$ on a pure magnetic dipole; (c) $y$ component of the optical torque $\Gamma_{\mathrm{Qm},y}$ on a pure magnetic quadrupole; (d) $z$ component of the optical torque $\Gamma_{\mathrm{Qm}, z}$ on a pure magnetic quadrupole.}
\label{fig:cpl_2wave_dQT}
\end{figure}

For an induced magnetic dipole with Mie coefficient $b_1$ in such a field, an optical torque $\mathbf{\Gamma}_{\mathrm{m}}=\frac{6\pi}{k^3}[\Re(b_1)-|b_1|^2]\mathbf{s}^{\mathrm{m}}$ will be exterted on the Mie particle. For an induced magnetic quadrupole with Mie coefficient $b_2$ in such a field, an optical torque $\mathbf{\Gamma}_{\mathrm{Qm}}=\frac{120\pi}{k^5}\left[\Re\{b_2\}-|b_2|^2\right]\mathbf{s}^{\mathrm{Qm}}$ is exterted on the Mie particle. In Fig. \ref{fig:ps_2wave_dQT} and Fig. \ref{fig:cpl_2wave_dQT}, we show a magnetic dipole and a magnetic quadrupole with equal Mie coefficients $b_1=b_2=0.802-0.047i$ in three types of EM field built up by two-wave interference: purely $p$-polarised plane waves ($E_{1}^{p}=E_{2}^{p}=1$ V/m, corresponding to an intensity of $1.3\times 10^{-3}\,\,\mathrm{W/m^2}$ in a single beam), purely $s$-polarised plane waves ($E_{1}^{s}=E_{2}^{s}=1$ V/m), and circularly polarised plane waves with the same helicity ($E_{1}^{p}=E_{2}^{p}=1$ V/m, $E_{1}^{s}=E_{2}^{s}=i$ V/m).

The differences between the field spin vector $\mathbf{s}^{\mathrm{m}}$ and the field gradient spin vector $\mathbf{s}^{\mathrm{Qm}}$ are shown in Eq. \ref{eq:twowavedQs} and clearly illustrated in Fig. \ref{fig:ps_2wave_dQT}. As shown in Eq. \ref{eq:twowavedQs} and Fig. \ref{fig:ps_2wave_dQT}(a), the electromagnetic field built up by two purely TE ($s$-)polarised plane waves introduces a transverse spin in the magnetic field, and results in an optical torque on an isotropic magnetic dipole in the transverse $y$ direction. The same EM field also exerts a transverse optical torque on a magnetic quadrupole. However, the optical torque acting on a magnetic quadrupole $\Gamma_{\mathrm{Qm},y}$ has a different dependence on incidence angle $\theta_{\mathrm{in}}$ (following the relations that $k_x=k\sin\theta_{\mathrm{in}}$ and $k_z=k\cos\theta_{\mathrm{in}}$) from its dipolar counterpart $\Gamma_{\mathrm{m},y}$. The different torques can be expressed analytically, together with their ratio.

\begin{equation}
\begin{aligned}\label{eq:twowavedQTs}
&\Gamma_{\mathrm{m},y}=\frac{6\pi}{k^3}\left[\Re(b_1)-|b_1|^2\right]\frac{2k_xk_z\varepsilon_0}{k^2}\Im\big\{E_{1}^{s*}E_{2}^{s}\mathrm{exp}(i\varphi_0-2ik_xx)\big\},\\
&\Gamma_{\mathrm{Qm},y}=\frac{120\pi}{k^5}\left[\Re\{b_2\}-|b_2|^2\right]\frac{2k_xk_z(k_z^2-k_x^2)\varepsilon_0}{3k^2}\Im\big\{E_{1}^{s*}E_{2}^{s}\mathrm{exp}(i\varphi_0-2ik_xx)\big\},\\
&\frac{\Gamma_{\mathrm{Qm},y}}{\Gamma_{m,y}}=\frac{20}{3}\frac{\left[\Re(b_2)-|b_2|^2\right]}{\left[\Re(b_1)-|b_1|^2\right]}\frac{(k_z^2-k_x^2)}{k^2}.
\end{aligned}
\end{equation}

When $\theta_{\mathrm{in}}$ is below $45^{\circ}$, the torque on a magnetic quadrupole $\Gamma_{\mathrm{Qm},y}$ and the dipolar torque $\Gamma_{\mathrm{m},y}$ are in phase, as $(k_z^2-k_x^2)/(k^2)>0$. Above $45^{\circ}$, the torque on a magnetic quadrupole $\Gamma_{\mathrm{Qm},y}$ and the dipolar torque $\Gamma_{\mathrm{m},y}$ are out of phase, as $(k_z^2-k_x^2)/(k^2)<0$. In other words, the optical torque acting on a magnetic quadrupole points along the opposite orientation of the local magnetic field spin vector $\mathbf{s}^{\mathrm{m}}$. If one uses the local field spin vector $\mathbf{s}^{\mathrm{m}}$ to indicate the orientation of the optical torque, this result might seem counter-intuitive and be interpreted as a `negative' torque \cite{JChenSC2014, MNVesperinasOL2015, MNVesperinasPRA2015j2, GTkachenkoOptica2020}. However, as we demonstrated, this counter-intuitive impression is due to the fact that the field spin vector shouldn't be used to predict the orientation of a quadrupolar torque in the first place. Instead, the quadrupolar torque is always aligned with the field gradient spin vector as the absorption cross section is always a positive value for an absorbing isotropic Mie particle, and thus a `positive' torque.

The difference between dipolar and quadrupolar torque is even starker in the case of an electromagnetic field built up by two purely TM ($p$-)polarised plane waves. This type of EM field does not have any magnetic field spin, and does not introduce optical torque on an isotropic magnetic dipole as shown in Fig. \ref{fig:ps_2wave_dQT}(b). However, a nonzero transverse spin of magnetic field gradient exists in such an electromagnetic field, and can introduce a strong optical torque to a magnetic quadrupole as shown in Fig. \ref{fig:ps_2wave_dQT}(d). It is also very interesting to notice the remarkable similarity between the dipolar torque for s-polarised waves and the quadrupolar torque for p-polarised waves (compare \ref{fig:ps_2wave_dQT}(a) with \ref{fig:ps_2wave_dQT}(d)).

The two types of EM fields based on purely $p$-polarised or purely $s$-polarised two-wave interference don't exhibit net spin densities when integrating the spin densities in either $x=0$, $y=0$ or $z=0$ plane. In Fig. \ref{fig:cpl_2wave_dQT}, we study an electromagnetic field set up by the interference of two circularly polarised plane waves ($E_{1}^{p}=E_{2}^{p}=1$ V/m, $E_{1}^{s}=E_{2}^{s}=i$ V/m), where there is a net non-zero spin density along the $z$ direction when integrating the EM field or field gradient spin in the $z=0$ plane. The $z$ component of the optical torque $\Gamma_{\mathrm{m},z}$ on a magnetic dipole shows the interference fringe pattern along $x$, but none of its values is is negative. Only positive values of torque are seen, aligned with the net magnetic field spin density of the incident waves along the $z$ direction, as shown in Fig. \ref{fig:cpl_2wave_dQT}(b). Meanwhile, the $z$ component of the optical torque $\Gamma_{\mathrm{Qm},z}$ on a magnetic quadrupole shows locally negative values compared to the net magnetic field spin density along $z$ direction as shown in Fig. \ref{fig:cpl_2wave_dQT}(d). They might be rightly interpreted as `negative' torques, with reference to the net spin densities of the magnetic field gradient integrated across the entire $z=0$ plane. Yet when referred to the magnetic field gradient spin vector, they are still `positive' torques for any absorbing isotropic Mie particle.

The results in Fig. \ref{fig:ps_2wave_dQT} and Fig. \ref{fig:cpl_2wave_dQT} have also been verified using numerical integration of the total angular momentum flux, and agree with the results of our analytical expressions.

\section{Conclusions}
In this report, we study the optical torque acting on an induced multipole in an isotropic Mie particle and a time harmonic general (inhomogeneous) electromagnetic field. 

We first try to address the confusion on the `extrinsic' part of the dipolar torque present in Ref.\cite{MNVesperinasPRA2015,MNVesperinasPRA2015j2}, including the prediction of a torque that is dependent on the gradient of helicity density. With detailed calculations in the appendices, we rederive the analytical expressions of the dipolar torque in a general electromagnetic field by exploring the separate conservation laws of the total, spin and orbital angular momenta, a method proposed by Ref.\cite{MNVesperinasPRA2015,MNVesperinasPRA2015j2}. We prove that the `gradient' torque does not exist in an isotropic dipole. Though the `extrinsic' type of terms exist in the analytical expressions of dipolar torque attributed separately to the spin and orbital parts of the angular momentum, they cancel exactly with each other in the total dipolar torque. 

In the second part of the work, we derive the analytical expression of the optical torque acting on an isotropic electromagnetic quadrupole in a general (inhomogeneous) electromagnetic field. With the analytical result, we show that the quadrupolar torque on an isotropic Mie particle is still proportional to the absorption cross section, confirming the transfer of the angular momentum to mechanical action through absorption. However, the relevant physical property of the incident beam that must be used to indicate the orientation of the optical torque is not the spin of the electromagnetic field, but the spin of the EM field gradient. Using the two-wave interference as an example, we show that there are significant differences between the spin of the electromagnetic field and the spin of the electromagnetic field gradient. As a result, simply using the local EM field spin can give rise to wrong predictions on the orientation of optical torque on large particles with strong quadrupole resonance. We show that an extraordinary transverse spin of the magnetic field gradient appears in a field formed by purely TM-polarised two wave interference, even though the magnetic field spin is zero. Furthermore, some of the non-intuitive ‘negative’ torque that arises in the two-wave interference is often due to an incorrect interpretation of the physical properties related to the optical torque acting on a quadrupole and other higher-order multipoles.

Considering the growing interest from the nanophotonics community on studying and understanding the mechanical interaction between structured light and structured materials, we felt the need to address some discrepancies in the literature. By using consistency checks, double-checking our results using different methods, and scrutinizing our expressions via examples, we have produced what we feel are trusty analytical expressions, which we hope serve the community as much as it served us to clear up some confusions and improve our understanding of the role of angular momentum in the mechanical interaction of light with matter.

\section*{Acknowledgement}
This work is supported by European Research Council Starting Grant No. ERC-2016-STG-714151-PSINFONI.

\newpage
\appendix
\section{Induced dipole and quadrupole moments in isotropic Mie particles}\label{appendixA}
For an isotropic Mie particle, the polarizabilities for electric dipole $\alpha_{\mathrm{e}}$, magnetic dipole $\alpha_{\mathrm{m}}$, electric quadrupole $\alpha_{\mathrm{Qe}}$ and magnetic quadrupole $\alpha_{\mathrm{Qm}}$ are defined by the corresponding Mie coefficients as follows: 
\begin{equation}
\begin{aligned}\label{eq:empolarizability} 
&\alpha_{\mathrm{e}}=i\frac{6\pi}{k^3}\varepsilon_0 a_1,\,\,\,\alpha_{\mathrm{m}}=i\frac{6\pi}{k^3}b_1,\\\nonumber
&\alpha_{\mathrm{Qe}}=i\frac{120\pi}{k^5}\varepsilon_0 a_2,\,\,\,\alpha_{\mathrm{Qm}}=i\frac{120\pi}{k^5}b_2.
\end{aligned}
\end{equation}
The induced electric and magnetic dipole moments are vectors determined by the EM dipole polarizablities and the incident EM field:
\begin{equation}
\begin{aligned}\label{eq:emdipolemoment} 
\mathbf{p}=\alpha_{\mathrm{e}}\mathbf{E}_{\mathrm{inc}},\,\,\, \mathbf{m}=\alpha_m\mathbf{H}_{\mathrm{inc}}.
\end{aligned}
\end{equation}
The induced electric and magnetic quadrupole moments are tensors determined by the EM quadrupole polarizablities and the incident EM field and the field gradients \cite{DZHanPRB2009, FArangoACSPhoton2011,TDasPRB2015}: 
\begin{equation}
\begin{aligned}\label{eq:emquadrupolemoment} 
\overleftrightarrow{\mathbf{Q}}^{\mathrm{e}}=\alpha_{\mathrm{Qe}}\overleftrightarrow{\mathbcal{D}}^\mathrm{e},\,\,\, \overleftrightarrow{\mathbf{Q}}^{\mathrm{m}}=\alpha_{\mathrm{Qm}}\overleftrightarrow{\mathbcal{D}}^{\mathrm{m}}, 
\end{aligned}
\end{equation}
where 
\begin{equation}
\begin{aligned}\label{eq:emquadrupolefield} 
\overleftrightarrow{\mathbcal{D}}^\mathrm{e}&=\frac{\nabla\otimes\mathbf{E}_{\mathrm{inc}}+\mathbf{E}_{\mathrm{inc}}\otimes\nabla}{2},\,\,\,\overleftrightarrow{\mathbcal{D}}^{\mathrm{m}}&=\frac{\nabla\otimes\mathbf{H}_{\mathrm{inc}}+\mathbf{H}_{\mathrm{inc}}\otimes\nabla}{2}.
\end{aligned}
\end{equation}

\section{Radiation fields of electromagnetic dipole and quadrupole}\label{appendixB}
Knowing the electromagnetic dipole and quadrupole moments, the corresponding radiation fields can be analytically expressed \cite{VEBabichevaPRB2019}. 
The radiation fields of the dipoles are expresed as, 
\begin{equation}
\begin{aligned}\label{eq:emdipoleradiation} 
\mathbf{E}_{\mathrm{p}}&=\frac{k^2}{\varepsilon_0}\left(\overleftrightarrow{\mathbf{G}}^{\mathrm{d}}\cdot\mathbf{p}\right),\\
\mathbf{H}_{\mathrm{p}}&=-ikc_0\mathbf{\nabla}\times\left(\overleftrightarrow{\mathbf{G}}^{\mathrm{d}}\cdot\mathbf{p}\right)=-ikc_0\left(\mathbf{g}^{\mathrm{d}}\times\mathbf{p}\right),\\
\mathbf{E}_{\mathrm{m}}&=\frac{ik}{c_0\varepsilon_0}\mathbf{\nabla}\times\left(\overleftrightarrow{\mathbf{G}}^{\mathrm{d}}\cdot\mathbf{m}\right)=\frac{ik}{c_0\varepsilon_0}\left(\mathbf{g}^{\mathrm{d}}\times\mathbf{m}\right),\\
\mathbf{H}_{\mathrm{m}}&=k^2\left(\overleftrightarrow{\mathbf{G}}^{\mathrm{d}}\cdot\mathbf{m}\right),
\end{aligned}
\end{equation}
where 
\begin{equation}
\begin{aligned}\label{eq:emdipoleGreen} 
\overleftrightarrow{\mathbf{G}}^{\mathrm{d}}=&\Bigg\{\left(1+\frac{i}{kr}-\frac{1}{(kr)^2}\right)\overleftrightarrow{\mathbf{I}}-\left(1+\frac{3i}{kr}-\frac{3}{(kr)^2}\right)\hat{\mathbf{r}}\otimes\hat{\mathbf{r}}\Bigg\}\frac{\mathrm{e}^{ikr}}{4\pi r},\\
\mathbf{g}^{\mathrm{d}}=&\hat{\mathbf{r}}\left(ik-\frac{1}{r}\right)\frac{\mathrm{e}^{ikr}}{4\pi r},
\end{aligned}
\end{equation}
with $\overleftrightarrow{\mathbf{I}}$ being the 3-by-3 identity tensor.

The radiation fields of the quadrupoles are expresed as,
\begin{equation}
\begin{aligned}\label{eq:emquadrupoleradiation} 
\mathbf{E}_{\mathrm{Qe}}&=\frac{k^2}{\varepsilon_0}\left[\overleftrightarrow{\mathbf{G}}^{\mathrm{Q}}\cdot\left(\overleftrightarrow{\mathbf{Q}}^{\mathrm{e}}\cdot\hat{\mathbf{r}}\right)\right],\\
\mathbf{H}_{\mathrm{Qe}}&=-ikc_0\mathbf{\nabla}\times\left[\overleftrightarrow{\mathbf{G}}^{\mathrm{Q}}\cdot\left(\overleftrightarrow{\mathbf{Q}}^{\mathrm{e}}\cdot\hat{\mathbf{r}}\right)\right]=-ikc_0\left[\mathbf{g}^{\mathrm{Q}}\times\left(\overleftrightarrow{\mathbf{Q}}^{\mathrm{e}}\cdot\hat{\mathbf{r}}\right)\right],\\
\mathbf{E}_{\mathrm{Qm}}&=\frac{ik}{c_0\varepsilon_0}\mathbf{\nabla}\times\left[\overleftrightarrow{\mathbf{G}}^{\mathrm{Q}}\cdot\left(\overleftrightarrow{\mathbf{Q}}^{\mathrm{m}}\cdot\hat{\mathbf{r}}\right)\right]=\frac{ik}{c_0\varepsilon_0}\left[\mathbf{g}^{\mathrm{Q}}\times\left(\overleftrightarrow{\mathbf{Q}}^{\mathrm{m}}\cdot\hat{\mathbf{r}}\right)\right],\\
\mathbf{H}_{\mathrm{Qm}}&=k^2\left[\overleftrightarrow{\mathbf{G}}^{\mathrm{Q}}\cdot\left(\overleftrightarrow{\mathbf{Q}}^{\mathrm{m}}\cdot\hat{\mathbf{r}}\right)\right],
\end{aligned}
\end{equation}
where 
\begin{equation}
\begin{aligned}\label{eq:emquadrupoleGreen} 
\overleftrightarrow{\mathbf{G}}^{\mathrm{Q}}=&\left\{\left(-1-\frac{3i}{kr}+\frac{6}{(kr)^2}+\frac{6i}{(kr)^3}\right)\overleftrightarrow{\mathbf{I}}+\left(1+\frac{6i}{kr}-\frac{15}{(kr)^2}-\frac{15i}{(kr)^3}\right)\hat{\mathbf{r}}\otimes\hat{\mathbf{r}}\right\}\frac{ik\mathrm{e}^{ikr}}{24\pi r},\\
\mathbf{g}^{\mathrm{Q}}=&\hat{\mathbf{r}}\left[1+\frac{3i}{kr}-\frac{3}{(kr)^2}\right]\frac{k^2\mathrm{e}^{ikr}}{24\pi r}.
\end{aligned}
\end{equation}

\section{Derivation of the analytical dipolar torque}\label{appendixC}
In this section, we will outline the main derivation steps for the analytical expressions of optical torque $\mathbf{\Gamma}_{\mathrm{p}}$ on an electric dipole in a general electromagnetic field. Assume an isotropic Mie particle is placed at the origin $\mathbf{r}_0$. As shown in Eq. \ref{eq:TorqueAM}, the total optical torque be calculated by integerating the total angular momentum over an enclosed spherical surface surrounding the object,  

\begin{equation}
\begin{aligned}
\mathbf{\Gamma}=&\int_0^{2\pi}\int_{0}^{\pi} \mathbf{r}\times\left(\langle\overleftrightarrow{\mathbf{T}}\rangle\cdot\hat{\mathbf{n}}\right) r^2\sin\theta \mathrm{d}\theta \mathrm{d}\phi,\\
=&\Re \int_0^{2\pi}\int_{0}^{\pi} (r\hat{\mathbf{n}})\times\left\{\frac{\varepsilon_0}{2}\mathbf{E}_{\mathrm{tot}}\left(\mathbf{E}_{\mathrm{tot}}^*\cdot\hat{\mathbf{n}}\right)+\frac{\mu_0}{2}\mathbf{H}_{\mathrm{tot}}\left(\mathbf{H}_{\mathrm{tot}}^*\cdot\hat{\mathbf{n}}\right)\right\}  r^2\sin\theta \mathrm{d}\theta \mathrm{d}\phi,
\end{aligned}
\end{equation}
where $\mathbf{r}=\mathbf{r}'-\mathbf{r}_0=r\hat{\mathbf{n}}$,  $\mathbf{r}'$ denotes a point on the spherical surface, $r$ is the radius of the spherical surface and $\hat{\mathbf{n}}$ is the outward unit vector normal to the surface, $\hat{\mathbf{n}}=\hat{\mathbf{r}}=n_x\hat{\mathbf{e}}_x+n_y\hat{\mathbf{e}}_y+n_z\hat{\mathbf{e}}_z$ ($n_x=\sin\theta\cos\phi$, $n_y=\sin\theta\sin\phi$ and $n_z=\cos\theta$). Here we only consider the torque contributed by the induced electric dipole and the incident field so that $\mathbf{E}_{\mathrm{tot}}(\mathbf{r}')=\mathbf{E}_{\mathrm{inc}}(\mathbf{r}')+\mathbf{E}_{\mathrm{p}}(\mathbf{r}')$ and $\mathbf{H}_{\mathrm{tot}}(\mathbf{r}')=\mathbf{H}_{\mathrm{inc}}(\mathbf{r}')+\mathbf{H}_{\mathrm{p}}(\mathbf{r}')$.
The analytical expressions of optical torque on an electric dipole can be derived based on the knowledge that the radiation field of an induced electric dipole can be analytically expression as in Appendix \ref{appendixA} and \ref{appendixB}.  

The optical torque $\mathbf{\Gamma}_{\mathrm{p}}$, attributed to the interaction between the induced electric dipole and incident EM field, can be decomposed into different parts as shown in Eq. \ref{eq:TorqueAM} and Eq. \ref{eq:edTorque_components}, namely the torque component purely dependent on incident EM field, 

\begin{equation}
\begin{aligned}
\mathbf{\Gamma}_{\mathrm{inc}}&=\frac{\varepsilon_0r^3}{2}\Re \int_0^{2\pi}\int_{0}^{\pi} \left(\hat{\mathbf{n}}\times\mathbf{E}_{\mathrm{inc}}\right)\left(\mathbf{E}_{\mathrm{inc}}^*\cdot\hat{\mathbf{n}}\right) \sin\theta \mathrm{d}\theta \mathrm{d}\phi\\
&+\frac{\mu_0r^3}{2}\Re \int_0^{2\pi}\int_{0}^{\pi} \left(\hat{\mathbf{n}}\times\mathbf{H}_{\mathrm{inc}}\right)\left(\mathbf{H}_{\mathrm{inc}}^*\cdot\hat{\mathbf{n}}\right) \sin\theta \mathrm{d}\theta \mathrm{d}\phi,
\end{aligned}
\end{equation}
the extinction torque $\mathbf{\Gamma}_{\mathrm{p, mix}}$ dependent on the interference between incident and radiation fields 
\begin{equation}
\begin{aligned}
\mathbf{\Gamma}_{\mathrm{p, mix}}&=\frac{\varepsilon_0r^3}{2}\Re \int_0^{2\pi}\int_{0}^{\pi} \left(\hat{\mathbf{n}}\times\mathbf{E}_{\mathrm{inc}}\right)\left(\mathbf{E}_{\mathrm{p}}^*\cdot\hat{\mathbf{n}}\right) \sin\theta \mathrm{d}\theta \mathrm{d}\phi \\
&+\frac{\varepsilon_0r^3}{2}\Re \int_0^{2\pi}\int_{0}^{\pi} \left(\hat{\mathbf{n}}\times\mathbf{E}_{\mathrm{p}}\right)\left(\mathbf{E}_{\mathrm{inc}}^*\cdot\hat{\mathbf{n}}\right) \sin\theta \mathrm{d}\theta \mathrm{d}\phi\\
&+\frac{\mu_0r^3}{2}\Re \int_0^{2\pi}\int_{0}^{\pi} \left(\hat{\mathbf{n}}\times\mathbf{H}_{\mathrm{inc}}\right)\left(\mathbf{H}_{\mathrm{p}}^*\cdot\hat{\mathbf{n}}\right) \sin\theta \mathrm{d}\theta \mathrm{d}\phi \\
&+\frac{\mu_0r^3}{2}\Re \int_0^{2\pi}\int_{0}^{\pi} \left(\hat{\mathbf{n}}\times\mathbf{H}_{\mathrm{p}}\right)\left(\mathbf{H}_{\mathrm{inc}}^*\cdot\hat{\mathbf{n}}\right) \sin\theta \mathrm{d}\theta \mathrm{d}\phi,
\end{aligned}
\end{equation}
and the recoil torque $\mathbf{\Gamma}_{\mathrm{p, recoil}}$ as a result of self-interaction of the induced electric dipole
\begin{equation}
\begin{aligned}
\mathbf{\Gamma}_{\mathrm{p, recoil}}&=\frac{\varepsilon_0r^3}{2}\Re \int_0^{2\pi}\int_{0}^{\pi} \left(\hat{\mathbf{n}}\times\mathbf{E}_{\mathrm{p}}\right)\left(\mathbf{E}_{\mathrm{p}}^*\cdot\hat{\mathbf{n}}\right) \sin\theta \mathrm{d}\theta \mathrm{d}\phi \\
&+ \frac{\mu_0r^3}{2}\Re \int_0^{2\pi}\int_{0}^{\pi}\left(\hat{\mathbf{n}}\times\mathbf{H}_{\mathrm{p}}\right)\left(\mathbf{H}_{\mathrm{p}}^*\cdot\hat{\mathbf{n}}\right) \sin\theta \mathrm{d}\theta \mathrm{d}\phi. \\
\end{aligned}
\end{equation}

Due to the fact that the angular momenta of the incident, radiation and total fields are separately conserved quantities, the corresponding contributions to the dipolar torque can be calculated independent of the radius of the enclosed surface,
\begin{equation}
\begin{aligned}
&\mathbf{\Gamma}_{\mathrm{p}}=\lim_{r\rightarrow \infty}\mathbf{\Gamma}_{\mathrm{p}}(r)=\lim_{r\rightarrow 0}\mathbf{\Gamma}_{\mathrm{p}}(r), \\
&\mathbf{\Gamma}_{\mathrm{inc}}=\lim_{r\rightarrow \infty}\mathbf{\Gamma}_{\mathrm{inc}}(r)=\lim_{r\rightarrow 0}\mathbf{\Gamma}_{\mathrm{inc}}(r), \\
&\mathbf{\Gamma}_{\mathrm{p, recoil}}=\lim_{r\rightarrow \infty}\mathbf{\Gamma}_{\mathrm{p, recoil}}(r)=\lim_{r\rightarrow 0}\mathbf{\Gamma}_{\mathrm{p, recoil}}(r), \\
&\mathbf{\Gamma}_{\mathrm{p, mix}}=\lim_{r\rightarrow \infty}\mathbf{\Gamma}_{\mathrm{p, mix}}(r)=\lim_{r\rightarrow 0}\mathbf{\Gamma}_{\mathrm{p, mix}}(r), 
\end{aligned}
\end{equation}

It follows from Eq. \ref{eq:emdipoleradiation} and Eq. \ref{eq:emdipoleGreen} that
\begin{equation}
\begin{aligned}
\mathbf{H}_{\mathrm{p}}^*\cdot\hat{\mathbf{n}}=0,
\end{aligned}
\end{equation}

It is thus easy to prove that 
\begin{equation}
\begin{aligned}
&\frac{\mu_0r^3}{2}\Re \int_0^{2\pi}\int_{0}^{\pi} \left(\hat{\mathbf{n}}\times\mathbf{H}_{\mathrm{inc}}\right)\big(\mathbf{H}_{\mathrm{p}}^*\cdot\hat{\mathbf{n}}\big) \sin\theta \mathrm{d}\theta \mathrm{d}\phi=0,\\
&\frac{\mu_0r^3}{2}\Re \int_0^{2\pi}\int_{0}^{\pi} \left(\hat{\mathbf{n}}\times\mathbf{H}_{\mathrm{p}}\right)\big(\mathbf{H}_{\mathrm{p}}^*\cdot\hat{\mathbf{n}}\big) \sin\theta \mathrm{d}\theta \mathrm{d}\phi=0
\end{aligned}
\end{equation}

In the small $r$ limit that $r\rightarrow0$, we can approximate the incident electric field on the integration spherical surface, to the first order as:
\begin{equation}
\begin{aligned}
&\mathbf{E}_{\mathrm{inc}}(\mathbf{r}')\approx\mathbf{E}_{\mathrm{0}}+r(\hat{\mathbf{n}}\cdot\nabla)\mathbf{E}_{\mathrm{0}}, \\
&\mathbf{H}_{\mathrm{inc}}(\mathbf{r}')\approx\mathbf{H}_{\mathrm{0}}+r(\hat{\mathbf{n}}\cdot\nabla)\mathbf{H}_{\mathrm{0}}, 
\end{aligned}
\end{equation}
where $\mathbf{E}_{\mathrm{0}}=\mathbf{E}_{\mathrm{inc}}(\mathbf{r}_{0})$ and $\mathbf{H}_{\mathrm{0}}=\mathbf{H}_{\mathrm{inc}}(\mathbf{r}_{0})$.

The optical torque component $\mathbf{\Gamma}_{\mathrm{inc}}$ attributed purely to the incident EM field can be proven to be zero as follows:
\begin{equation}
\begin{aligned}
\mathbf{\Gamma}_{\mathrm{inc}}&=\lim_{r\rightarrow 0}\frac{\varepsilon_0r^3}{2}\Re \int_0^{2\pi}\int_{0}^{\pi} \left(\hat{\mathbf{n}}\times\mathbf{E}_{\mathrm{inc}}\right)\left(\mathbf{E}_{\mathrm{inc}}^*\cdot\hat{\mathbf{n}}\right) \sin\theta \mathrm{d}\theta \mathrm{d}\phi+\lim_{r\rightarrow 0}\frac{\mu_0r^3}{2}\Re \int_0^{2\pi}\int_{0}^{\pi} \left(\hat{\mathbf{n}}\times\mathbf{H}_{\mathrm{inc}}\right)\left(\mathbf{H}_{\mathrm{inc}}^*\cdot\hat{\mathbf{n}}\right) \sin\theta \mathrm{d}\theta \mathrm{d}\phi\\
&=\lim_{r\rightarrow 0}\frac{\varepsilon_0r^3}{2}\Re \int_0^{2\pi}\int_{0}^{\pi} \left(\hat{\mathbf{n}}\times\mathbf{E}_{\mathrm{0}}\right)\left(\mathbf{E}_{\mathrm{0}}^*\cdot\hat{\mathbf{n}}\right) \sin\theta \mathrm{d}\theta \mathrm{d}\phi+\lim_{r\rightarrow 0}\frac{\mu_0r^3}{2}\Re \int_0^{2\pi}\int_{0}^{\pi} \left(\hat{\mathbf{n}}\times\mathbf{H}_{\mathrm{0}}\right)\left(\mathbf{H}_{\mathrm{0}}^*\cdot\hat{\mathbf{n}}\right) \sin\theta \mathrm{d}\theta \mathrm{d}\phi\\
&=0.\\
\end{aligned}
\end{equation}

We then caculate the remaining non-zero terms of $\mathbf{\Gamma}_{\mathrm{p}}$. The following properties related to $\hat{\mathbf{n}}$ have been used in the derivations:
\begin{equation}
\begin{aligned}
&\int_0^{2\pi}\int_{0}^{\pi} n_l \sin\theta \mathrm{d}\theta \mathrm{d}\phi=0, \\
&\int_0^{2\pi}\int_{0}^{\pi} n_u n_v \sin\theta \mathrm{d}\theta \mathrm{d}\phi=\frac{4\pi}{3}\delta_{uv}, \\
&\int_0^{2\pi}\int_{0}^{\pi} n_l n_u n_v \sin\theta \mathrm{d}\theta \mathrm{d}\phi=0, 
\end{aligned}
\end{equation}
and 
\begin{equation}
\begin{aligned}
&\int_0^{2\pi}\int_{0}^{\pi} n_l n_u n_v n_m \sin\theta \mathrm{d}\theta \mathrm{d}\phi=0, \,\,\,\mathrm{except}\\
&\int_0^{2\pi}\int_{0}^{\pi} (n_u)^4\sin\theta \mathrm{d}\theta \mathrm{d}\phi=\frac{4\pi}{5}, \\
&\int_0^{2\pi}\int_{0}^{\pi} (n_u)^2(n_v)^2\sin\theta \mathrm{d}\theta \mathrm{d}\phi=\frac{4\pi}{15}, \,\, u\ne v. \\
\end{aligned}
\end{equation}

It follows from Eq. \ref{eq:emdipoleradiation} and Eq. \ref{eq:emdipoleGreen} that
\begin{equation}
\begin{aligned}
&\mathbf{E}_{\mathrm{p}}^*\cdot\hat{\mathbf{n}}=\frac{k^2 e^{-ikr}}{4\pi \varepsilon_0 r}\left[\frac{2i}{kr}+\frac{2}{(kr)^2}\right]\left(\mathbf{p}^*\cdot\hat{\mathbf{n}}\right),\\
\end{aligned}
\end{equation}
and using aforementioned properties of $\hat{\mathbf{n}}$, we can derive that
\begin{equation}
\begin{aligned}
&\lim_{r\rightarrow0}\frac{\varepsilon_0r^3}{2}\Re \int_0^{2\pi}\int_{0}^{\pi} \left(\hat{\mathbf{n}}\times\mathbf{E}_{\mathrm{inc}}\right)\left(\mathbf{E}_{\mathrm{p}}^*\cdot\hat{\mathbf{n}}\right) \sin\theta \mathrm{d}\theta \mathrm{d}\phi\\
&=\lim_{r\rightarrow0}\frac{\varepsilon_0r^3}{2}\Re \int_0^{2\pi}\int_{0}^{\pi} \Big(\hat{\mathbf{n}}\times\mathbf{E}_{\mathrm{inc}}\Big)\frac{k^2 e^{-ikr}}{4\pi \varepsilon_0 r}\Big[\frac{2i}{kr}+\frac{2}{(kr)^2}\Big]\left(\mathbf{p}^*\cdot\hat{\mathbf{n}}\right) \sin\theta \mathrm{d}\theta \mathrm{d}\phi\\
&=\lim_{r\rightarrow0}\Re \int_0^{2\pi}\int_{0}^{\pi} \Big(\hat{\mathbf{n}}\times\mathbf{E}_{\mathrm{inc}}\Big)\frac{e^{-ikr}}{4\pi}\Big[ikr+1\Big]\left(\mathbf{p}^*\cdot\hat{\mathbf{n}}\right) \sin\theta \mathrm{d}\theta \mathrm{d}\phi\\
&=\frac{1}{4\pi}\Re \int_0^{2\pi}\int_{0}^{\pi} \left(\hat{\mathbf{n}}\times\mathbf{E}_{\mathrm{0}}\right)\left(\mathbf{p}^*\cdot\hat{\mathbf{n}}\right)\sin\theta \mathrm{d}\theta \mathrm{d}\phi\\
&=\frac{1}{3}\Re\{\mathbf{p}^*\times\mathbf{E}_{\mathrm{0}}\},
\end{aligned}
\end{equation}

It follows from Eq. \ref{eq:emdipoleradiation} and Eq. \ref{eq:emdipoleGreen} that
\begin{equation}
\begin{aligned}
\hat{\mathbf{n}}\times\mathbf{E}_{\mathrm{p}}&=\frac{k^2 e^{ikr}}{4\pi \varepsilon_0 r}\Bigg[1+\frac{i}{kr}-\frac{1}{(kr)^2}\Bigg]\Big(\hat{\mathbf{n}}\times\mathbf{p}\Big),
\end{aligned}
\end{equation}
and using aforementioned properties of $\hat{\mathbf{n}}$, we can derive that
\begin{equation}
\begin{aligned}
&\lim_{r\rightarrow0}\frac{\varepsilon_0r^3}{2}\Re \int_0^{2\pi}\int_{0}^{\pi} \Big(\hat{\mathbf{n}}\times\mathbf{E}_{\mathrm{p}}\Big)\left(\mathbf{E}_{\mathrm{inc}}^*\cdot\hat{\mathbf{n}}\right) \sin\theta \mathrm{d}\theta \mathrm{d}\phi\\
&=\lim_{r\rightarrow0}\frac{\varepsilon_0r^3}{2}\Re \int_0^{2\pi}\int_{0}^{\pi}\frac{k^2 e^{ikr}}{4\pi \varepsilon_0 r}\Bigg[1+\frac{i}{kr}-\frac{1}{(kr)^2}\Bigg]\Big(\hat{\mathbf{n}}\times\mathbf{p}\Big)\left(\mathbf{E}_{\mathrm{inc}}^*\cdot\hat{\mathbf{n}}\right) \sin\theta \mathrm{d}\theta \mathrm{d}\phi\\
&=\lim_{r\rightarrow0}\frac{}{}\Re \int_0^{2\pi}\int_{0}^{\pi}\frac{e^{ikr}}{8\pi}\Big(k^2r^2+ikr-1\Big)\Big(\hat{\mathbf{n}}\times\mathbf{p}\Big)\left(\mathbf{E}_{\mathrm{inc}}^*\cdot\hat{\mathbf{n}}\right) \sin\theta \mathrm{d}\theta \mathrm{d}\phi\\
&=-\frac{1}{8\pi}\Re \int_0^{2\pi}\int_{0}^{\pi}\Big(\hat{\mathbf{n}}\times\mathbf{p}\Big)\Big(\mathbf{E}_{\mathrm{0}}^*\cdot\hat{\mathbf{n}}\Big) \sin\theta \mathrm{d}\theta \mathrm{d}\phi\\
&=\frac{1}{6}\Re\{\mathbf{p}^*\times\mathbf{E}_{\mathrm{0}}\},
\end{aligned}
\end{equation}

It follows from Eq. \ref{eq:emdipoleradiation} and Eq. \ref{eq:emdipoleGreen} that 
\begin{equation}
\begin{aligned}
\hat{\mathbf{n}}\times\mathbf{H}_{\mathrm{p}}&=\frac{-i k c_0 e^{ikr}}{4\pi r}\Big[ik-\frac{1}{r}\Big]\Big[\hat{\mathbf{n}}\times\big(\hat{\mathbf{n}}\times\mathbf{p}\big)\Big],
\end{aligned}
\end{equation}
it is easy to prove that 
\begin{equation}
\begin{aligned}
&\lim_{r\rightarrow0}\frac{\mu_0r^3}{2}\Re \int_0^{2\pi}\int_{0}^{\pi} \left(\hat{\mathbf{n}}\times\mathbf{H}_{\mathrm{p}}\right)\left(\mathbf{H}_{\mathrm{inc}}^*\cdot\hat{\mathbf{n}}\right) \sin\theta \mathrm{d}\theta \mathrm{d}\phi\\
&=\lim_{r\rightarrow0}\frac{\mu_0r^3}{2}\Re \int_0^{2\pi}\int_{0}^{\pi} \frac{-i k c_0 e^{ikr}}{4\pi r}\Big[ik-\frac{1}{r}\Big]\Big[\hat{\mathbf{n}}\times\big(\hat{\mathbf{n}}\times\mathbf{p}\big)\Big]\left(\mathbf{H}_{\mathrm{inc}}^*\cdot\hat{\mathbf{n}}\right) \sin\theta \mathrm{d}\theta \mathrm{d}\phi\\
&=\lim_{r\rightarrow0}\Re\int_0^{2\pi}\int_{0}^{\pi}\frac{\mu_0c_0e^{ikr}}{8\pi}\big(k^2r^2+ikr\big)\Big[\hat{\mathbf{n}}\times\big(\hat{\mathbf{n}}\times\mathbf{p}\big)\Big]\left(\mathbf{H}_{\mathrm{inc}}^*\cdot\hat{\mathbf{n}}\right) \sin\theta \mathrm{d}\theta \mathrm{d}\phi\\
&=0,
\end{aligned}
\end{equation}

The last nonzero term contributing to the electric dipolar torque is the so called `recoil' torque. Using the relations of $\left(\hat{\mathbf{n}}\times\mathbf{E}_{\mathrm{p}}\right)$ and $\left(\mathbf{E}_{\mathrm{p}}^*\cdot\hat{\mathbf{n}}\right)$ just derived, we can arrive at the analytical expression of $\mathbf{\Gamma}_{\mathrm{p, recoil}}$:
\begin{equation}
\begin{aligned}
\mathbf{\Gamma}_{\mathrm{p, recoil}}&=\lim_{r\to\infty}\frac{\varepsilon_0r^3}{2}\Re \int_0^{2\pi}\int_{0}^{\pi} \left(\hat{\mathbf{n}}\times\mathbf{E}_{\mathrm{p}}\right)\left(\mathbf{E}_{\mathrm{p}}^*\cdot\hat{\mathbf{n}}\right) \sin\theta \mathrm{d}\theta \mathrm{d}\phi\\
&=\lim_{r\to\infty}\frac{\varepsilon_0r^3}{2}\Re \int_0^{2\pi}\int_{0}^{\pi}\frac{k^2 e^{ikr}}{4\pi \varepsilon_0 r}\left[1+\frac{i}{kr}-\frac{1}{(kr)^2}\right]\left(\hat{\mathbf{n}}\times\mathbf{p}\right)\frac{k^2 e^{-ikr}}{4\pi \varepsilon_0 r}\left[\frac{2i}{kr}+\frac{2}{(kr)^2}\right]\left(\mathbf{p}^*\cdot\hat{\mathbf{n}}\right) \sin\theta \mathrm{d}\theta \mathrm{d}\phi\\
&=\Re \int_0^{2\pi}\int_{0}^{\pi}\frac{ik^3}{16\pi^2 \varepsilon_0 }\left[\hat{\mathbf{n}}\times\mathbf{p}\right]\left(\mathbf{p}^*\cdot\hat{\mathbf{n}}\right) \sin\theta \mathrm{d}\theta \mathrm{d}\phi\\
&=\Re \frac{ik^3}{16\pi^2 \varepsilon_0}\frac{4\pi}{3}\{\mathbf{p}^*\times\mathbf{p}\}\\
&=-\frac{k^3}{12\pi \varepsilon_0}\Im\{\mathbf{p}^*\times\mathbf{p}\}.
\end{aligned}
\end{equation}

The analytical expression of optical torque corresponding to an induced electric dipole of a particle positioned in a general electromagnetic field is thus derived from the total angular momentum flux method. The analytical expression of a magnetic dipole torque can be easily derived in a similar manner. Both analytical expressions can be given as,
\begin{equation}
\begin{aligned}
\mathbf{\Gamma}_{\mathrm{p}}=&\frac{1}{2}\Re\{\mathbf{p}^*\times\mathbf{E}_{\mathrm{0}}\}-\frac{k^3}{12\pi\varepsilon_0}\Im\{\mathbf{p}^*\times\mathbf{p}\}\\
=&\frac{1}{2}\Re\{\mathbf{p}^*\times\mathbf{E}_{\mathrm{inc}}\}-\frac{k^3}{12\pi\varepsilon_0}\Im\{\mathbf{p}^*\times\mathbf{p}\},\\
\mathbf{\Gamma}_{\mathrm{m}}=&\frac{1}{2}\Re\{\mathbf{m}^*\times\mu_0\mathbf{H}_{\mathrm{0}}\}-\frac{k^3\mu_0}{12\pi}\Im\{\mathbf{m}^*\times\mathbf{m}\}\\
=&\frac{1}{2}\Re\{\mathbf{m}^*\times\mu_0\mathbf{H}_{\mathrm{inc}}\}-\frac{k^3\mu_0}{12\pi}\Im\{\mathbf{m}^*\times\mathbf{m}\}.
\end{aligned}
\end{equation}

For a combined electric and magnetic dipole, additional terms due to the interference of the electric and magnetic dipole could arise. However, these are shown to be zero, following Eq. \ref{eq:emdipoleradiation}, Eq. \ref{eq:emdipoleGreen} and the aforementioned properties of $\hat{\mathbf{n}}$, it is easy to prove that:
\begin{equation}
\begin{aligned}
\mathbf{\Gamma}_{\mathrm{pm, recoil}}=&\frac{\varepsilon_0r^3}{2}\Re \int_0^{2\pi}\int_{0}^{\pi} \left(\hat{\mathbf{n}}\times\mathbf{E}_{\mathrm{p}}\right)\left(\mathbf{E}_{\mathrm{m}}^*\cdot\hat{\mathbf{n}}\right) \sin\theta \mathrm{d}\theta \mathrm{d}\phi\\
&+\frac{\varepsilon_0r^3}{2}\Re \int_0^{2\pi}\int_{0}^{\pi} \left(\hat{\mathbf{n}}\times\mathbf{E}_{\mathrm{m}}\right)\left(\mathbf{E}_{\mathrm{p}}^*\cdot\hat{\mathbf{n}}\right) \sin\theta \mathrm{d}\theta \mathrm{d}\phi\\
&+\frac{\varepsilon_0r^3}{2}\Re \int_0^{2\pi}\int_{0}^{\pi} \left(\hat{\mathbf{n}}\times\mathbf{H}_{\mathrm{p}}\right)\left(\mathbf{H}_{\mathrm{m}}^*\cdot\hat{\mathbf{n}}\right) \sin\theta \mathrm{d}\theta \mathrm{d}\phi\\
&+\frac{\varepsilon_0r^3}{2}\Re \int_0^{2\pi}\int_{0}^{\pi} \left(\hat{\mathbf{n}}\times\mathbf{H}_{\mathrm{m}}\right)\left(\mathbf{H}_{\mathrm{p}}^*\cdot\hat{\mathbf{n}}\right) \sin\theta \mathrm{d}\theta \mathrm{d}\phi\\
=&\lim_{r\to\infty}\frac{\varepsilon_0r^3}{2}\Re \int_0^{2\pi}\int_{0}^{\pi} \left(\hat{\mathbf{n}}\times\mathbf{E}_{\mathrm{m}}\right)\left(\mathbf{E}_{\mathrm{p}}^*\cdot\hat{\mathbf{n}}\right) \sin\theta \mathrm{d}\theta \mathrm{d}\phi\\
&+\lim_{r\to\infty}\frac{\varepsilon_0r^3}{2}\Re \int_0^{2\pi}\int_{0}^{\pi} \left(\hat{\mathbf{n}}\times\mathbf{H}_{\mathrm{p}}\right)\left(\mathbf{H}_{\mathrm{m}}^*\cdot\hat{\mathbf{n}}\right) \sin\theta \mathrm{d}\theta \mathrm{d}\phi\\
=&0.
\end{aligned}
\end{equation}

To summarise, the dipolar optical torque in a general electromagnetic field can be expressed as,
\begin{equation}
\begin{aligned}
\mathbf{\Gamma}_{\mathrm{d}}=\mathbf{\Gamma}_{\mathrm{p}}+\mathbf{\Gamma}_{\mathrm{m}}.
\end{aligned}
\end{equation}

\section{Derivation of the analytical `spin' and `orbital' dipolar torque}\label{appendixD}
In this section, a brief derivation of the optical torque acting on an electric dipole using the spin and orbital angular momentum flux method is shown, and the optical torque acting on a magnetic dipole can be derived in a similar way. The optical torque attributed to the spin angular momentum flux $\langle\overleftrightarrow{\mathbf{M}}^{s}\rangle$ can be derived by integrating the SAM flux over a spherical surface with the object at its centre as defined in previous section,
\begin{equation}
\begin{aligned}
\mathbf{\Gamma}^s_{\mathrm{p}}=&\int_0^{2\pi}\int_0^{\pi} \left(\langle\overleftrightarrow{\mathbf{M}}^{s}\rangle\cdot \hat{\mathbf{n}}\right)r^2\sin\theta\mathrm{d}\theta\mathrm{d}\phi,\\
=&\int_0^{2\pi}\int_0^{\pi} \frac{1}{2\omega}\Im\left\{\mathbf{E}_{\mathrm{tot}}\left(\mathbf{H}_{\mathrm{tot}}^*\cdot\hat{\mathbf{n}}\right)+\mathbf{H}_{\mathrm{tot}}^*\left(\mathbf{E}_{\mathrm{tot}}\cdot\hat{\mathbf{n}}\right)-\left(\mathbf{E}_{\mathrm{tot}}\cdot\mathbf{H}_{\mathrm{tot}}^*\right)\hat{\mathbf{n}}\right\} r^2\sin\theta\mathrm{d}\theta\mathrm{d}\phi,
\end{aligned}
\end{equation}
where  $\mathbf{E}_{\mathrm{tot}}(\mathbf{r}')=\mathbf{E}_{\mathrm{inc}}(\mathbf{r}')+\mathbf{E}_{\mathrm{p}}(\mathbf{r}')$ and $\mathbf{H}_{\mathrm{tot}}(\mathbf{r}')=\mathbf{H}_{\mathrm{inc}}(\mathbf{r}')+\mathbf{H}_{\mathrm{p}}(\mathbf{r}')$. 

The SAM related torque on an electric dipole $\mathbf{\Gamma}^s_{\mathrm{p}}$ can also be separated into three parts, 
\begin{equation}
\begin{aligned}
\mathbf{\Gamma}^s_{\mathrm{p}}=\mathbf{\Gamma}^s_{\mathrm{inc}}+\mathbf{\Gamma}^s_{\mathrm{p,mix}}+\mathbf{\Gamma}^s_{\mathrm{p,recoil}},
\end{aligned}
\end{equation}
where $\mathbf{\Gamma}^s_{\mathrm{inc}}$ depends purely on the incident field, 
\begin{equation}
\begin{aligned}
\mathbf{\Gamma}^s_{\mathrm{inc}}=\int_0^{2\pi}\int_0^{\pi} \frac{1}{2\omega}\Im\left\{\mathbf{E}_{\mathrm{inc}}\left(\mathbf{H}_{\mathrm{inc}}^*\cdot\hat{\mathbf{n}}\right)+\mathbf{H}_{\mathrm{inc}}^*\left(\mathbf{E}_{\mathrm{inc}}\cdot\hat{\mathbf{n}}\right)-\left(\mathbf{E}_{\mathrm{inc}}\cdot\mathbf{H}_{\mathrm{inc}}^*\right)\hat{\mathbf{n}}\right\} r^2\sin\theta\mathrm{d}\theta\mathrm{d}\phi,
\end{aligned}
\end{equation}

$\mathbf{\Gamma}^s_{\mathrm{p,mix}}$ relies on the interference between the incident and radiation field, 
\begin{equation}
\begin{aligned}
\mathbf{\Gamma}^s_{\mathrm{p,mix}}=\int_0^{2\pi}\int_0^{\pi} \frac{1}{2\omega}\Im\Big\{&\mathbf{E}_{\mathrm{inc}}\big(\mathbf{H}_{\mathrm{p}}^*\cdot\hat{\mathbf{n}}\big)+\mathbf{E}_{\mathrm{p}}\big(\mathbf{H}_{\mathrm{inc}}^*\cdot\hat{\mathbf{n}}\big)+\mathbf{H}_{\mathrm{inc}}^*\big(\mathbf{E}_{\mathrm{p}}\cdot\hat{\mathbf{n}}\big)+\mathbf{H}_{\mathrm{p}}^*\big(\mathbf{E}_{\mathrm{inc}}\cdot\hat{\mathbf{n}}\big)\\
&-\big(\mathbf{E}_{\mathrm{inc}}\cdot\mathbf{H}_{\mathrm{p}}^*+\mathbf{E}_{\mathrm{p}}\cdot\mathbf{H}_{\mathrm{inc}}^*\big)\hat{\mathbf{n}}\Big\} r^2\sin\theta\mathrm{d}\theta\mathrm{d}\phi,
\end{aligned}
\end{equation}

while $\mathbf{\Gamma}^s_{\mathrm{p,recoil}}$ is attributed to the radiation field only,
\begin{equation}
\begin{aligned}
\mathbf{\Gamma}^s_{\mathrm{p,recoil}}=\int_0^{2\pi}\int_0^{\pi} \frac{1}{2\omega}\Im\Big\{\mathbf{E}_{\mathrm{p}}\big(\mathbf{H}_{\mathrm{p}}^*\cdot\hat{\mathbf{n}}\big)+\mathbf{H}_{\mathrm{p}}^*\big(\mathbf{E}_{\mathrm{p}}\cdot\hat{\mathbf{n}}\big)-\big(\mathbf{E}_{\mathrm{p}}\cdot\mathbf{H}_{\mathrm{p}}^*\big)\hat{\mathbf{n}}\Big\} r^2\sin\theta\mathrm{d}\theta\mathrm{d}\phi.
\end{aligned}
\end{equation}

Due to the fact that the spin angular momenta of the incident, radiation and total fields are separately conserved quantities, the corresponding contributions to the dipolar torque can be calculated independent of the radius of the enclosed surface,
\begin{equation}
\begin{aligned}
&\mathbf{\Gamma}^s_{\mathrm{p}}=\lim_{r\rightarrow \infty}\mathbf{\Gamma}^s_{\mathrm{p}}(r)=\lim_{r\rightarrow 0}\mathbf{\Gamma}^s_{\mathrm{p}}(r), \\
&\mathbf{\Gamma}^s_{\mathrm{inc}}=\lim_{r\rightarrow \infty}\mathbf{\Gamma}^s_{\mathrm{inc}}(r)=\lim_{r\rightarrow 0}\mathbf{\Gamma}^s_{\mathrm{inc}}(r), \\
&\mathbf{\Gamma}^s_{\mathrm{p, recoil}}=\lim_{r\rightarrow \infty}\mathbf{\Gamma}^s_{\mathrm{p, recoil}}(r)=\lim_{r\rightarrow 0}\mathbf{\Gamma}^s_{\mathrm{p, recoil}}(r), \\
&\mathbf{\Gamma}^s_{\mathrm{p, mix}}=\lim_{r\rightarrow \infty}\mathbf{\Gamma}^s_{\mathrm{p, mix}}(r)=\lim_{r\rightarrow 0}\mathbf{\Gamma}^s_{\mathrm{p, mix}}(r), 
\end{aligned}
\end{equation}

In the small $r$ limit that $r\rightarrow0$, using the approximation that $\mathbf{E}_{\mathrm{inc}}(\mathbf{r}')\approx\mathbf{E}_{\mathrm{0}}+r(\hat{\mathbf{n}}\cdot\nabla)\mathbf{E}_{\mathrm{0}}$ and $\mathbf{H}_{\mathrm{inc}}(\mathbf{r}')\approx\mathbf{H}_{\mathrm{0}}+r(\hat{\mathbf{n}}\cdot\nabla)\mathbf{H}_{\mathrm{0}}$, it is easy to prove that the optical torque component $\mathbf{\Gamma}^s_{\mathrm{inc}}$ attributed purely to the incident field is zero:
\begin{equation}
\begin{aligned}
\mathbf{\Gamma}^s_{\mathrm{inc}}&=\lim_{r\rightarrow 0}\int_0^{2\pi}\int_0^{\pi} \frac{1}{2\omega}\Im\left\{\mathbf{E}_{\mathrm{inc}}\left(\mathbf{H}_{\mathrm{inc}}^*\cdot\hat{\mathbf{n}}\right)+\mathbf{H}_{\mathrm{inc}}^*\left(\mathbf{E}_{\mathrm{inc}}\cdot\hat{\mathbf{n}}\right)-\left(\mathbf{E}_{\mathrm{inc}}\cdot\mathbf{H}_{\mathrm{inc}}^*\right)\hat{\mathbf{n}}\right\} r^2\sin\theta\mathrm{d}\theta\mathrm{d}\phi\\
&=\lim_{r\rightarrow 0} r^2\int_0^{2\pi}\int_0^{\pi} \frac{1}{2\omega}\Im\left\{\mathbf{E}_{\mathrm{0}}\left(\mathbf{H}_{\mathrm{0}}^*\cdot\hat{\mathbf{n}}\right)+\mathbf{H}_{\mathrm{0}}^*\left(\mathbf{E}_{\mathrm{0}}\cdot\hat{\mathbf{n}}\right)-\left(\mathbf{E}_{\mathrm{0}}\cdot\mathbf{H}_{\mathrm{0}}^*\right)\hat{\mathbf{n}}\right\} \sin\theta\mathrm{d}\theta\mathrm{d}\phi\\
&=0.\\
\end{aligned}
\end{equation}

The various components of $\mathbf{\Gamma}^s_{\mathrm{p,mix}}$ can be derived following a similar method as in the previous section, using the dipolar radiation field properties in Eq. \ref{eq:emdipoleradiation} and Eq. \ref{eq:emdipoleGreen}, the aforementioned integration properties of the unit vector $\hat{\mathbf{n}}$, and the approximation of the incident field in the small $r$ limit $\mathbf{E}_{\mathrm{inc}}(\mathbf{r}')\approx\mathbf{E}_{\mathrm{0}}+r(\hat{\mathbf{n}}\cdot\nabla)\mathbf{E}_{\mathrm{0}}$ and $\mathbf{H}_{\mathrm{inc}}(\mathbf{r}')\approx\mathbf{H}_{\mathrm{0}}+r(\hat{\mathbf{n}}\cdot\nabla)\mathbf{H}_{\mathrm{0}}$. Their analytical results are listed below,
\begin{equation}
\begin{aligned}
\int_0^{2\pi}\int_0^{\pi} \frac{1}{2\omega}\Im\left\{\mathbf{E}_{\mathrm{inc}}\left(\mathbf{H}_{\mathrm{p}}^*\cdot\hat{\mathbf{n}}\right)\right\} r^2\sin\theta\mathrm{d}\theta\mathrm{d}\phi=0,
\end{aligned}
\end{equation}

\begin{equation}
\begin{aligned}
\lim_{r\to0}\int_0^{2\pi}\int_0^{\pi} \frac{1}{2\omega}\Im\left\{\mathbf{H}_{\mathrm{p}}^*\left(\mathbf{E}_{\mathrm{inc}}\cdot\hat{\mathbf{n}}\right)\right\} r^2\sin\theta\mathrm{d}\theta\mathrm{d}\phi&=\Im\int_0^{2\pi}\int_0^{\pi}\left\{\frac{-i}{8\pi} \left(\hat{\mathbf{n}}\times\mathbf{p}^*\right)\left(\mathbf{E}_{\mathrm{0}}\cdot\hat{\mathbf{n}}\right)\right\} \sin\theta\mathrm{d}\theta\mathrm{d}\phi\\
&=\frac{1}{6}\Re\left\{\mathbf{p}^*\times\mathbf{E}_{\mathrm{0}}\right\},
\end{aligned}
\end{equation}

\begin{equation}
\begin{aligned}
\lim_{r\to0}\int_0^{2\pi}\int_0^{\pi} \frac{1}{2\omega}\Im\left\{-\left(\mathbf{E}_{\mathrm{inc}}\cdot\mathbf{H}_{\mathrm{p}}^*\right)\hat{\mathbf{n}}\right\} r^2\sin\theta\mathrm{d}\theta\mathrm{d}\phi&=\Im\int_0^{2\pi}\int_0^{\pi}\left\{\frac{i}{8\pi} \left[\left(\hat{\mathbf{n}}\times\mathbf{p}^*\right)\cdot\mathbf{E}_{\mathrm{0}}\right]\hat{\mathbf{n}}\right\} \sin\theta\mathrm{d}\theta\mathrm{d}\phi\\
&=\frac{1}{6}\Re\left\{\mathbf{p}^*\times\mathbf{E}_{\mathrm{0}}\right\},
\end{aligned}
\end{equation}

\begin{equation}
\begin{aligned}
\lim_{r\to0}\int_0^{2\pi}\int_0^{\pi} \frac{1}{2\omega}\Im\left\{\mathbf{E}_{\mathrm{p}}\left(\mathbf{H}^*_{\mathrm{inc}}\cdot\hat{\mathbf{n}}\right)\right\} r^2\sin\theta\mathrm{d}\theta\mathrm{d}\phi&=\Im\int_0^{2\pi}\int_0^{\pi}\frac{-1}{8\pi\omega\varepsilon_0}\left\{\left[\left(\hat{\mathbf{n}}\cdot\nabla\right)\mathbf{H}_{\mathrm{0}}^*\right]\cdot\hat{\mathbf{n}}\right\}\mathbf{p} \sin\theta\mathrm{d}\theta\mathrm{d}\phi\\
&+\Im\int_0^{2\pi}\int_0^{\pi}\frac{3}{8\pi\omega\varepsilon_0}\left\{\left[\left(\hat{\mathbf{n}}\cdot\nabla\right)\mathbf{H}_{\mathrm{0}}^*\right]\cdot\hat{\mathbf{n}}\right\}(\hat{\mathbf{n}}\cdot\mathbf{p})\hat{\mathbf{n}} \sin\theta\mathrm{d}\theta\mathrm{d}\phi\\
&=\Im\int_0^{2\pi}\int_0^{\pi}\frac{3}{8\pi\omega\varepsilon_0}\left\{\left[\left(\hat{\mathbf{n}}\cdot\nabla\right)\mathbf{H}_{\mathrm{0}}^*\right]\cdot\hat{\mathbf{n}}\right\}(\hat{\mathbf{n}}\cdot\mathbf{p})\hat{\mathbf{n}} \sin\theta\mathrm{d}\theta\mathrm{d}\phi\\
&=\frac{1}{10}\Re\left\{\mathbf{p}^*\times\mathbf{E}_{\mathrm{0}}\right\}+\frac{1}{5\omega\varepsilon_0}\Im\left\{(\mathbf{p}\cdot\nabla)\mathbf{H}_{\mathrm{0}}^*\right\},
\end{aligned}
\end{equation}

\begin{equation}
\begin{aligned}
\lim_{r\to0}\int_0^{2\pi}\int_0^{\pi} \frac{1}{2\omega}\Im\left\{\mathbf{H}_{\mathrm{inc}}^*\left(\mathbf{E}_{\mathrm{p}}\cdot\hat{\mathbf{n}}\right)\right\} r^2\sin\theta\mathrm{d}\theta\mathrm{d}\phi&=\Im\int_0^{2\pi}\int_0^{\pi}\left\{\frac{1}{4\pi\omega\varepsilon_0}(\hat{\mathbf{n}}\cdot\mathbf{p}) \left[\left(\hat{\mathbf{n}}\cdot\nabla\right)\mathbf{H}_{\mathrm{0}}^*\right]\right\} \sin\theta\mathrm{d}\theta\mathrm{d}\phi\\
&=\frac{1}{3\omega\varepsilon_0}\Im\left\{(\mathbf{p}\cdot\nabla)\mathbf{H}_{\mathrm{0}}^*\right\},
\end{aligned}
\end{equation}

\begin{equation}
\begin{aligned}
\lim_{r\to0}\int_0^{2\pi}\int_0^{\pi} \frac{1}{2\omega}\Im\left\{-\left(\mathbf{E}_{\mathrm{p}}\cdot\mathbf{H}_{\mathrm{inc}}^*\right)\hat{\mathbf{n}}\right\} r^2\sin\theta\mathrm{d}\theta\mathrm{d}\phi&=\Im\int_0^{2\pi}\int_0^{\pi}\left\{\frac{1}{8\pi\omega\varepsilon_0} \left\{\left[\left(\hat{\mathbf{n}}\cdot\nabla\right)\mathbf{H}_{\mathrm{0}}^*\right]\cdot\mathbf{p}\right\}\hat{\mathbf{n}}\right\} \sin\theta\mathrm{d}\theta\mathrm{d}\phi\\
&+\Im\int_0^{2\pi}\int_0^{\pi}\frac{-3}{8\pi\omega\varepsilon_0}\left\{\left[\left(\hat{\mathbf{n}}\cdot\nabla\right)\mathbf{H}_{\mathrm{0}}^*\right]\cdot\hat{\mathbf{n}}\right\}(\hat{\mathbf{n}}\cdot\mathbf{p})\hat{\mathbf{n}} \sin\theta\mathrm{d}\theta\mathrm{d}\phi\\
&=\frac{1}{6}\Re\left(\mathbf{p}^*\times\mathbf{E}_{\mathrm{0}}\right)+\frac{1}{6\omega\varepsilon_0}\Im\left[(\mathbf{p}\cdot\nabla)\mathbf{H}_{\mathrm{0}}^*\right]\\
&-\left\{\frac{1}{10}\Re\left(\mathbf{p}^*\times\mathbf{E}_{\mathrm{0}}\right)+\frac{1}{5\omega\varepsilon_0}\Im\left[(\mathbf{p}\cdot\nabla)\mathbf{H}_{\mathrm{0}}^*\right]\right\}.
\end{aligned}
\end{equation}

From the above results, we can get the analytical expression for $\mathbf{\Gamma}^s_{\mathrm{p,mix}}$ 
\begin{equation}
\begin{aligned}
\mathbf{\Gamma}^s_{\mathrm{p,mix}}=\frac{1}{2}\Re\left\{\mathbf{p}^*\times\mathbf{E}_{\mathrm{inc}}\right\}+\frac{1}{2\omega\varepsilon_0}\Im\left\{(\mathbf{p}\cdot\nabla)\mathbf{H}_{\mathrm{inc}}^*\right\}.
\end{aligned}
\end{equation}

Similarly, the various components of $\mathbf{\Gamma}^s_{\mathrm{p,recoil}}$ can be derived using the dipolar radiation field properties in Eq. \ref{eq:emdipoleradiation} and Eq. \ref{eq:emdipoleGreen} and the aforementioned integration properties of the unit vector $\hat{\mathbf{n}}$ in the large $r$ limit as $r\rightarrow\infty$,
\begin{equation}
\begin{aligned}
&\lim_{r\to\infty}\int_0^{2\pi}\int_0^{\pi} \frac{1}{2\omega}\Im\left\{\mathbf{H}_{\mathrm{p}}^*\left(\mathbf{E}_{\mathrm{p}}\cdot\hat{\mathbf{n}}\right)\right\}r^2\sin\theta\mathrm{d}\theta\mathrm{d}\phi=0,\\
&\int_0^{2\pi}\int_0^{\pi} \frac{1}{2\omega}\Im\left\{\mathbf{E}_{\mathrm{p}}\left(\mathbf{H}_{\mathrm{p}}^*\cdot\hat{\mathbf{n}}\right)\right\}r^2\sin\theta\mathrm{d}\theta\mathrm{d}\phi=0,
\end{aligned}
\end{equation}

\begin{equation}
\begin{aligned}
&\lim_{r\to\infty}\int_0^{2\pi}\int_0^{\pi} \frac{1}{2\omega}\Im\left\{-\left(\mathbf{E}_{\mathrm{p}}\cdot\mathbf{H}_{\mathrm{p}}^*\right)\hat{\mathbf{n}}\right\} r^2\sin\theta\mathrm{d}\theta\mathrm{d}\phi\\
&=\Im\int_0^{2\pi}\int_0^{\pi}\left\{-\frac{k^3}{32\pi^2\varepsilon_0} \left[\left(\hat{\mathbf{n}}\times\mathbf{p}^*\right)\cdot\mathbf{p}\right]\hat{\mathbf{n}}\right\} \sin\theta\mathrm{d}\theta\mathrm{d}\phi\\
&=-\frac{k^3}{24\pi^2\varepsilon_0}\Im\left\{\mathbf{p}^*\times\mathbf{p}\right\}, 
\end{aligned}
\end{equation}
and we can get the analytical expression for $\mathbf{\Gamma}^s_{\mathrm{p,recoil}}$ as
\begin{equation}
\begin{aligned}
\mathbf{\Gamma}^s_{\mathrm{p,recoil}}=-\frac{k^3}{24\pi^2\varepsilon_0}\Im\left\{\mathbf{p}^*\times\mathbf{p}\right\}.
\end{aligned}
\end{equation}

The optical torque on an electric dipole attributed to the orbital angular momentum flux can be calculated from
\begin{equation}
\begin{aligned}
\mathbf{\Gamma}^o=&\int_0^{2\pi}\int_0^{\pi} \left(\langle\overleftrightarrow{\mathbf{M}}^{o}\rangle\cdot \hat{\mathbf{n}}\right)r^2\sin\theta\mathrm{d}\theta\mathrm{d}\phi,\\
=&\int_0^{2\pi}\int_0^{\pi} \frac{-1}{4\omega}\Im\left\{\mathbf{E}_{\mathrm{tot}}\left(\mathbf{H}_{\mathrm{tot}}^*\cdot\hat{\mathbf{n}}\right)+\mathbf{H}_{\mathrm{tot}}^*\left(\mathbf{E}_{\mathrm{tot}}\cdot\hat{\mathbf{n}}\right)\right\}r^2\sin\theta\mathrm{d}\theta\mathrm{d}\phi\\
&+\int_0^{2\pi}\int_0^{\pi}\frac{1}{4\omega}\Im\left\{\left[(r\hat{\mathbf{n}}\times\nabla)\otimes\mathbf{E}_{\mathrm{tot}}^*\right]\times\mathbf{H}_{\mathrm{tot}}\right\}\cdot \hat{\mathbf{n}}r^2\sin\theta\mathrm{d}\theta\mathrm{d}\phi\\
&+\int_0^{2\pi}\int_0^{\pi}\frac{1}{4\omega}\Im\left\{\left[(r\hat{\mathbf{n}}\times\nabla)\otimes\mathbf{H}_{\mathrm{tot}}\right]\times\mathbf{E}_{\mathrm{tot}}^*\right\}\cdot \hat{\mathbf{n}}r^2\sin\theta\mathrm{d}\theta\mathrm{d}\phi,
\end{aligned}
\end{equation}

In the following derivation, both expressions of $\nabla$ in the Cartesian and spherical coordiates are applied such that
\begin{equation}
\begin{aligned}
r\hat{\mathbf{n}}\times\nabla=&r\Big(\hat{\mathbf{e}}_xn_x+\hat{\mathbf{e}}_yn_y+\hat{\mathbf{e}}_zn_z\Big)\times\Big(\hat{\mathbf{e}}_x\frac{\partial}{\partial_x}+\hat{\mathbf{e}}_y\frac{\partial}{\partial_y}+\hat{\mathbf{e}}_z\frac{\partial}{\partial_z}\Big)\\
=&r\hat{\mathbf{r}}\times\nabla\\
=&\hat{\mathbf{\theta}}\frac{\partial}{\partial_{\theta}}+\hat{\mathbf{\phi}}\frac{1}{\sin \theta}\frac{\partial}{\partial_{\phi}}.
\end{aligned}
\end{equation}

The OAM related torque on an electric dipole $\mathbf{\Gamma}^o_{\mathrm{p}}$ can be separated into three parts, 
\begin{equation}
\begin{aligned}
\mathbf{\Gamma}^o_{\mathrm{p}}=\mathbf{\Gamma}^o_{\mathrm{inc}}+\mathbf{\Gamma}^o_{\mathrm{p,mix}}+\mathbf{\Gamma}^o_{\mathrm{p,recoil}},
\end{aligned}
\end{equation}
where $\mathbf{\Gamma}^o_{\mathrm{inc}}$ depends purely on the incident field, 
\begin{equation}
\begin{aligned}
\mathbf{\Gamma}^o_{\mathrm{inc}}=&\int_0^{2\pi}\int_0^{\pi} \frac{-1}{4\omega}\Im\left\{\mathbf{E}_{\mathrm{inc}}\left(\mathbf{H}_{\mathrm{inc}}^*\cdot\hat{\mathbf{n}}\right)+\mathbf{H}_{\mathrm{inc}}^*\left(\mathbf{E}_{\mathrm{inc}}\cdot\hat{\mathbf{n}}\right)\right\}r^2\sin\theta\mathrm{d}\theta\mathrm{d}\phi\\
&+\int_0^{2\pi}\int_0^{\pi}\frac{1}{4\omega}\Im\left\{\left[(r\hat{\mathbf{n}}\times\nabla)\otimes\mathbf{E}_{\mathrm{inc}}^*\right]\times\mathbf{H}_{\mathrm{inc}}\right\}\cdot \hat{\mathbf{n}}r^2\sin\theta\mathrm{d}\theta\mathrm{d}\phi\\
&+\int_0^{2\pi}\int_0^{\pi}\frac{1}{4\omega}\Im\left\{\left[(r\hat{\mathbf{n}}\times\nabla)\otimes\mathbf{H}_{\mathrm{inc}}\right]\times\mathbf{E}_{\mathrm{inc}}^*\right\}\cdot \hat{\mathbf{n}}r^2\sin\theta\mathrm{d}\theta\mathrm{d}\phi,
\end{aligned}
\end{equation}

$\mathbf{\Gamma}^o_{\mathrm{p,mix}}$ relies on the interference between the incident and radiation field, 
\begin{equation}
\begin{aligned}
\mathbf{\Gamma}^o_{\mathrm{p,mix}}=&\int_0^{2\pi}\int_0^{\pi} \frac{-1}{4\omega}\Im\Big\{\mathbf{E}_{\mathrm{inc}}\left(\mathbf{H}_{\mathrm{p}}^*\cdot\hat{\mathbf{n}}\right)+\mathbf{H}_{\mathrm{p}}^*\left(\mathbf{E}_{\mathrm{inc}}\cdot\hat{\mathbf{n}}\right)+\mathbf{E}_{\mathrm{p}}\left(\mathbf{H}_{\mathrm{inc}}^*\cdot\hat{\mathbf{n}}\right)+\mathbf{H}_{\mathrm{inc}}^*\left(\mathbf{E}_{\mathrm{p}}\cdot\hat{\mathbf{n}}\right)\Big\}r^2\sin\theta\mathrm{d}\theta\mathrm{d}\phi\\
&+\int_0^{2\pi}\int_0^{\pi}\frac{1}{4\omega}\Im\Big\{\big[(r\hat{\mathbf{n}}\times\nabla)\otimes\mathbf{E}_{\mathrm{inc}}^*\big]\times\mathbf{H}_{\mathrm{p}}+\big[(r\hat{\mathbf{n}}\times\nabla)\otimes\mathbf{E}_{\mathrm{p}}^*\big]\times\mathbf{H}_{\mathrm{inc}}\Big\}\cdot \hat{\mathbf{n}}r^2\sin\theta\mathrm{d}\theta\mathrm{d}\phi\\
&+\int_0^{2\pi}\int_0^{\pi}\frac{1}{4\omega}\Im\Big\{\big[(r\hat{\mathbf{n}}\times\nabla)\otimes\mathbf{H}_{\mathrm{p}}\big]\times\mathbf{E}_{\mathrm{inc}}^*+\big[(r\hat{\mathbf{n}}\times\nabla)\otimes\mathbf{H}_{\mathrm{inc}}\big]\times\mathbf{E}_{\mathrm{p}}^*\Big\}\cdot \hat{\mathbf{n}}r^2\sin\theta\mathrm{d}\theta\mathrm{d}\phi,
\end{aligned}
\end{equation}
while $\mathbf{\Gamma}^o_{\mathrm{p,recoil}}$ is attributed to the radiation field only,

\begin{equation}
\begin{aligned}
\mathbf{\Gamma}^o_{\mathrm{p,recoil}}=&\int_0^{2\pi}\int_0^{\pi} \frac{-1}{4\omega}\Im\left\{\mathbf{E}_{\mathrm{p}}\left(\mathbf{H}_{\mathrm{p}}^*\cdot\hat{\mathbf{n}}\right)+\mathbf{H}_{\mathrm{p}}^*\left(\mathbf{E}_{\mathrm{p}}\cdot\hat{\mathbf{n}}\right)\right\}r^2\sin\theta\mathrm{d}\theta\mathrm{d}\phi\\
&+\int_0^{2\pi}\int_0^{\pi}\frac{1}{4\omega}\Im\left\{\left[(r\hat{\mathbf{n}}\times\nabla)\otimes\mathbf{E}_{\mathrm{p}}^*\right]\times\mathbf{H}_{\mathrm{p}}\right\}\cdot \hat{\mathbf{n}}r^2\sin\theta\mathrm{d}\theta\mathrm{d}\phi\\
&+\int_0^{2\pi}\int_0^{\pi}\frac{1}{4\omega}\Im\left\{\left[(r\hat{\mathbf{n}}\times\nabla)\otimes\mathbf{H}_{\mathrm{p}}\right]\times\mathbf{E}_{\mathrm{p}}^*\right\}\cdot \hat{\mathbf{n}}r^2\sin\theta\mathrm{d}\theta\mathrm{d}\phi.
\end{aligned}
\end{equation}

Due to the fact that the orbital angular momenta of the incident, radiation and total fields are separately conserved quantities, the corresponding contributions to the dipolar torque can be calculated independent of the radius of the enclosed surface,
\begin{equation}
\begin{aligned}
&\mathbf{\Gamma}^o_{\mathrm{p}}=\lim_{r\rightarrow \infty}\mathbf{\Gamma}^o_{\mathrm{p}}(r)=\lim_{r\rightarrow 0}\mathbf{\Gamma}^o_{\mathrm{p}}(r), \\
&\mathbf{\Gamma}^o_{\mathrm{inc}}=\lim_{r\rightarrow \infty}\mathbf{\Gamma}^o_{\mathrm{inc}}(r)=\lim_{r\rightarrow 0}\mathbf{\Gamma}^o_{\mathrm{inc}}(r), \\
&\mathbf{\Gamma}^o_{\mathrm{p, recoil}}=\lim_{r\rightarrow \infty}\mathbf{\Gamma}^o_{\mathrm{p, recoil}}(r)=\lim_{r\rightarrow 0}\mathbf{\Gamma}^o_{\mathrm{p, recoil}}(r), \\
&\mathbf{\Gamma}^o_{\mathrm{p, mix}}=\lim_{r\rightarrow \infty}\mathbf{\Gamma}^o_{\mathrm{p, mix}}(r)=\lim_{r\rightarrow 0}\mathbf{\Gamma}^o_{\mathrm{p, mix}}(r), 
\end{aligned}
\end{equation}

In the small $r$ limit that $r\rightarrow0$, using the approximation that $\mathbf{E}_{\mathrm{inc}}(\mathbf{r}')\approx\mathbf{E}_{\mathrm{0}}+r(\hat{\mathbf{n}}\cdot\nabla)\mathbf{E}_{\mathrm{0}}$ and $\mathbf{H}_{\mathrm{inc}}(\mathbf{r}')\approx\mathbf{H}_{\mathrm{0}}+r(\hat{\mathbf{n}}\cdot\nabla)\mathbf{H}_{\mathrm{0}}$, it is easy to prove that the optical torque component $\mathbf{\Gamma}^s_{\mathrm{inc}}$ attributed purely to the incident field is zero:
\begin{equation}
\begin{aligned}
\mathbf{\Gamma}^o_{\mathrm{inc}}=&\lim_{r\rightarrow 0}r^2\int_0^{2\pi}\int_0^{\pi} \frac{-1}{4\omega}\Im\left\{\mathbf{E}_{\mathrm{0}}\left(\mathbf{H}_{\mathrm{0}}^*\cdot\hat{\mathbf{n}}\right)+\mathbf{H}_{\mathrm{0}}^*\left(\mathbf{E}_{\mathrm{0}}\cdot\hat{\mathbf{n}}\right)\right\}\sin\theta\mathrm{d}\theta\mathrm{d}\phi\\
&+\lim_{r\rightarrow 0}r^2\int_0^{2\pi}\int_0^{\pi}\frac{1}{4\omega}\Im\left\{\left[(r\hat{\mathbf{n}}\times\nabla)\otimes\mathbf{E}_{\mathrm{0}}^*\right]\times\mathbf{H}_{\mathrm{0}}\right\}\cdot \hat{\mathbf{n}}\sin\theta\mathrm{d}\theta\mathrm{d}\phi\\
&+\lim_{r\rightarrow 0}r^2\int_0^{2\pi}\int_0^{\pi}\frac{1}{4\omega}\Im\left\{\left[(r\hat{\mathbf{n}}\times\nabla)\otimes\mathbf{H}_{\mathrm{0}}\right]\times\mathbf{E}_{\mathrm{0}}^*\right\}\cdot \hat{\mathbf{n}}\sin\theta\mathrm{d}\theta\mathrm{d}\phi,\\
=&0.
\end{aligned}
\end{equation}

The various components of $\mathbf{\Gamma}^o_{\mathrm{p,mix}}$ can be derived as follows. First of all, using the analytical expressions of integrals developed when deriving $\mathbf{\Gamma}^s_{\mathrm{p,mix}}$, one can easily arrive at the result that
\begin{equation}
\begin{aligned}
&\int_0^{2\pi}\int_0^{\pi} \frac{-1}{4\omega}\Im\Big\{\mathbf{E}_{\mathrm{inc}}\left(\mathbf{H}_{\mathrm{p}}^*\cdot\hat{\mathbf{n}}\right)+\mathbf{H}_{\mathrm{p}}^*\left(\mathbf{E}_{\mathrm{inc}}\cdot\hat{\mathbf{n}}\right)+\mathbf{E}_{\mathrm{p}}\left(\mathbf{H}_{\mathrm{inc}}^*\cdot\hat{\mathbf{n}}\right)+\mathbf{H}_{\mathrm{inc}}^*\left(\mathbf{E}_{\mathrm{p}}\cdot\hat{\mathbf{n}}\right)\Big\}r^2\sin\theta\mathrm{d}\theta\mathrm{d}\phi\\
&=\lim_{r\to0}\int_0^{2\pi}\int_0^{\pi} \frac{-1}{4\omega}\Im\Big\{\mathbf{H}_{\mathrm{p}}^*\left(\mathbf{E}_{\mathrm{inc}}\cdot\hat{\mathbf{n}}\right)+\mathbf{E}_{\mathrm{p}}\left(\mathbf{H}_{\mathrm{inc}}^*\cdot\hat{\mathbf{n}}\right)+\mathbf{H}_{\mathrm{inc}}^*\left(\mathbf{E}_{\mathrm{p}}\cdot\hat{\mathbf{n}}\right)\Big\}r^2\sin\theta\mathrm{d}\theta\mathrm{d}\phi\\
&=-\frac{2}{15}\Re\left\{\mathbf{p}\times\mathbf{E}^*_{\mathrm{0}}\right\}-\frac{4}{15\omega\varepsilon_0}\Im\left\{(\mathbf{p}\cdot\nabla)\mathbf{H}_{\mathrm{0}}^*\right\},
\end{aligned}
\end{equation}

Using the dipolar radiation field properties in Eq. \ref{eq:emdipoleradiation} and Eq. \ref{eq:emdipoleGreen} and the aforementioned integration properties of the unit vector $\hat{\mathbf{n}}$, one can arrive at the following result,
\begin{equation}
\begin{aligned}
&\lim_{r\to0}\int_0^{2\pi}\int_0^{\pi}\frac{1}{4\omega}\Im\Big\{\big[(r\hat{\mathbf{n}}\times\nabla)\otimes\mathbf{E}_{\mathrm{inc}}^*\big]\times\mathbf{H}_{\mathrm{p}}\Big\}\cdot \hat{\mathbf{n}}r^2\sin\theta\mathrm{d}\theta\mathrm{d}\phi\\
&=\lim_{r\to0}\Im\int_0^{2\pi}\int_0^{\pi}\frac{i}{16\pi}\Big\{\big[(r\hat{\mathbf{n}}\times\nabla)\otimes\mathbf{E}_{\mathrm{inc}}^*\big]\times\left(\hat{\mathbf{n}}\times\mathbf{p}\right)\Big\}\cdot \hat{\mathbf{n}}\sin\theta\mathrm{d}\theta\mathrm{d}\phi\\
&=0,
\end{aligned}
\end{equation}

Using the dipolar radiation field properties in Eq. \ref{eq:emdipoleradiation} and Eq. \ref{eq:emdipoleGreen}, the aforementioned integration properties of the unit vector $\hat{\mathbf{n}}$, and the approximation of the incident field in the small $r$ limit $\mathbf{E}_{\mathrm{inc}}(\mathbf{r}')\approx\mathbf{E}_{\mathrm{0}}+r(\hat{\mathbf{n}}\cdot\nabla)\mathbf{E}_{\mathrm{0}}$ and $\mathbf{H}_{\mathrm{inc}}(\mathbf{r}')\approx\mathbf{H}_{\mathrm{0}}+r(\hat{\mathbf{n}}\cdot\nabla)\mathbf{H}_{\mathrm{0}}$, one can derive that
\begin{equation}
\begin{aligned}\label{eq:appDOmix3}
&\lim_{r\to0}\int_0^{2\pi}\int_0^{\pi}\frac{1}{4\omega}\Im\Big\{\big[(r\hat{\mathbf{n}}\times\nabla)\otimes\mathbf{E}_{\mathrm{p}}^*\big]\times\mathbf{H}_{\mathrm{inc}}\Big\}\cdot \hat{\mathbf{n}}r^2\sin\theta\mathrm{d}\theta\mathrm{d}\phi\\
&=\Im\int_0^{2\pi}\int_0^{\pi}\frac{i3k}{16\pi\omega\varepsilon_0}\Big\{\big\{(r\hat{\mathbf{n}}\times\nabla)\otimes\left[\left(\mathbf{p}^*\cdot \hat{\mathbf{n}}\right)\hat{\mathbf{n}}\right]\big\}\times\mathbf{H}_{\mathrm{0}}\Big\}\cdot \hat{\mathbf{n}}\sin\theta\mathrm{d}\theta\mathrm{d}\phi\\
&+\Im\int_0^{2\pi}\int_0^{\pi}\frac{3}{16\pi\omega\varepsilon_0}\Big\{\big\{(r\hat{\mathbf{n}}\times\nabla)\otimes\left[\left(\mathbf{p}^*\cdot \hat{\mathbf{n}}\right)\hat{\mathbf{n}}\right]\big\}\times\left[\left(\hat{\mathbf{n}}\cdot\nabla\right)\mathbf{H}_{\mathrm{0}}\right]\Big\}\cdot \hat{\mathbf{n}}\sin\theta\mathrm{d}\theta\mathrm{d}\phi\\
&=\Im\int_0^{2\pi}\int_0^{\pi}\frac{3}{16\pi\omega\varepsilon_0}\Big\{\big\{(r\hat{\mathbf{n}}\times\nabla)\otimes\left[\left(\mathbf{p}^*\cdot \hat{\mathbf{n}}\right)\hat{\mathbf{n}}\right]\big\}\times\left[\left(\hat{\mathbf{n}}\cdot\nabla\right)\mathbf{H}_{\mathrm{0}}\right]\Big\}\cdot \hat{\mathbf{n}}\sin\theta\mathrm{d}\theta\mathrm{d}\phi\\
&=\frac{1}{20}\Re\left\{\mathbf{p}^*\times\mathbf{E}_{\mathrm{0}}\right\}-\frac{3}{20\omega\varepsilon_0}\Im\left\{(\mathbf{p}\cdot\nabla)\mathbf{H}_{\mathrm{0}}^*\right\}.
\end{aligned}
\end{equation}
In the derivation of Eq. \ref{eq:appDOmix3}, the following identity relations are applied,
\begin{equation}
\begin{aligned}
&(r\hat{\mathbf{n}}\times\nabla)\otimes\left[\left(\mathbf{p}^*\cdot \hat{\mathbf{n}}\right)\hat{\mathbf{n}}\right]\\
&=\begin{pmatrix}
-p_y^*n_x n_z+p_z^*n_xn_y & -p_x^*n_xn_z-2p_y^*n_yn_z-p_z^*(n_z^2-n_y^2) & p_x^*n_xn_y+p_y^*(n_y^2-n_z^2)+2p_z^*n_yn_z \\
2p_x^*n_xn_z+p_y^*n_yn_z+p_z^*(n_z^2-n_x^2) & p_x^*n_y n_z-p_z^*n_xn_y & -p_x^*(n_x^2-n_z^2)-p_y^*n_xn_y-2p_z^*n_xn_z\\
-2p_x^*n_xn_y-p_y^*(n_y^2-n_x^2)-p_z^*n_yn_z & p_x^*(n_x^2-n_y^2)+2p_y^*n_xn_y+p_z^*n_xn_z & -p_x^*n_y n_z+p_y^*n_xn_z
\end{pmatrix}\\
&=\begin{pmatrix}
n_x\left[\hat{\mathbf{e}}_x\cdot(\hat{\mathbf{n}}\times\mathbf{p}^*)\right] & -n_z(\mathbf{p}^*\cdot \hat{\mathbf{n}})+n_y\big[\hat{\mathbf{e}}_x\cdot(\hat{\mathbf{n}}\times\mathbf{p}^*)\big] & n_y(\mathbf{p}^*\cdot \hat{\mathbf{n}})+n_z\left[\hat{\mathbf{e}}_x\cdot(\hat{\mathbf{n}}\times\mathbf{p}^*)\right] \\
n_z(\mathbf{p}^*\cdot \hat{\mathbf{n}})+n_x\left[\hat{\mathbf{e}}_y\cdot(\hat{\mathbf{n}}\times\mathbf{p}^*)\right] & n_y\left[\hat{\mathbf{e}}_y\cdot(\hat{\mathbf{n}}\times\mathbf{p}^*)\right] & -n_x(\mathbf{p}^*\cdot \hat{\mathbf{n}})+n_z\left[\hat{\mathbf{e}}_y\cdot(\hat{\mathbf{n}}\times\mathbf{p}^*)\right]\\
-n_y(\mathbf{p}^*\cdot \hat{\mathbf{n}})+n_x\left[\hat{\mathbf{e}}_z\cdot(\hat{\mathbf{n}}\times\mathbf{p}^*)\right] & n_x(\mathbf{p}^*\cdot \hat{\mathbf{n}}) +n_y\left[\hat{\mathbf{e}}_z\cdot(\hat{\mathbf{n}}\times\mathbf{p}^*)\right] & n_z\left[\hat{\mathbf{e}}_z\cdot(\hat{\mathbf{n}}\times\mathbf{p}^*)\right]
\end{pmatrix}\\
&=\left(\mathbf{p}^*\cdot \hat{\mathbf{n}}\right)\left[(r\hat{\mathbf{n}}\times\nabla)\otimes\hat{\mathbf{n}}\right]+(\hat{\mathbf{n}}\times\mathbf{p}^*)\otimes\hat{\mathbf{n}},
\end{aligned}
\end{equation}
and
\begin{equation}
\begin{aligned}
\left[\overleftrightarrow{\mathbf{A}}\times\mathbf{a}\right]_{il}=\sum_{jk}\epsilon_{jkl}\left[\overleftrightarrow{\mathbf{A}}\right]_{ij} \left[\mathbf{a}\right]_k,
\end{aligned}
\end{equation}
where $\epsilon_{jkl}$ is the Levi-Civita symbol.

Using the dipolar radiation field properties in Eq. \ref{eq:emdipoleradiation} and Eq. \ref{eq:emdipoleGreen}, the aforementioned integration properties of the unit vector $\hat{\mathbf{n}}$, and the approximation of the incident field in the small $r$ limit $\mathbf{E}_{\mathrm{inc}}(\mathbf{r}')\approx\mathbf{E}_{\mathrm{0}}+r(\hat{\mathbf{n}}\cdot\nabla)\mathbf{E}_{\mathrm{0}}$ and $\mathbf{H}_{\mathrm{inc}}(\mathbf{r}')\approx\mathbf{H}_{\mathrm{0}}+r(\hat{\mathbf{n}}\cdot\nabla)\mathbf{H}_{\mathrm{0}}$, one can derive that
\begin{equation}
\begin{aligned}
&\lim_{r\to0}\int_0^{2\pi}\int_0^{\pi}\frac{1}{4\omega}\Im\left\{\left[(r\hat{\mathbf{n}}\times\nabla)\otimes\mathbf{H}_{\mathrm{inc}}\right]\times\mathbf{E}_{\mathrm{p}}^*\right\}\cdot \hat{\mathbf{n}}r^2\sin\theta\mathrm{d}\theta\mathrm{d}\phi\\
&=\Im\int_0^{2\pi}\int_0^{\pi}\frac{-1}{16\pi\omega\varepsilon_0}\left\{\left[(\hat{\mathbf{n}}\times\nabla)\otimes\mathbf{H}_{\mathrm{0}}\right]\times\mathbf{p}^*\right\}\cdot \hat{\mathbf{n}}\sin\theta\mathrm{d}\theta\mathrm{d}\phi\\
&+\Im\int_0^{2\pi}\int_0^{\pi}\frac{3}{16\pi\omega\varepsilon_0}\left\{\left[(\hat{\mathbf{n}}\times\nabla)\otimes\mathbf{H}_{\mathrm{0}}\right]\times\left[\left(\mathbf{p}^*\cdot \hat{\mathbf{n}}\right)\hat{\mathbf{n}}\right]\right\}\cdot \hat{\mathbf{n}}\sin\theta\mathrm{d}\theta\mathrm{d}\phi\\
&=\Im\int_0^{2\pi}\int_0^{\pi}\frac{-1}{16\pi\omega\varepsilon_0}\left\{\left[(\hat{\mathbf{n}}\times\nabla)\otimes\mathbf{H}_{\mathrm{0}}\right]\times\mathbf{p}^*\right\}\cdot \hat{\mathbf{n}}\sin\theta\mathrm{d}\theta\mathrm{d}\phi\\
&=-\frac{1}{12\omega\varepsilon_0}\Im\left\{(\mathbf{p}\cdot\nabla)\mathbf{H}_{\mathrm{0}}^*\right\},
\end{aligned}
\end{equation}

Using the dipolar radiation field properties in Eq. \ref{eq:emdipoleradiation} and Eq. \ref{eq:emdipoleGreen}, the aforementioned integration properties of the unit vector $\hat{\mathbf{n}}$, and the approximation of the incident field in the small $r$ limit $\mathbf{E}_{\mathrm{inc}}(\mathbf{r}')\approx\mathbf{E}_{\mathrm{0}}+r(\hat{\mathbf{n}}\cdot\nabla)\mathbf{E}_{\mathrm{0}}$ and $\mathbf{H}_{\mathrm{inc}}(\mathbf{r}')\approx\mathbf{H}_{\mathrm{0}}+r(\hat{\mathbf{n}}\cdot\nabla)\mathbf{H}_{\mathrm{0}}$, one can derive that
\begin{equation}
\begin{aligned}
&\lim_{r\to0}\int_0^{2\pi}\int_0^{\pi}\frac{1}{4\omega}\Im\left\{\left[(r\hat{\mathbf{n}}\times\nabla)\otimes\mathbf{H}_{\mathrm{p}}\right]\times\mathbf{E}_{\mathrm{inc}}^*\right\}\cdot \hat{\mathbf{n}}r^2\sin\theta\mathrm{d}\theta\mathrm{d}\phi\\
&=\Im\int_0^{2\pi}\int_0^{\pi}\frac{i}{16\pi}\left\{\left[(r\hat{\mathbf{n}}\times\nabla)\otimes\left(\hat{\mathbf{n}}\times\mathbf{p}\right)\right]\times\mathbf{E}_{\mathrm{0}}^*\right\}\cdot \hat{\mathbf{n}}\sin\theta\mathrm{d}\theta\mathrm{d}\phi\\
&=\Im\int_0^{2\pi}\int_0^{\pi}\frac{i}{16\pi}\Big\{\big\{\left[(r\hat{\mathbf{n}}\times\nabla)\otimes\hat{\mathbf{n}}\right]\times\mathbf{p}\big\}\times\mathbf{E}_{\mathrm{0}}^*\Big\}\cdot \hat{\mathbf{n}}\sin\theta\mathrm{d}\theta\mathrm{d}\phi\\
&=\frac{1}{12}\Re\left\{\mathbf{p}^*\times\mathbf{E}_{\mathrm{0}}\right\},
\end{aligned}
\end{equation}
where the following identity relation is applied in the derivation: 
\begin{equation}
\begin{aligned}
(r\hat{\mathbf{n}}\times\nabla)\otimes\hat{\mathbf{n}}=\begin{pmatrix}
0 & -n_z & n_y\\
n_z & 0 & -n_x\\
-n_y & n_x & 0
\end{pmatrix}.
\end{aligned}
\end{equation}

From above results, we can get the analytical expression for $\mathbf{\Gamma}^o_{\mathrm{p,mix}}$ 
\begin{equation}
\begin{aligned}
\mathbf{\Gamma}^o_{\mathrm{p,mix}}=-\frac{1}{2\omega\varepsilon_0}\Im\left[(\mathbf{p}\cdot\nabla)\mathbf{H}_{\mathrm{inc}}^*\right],\\
\end{aligned}
\end{equation}

The components of $\mathbf{\Gamma}^o_{\mathrm{p,recoil}}$ can be derived using the dipolar radiation field properties in Eq. \ref{eq:emdipoleradiation} and Eq. \ref{eq:emdipoleGreen} and the aforementioned integration properties of the unit vector $\hat{\mathbf{n}}$ in the large $r$ limit as $r\rightarrow\infty$,

\begin{equation}
\begin{aligned}
&\int_0^{2\pi}\int_0^{\pi} \frac{-1}{4\omega}\Im\left\{\mathbf{E}_{\mathrm{p}}\left(\mathbf{H}_{\mathrm{p}}^*\cdot\hat{\mathbf{n}}\right)+\mathbf{H}_{\mathrm{p}}^*\left(\mathbf{E}_{\mathrm{p}}\cdot\hat{\mathbf{n}}\right)\right\}r^2\sin\theta\mathrm{d}\theta\mathrm{d}\phi\\
&=\lim_{r\to\infty}\int_0^{2\pi}\int_0^{\pi} \frac{-1}{4\omega}\Im\big\{\mathbf{H}_{\mathrm{p}}^*\left(\mathbf{E}_{\mathrm{p}}\cdot\hat{\mathbf{n}}\right)\big\}r^2\sin\theta\mathrm{d}\theta\mathrm{d}\phi\\
&=0,
\end{aligned}
\end{equation}

\begin{equation}
\begin{aligned}
&\lim_{r\to\infty}\int_0^{2\pi}\int_0^{\pi}\frac{1}{4\omega}\Im\left\{\left[(r\hat{\mathbf{n}}\times\nabla)\otimes\mathbf{E}_{\mathrm{p}}^*\right]\times\mathbf{H}_{\mathrm{p}}\right\}\cdot \hat{\mathbf{n}}r^2\sin\theta\mathrm{d}\theta\mathrm{d}\phi\\
&=\Im\int_0^{2\pi}\int_0^{\pi}\frac{k^3}{64\pi^2\varepsilon_0}\left\{\left\{(r\hat{\mathbf{n}}\times\nabla)\otimes\left[\mathbf{p}^*-\left(\mathbf{p}^*\cdot \hat{\mathbf{n}}\right)\hat{\mathbf{n}}\right]\right\}\times\left(\hat{\mathbf{n}}\times\mathbf{p}\right)\right\}\cdot \hat{\mathbf{n}}\sin\theta\mathrm{d}\theta\mathrm{d}\phi\\
&=\Im\int_0^{2\pi}\int_0^{\pi}\frac{k^3}{64\pi^2\varepsilon_0}\left\{\left\{(r\hat{\mathbf{n}}\times\nabla)\otimes\left[-\left(\mathbf{p}^*\cdot \hat{\mathbf{n}}\right)\hat{\mathbf{n}}\right]\right\}\times\left(\hat{\mathbf{n}}\times\mathbf{p}\right)\right\}\cdot \hat{\mathbf{n}}\sin\theta\mathrm{d}\theta\mathrm{d}\phi\\
&=-\frac{k^3}{48\pi^2\varepsilon_0}\Im\left(\mathbf{p}^*\times\mathbf{p}\right),
\end{aligned}
\end{equation}

\begin{equation}
\begin{aligned}
&\lim_{r\to\infty}\int_0^{2\pi}\int_0^{\pi}\frac{1}{4\omega}\Im\left\{\left[(r\hat{\mathbf{n}}\times\nabla)\otimes\mathbf{H}_{\mathrm{p}}\right]\times\mathbf{E}_{\mathrm{p}}^*\right\}\cdot \hat{\mathbf{n}}r^2\sin\theta\mathrm{d}\theta\mathrm{d}\phi\\
&=\Im\int_0^{2\pi}\int_0^{\pi}\frac{k^3}{64\pi^2\varepsilon_0}\left\{\left\{(r\hat{\mathbf{n}}\times\nabla)\otimes \left(\hat{\mathbf{n}}\times\mathbf{p}\right)\right\}\times \left[\mathbf{p}^*-\left(\mathbf{p}^*\cdot \hat{\mathbf{n}}\right)\hat{\mathbf{n}}\right] \right\}\cdot \hat{\mathbf{n}}\sin\theta\mathrm{d}\theta\mathrm{d}\phi\\
&=\Im\int_0^{2\pi}\int_0^{\pi}\frac{k^3}{64\pi^2\varepsilon_0}\left\{\left\{(r\hat{\mathbf{n}}\times\nabla)\otimes \left(\hat{\mathbf{n}}\times\mathbf{p}\right)\right\}\times \mathbf{p}^* \right\}\cdot \hat{\mathbf{n}}\sin\theta\mathrm{d}\theta\mathrm{d}\phi\\
&=-\frac{k^3}{48\pi^2\varepsilon_0}\Im\left(\mathbf{p}^*\times\mathbf{p}\right),
\end{aligned}
\end{equation}
and we can get the analytical expression for $\mathbf{\Gamma}^o_{\mathrm{p,recoil}}$ as
\begin{equation}
\begin{aligned}
\mathbf{\Gamma}^o_{\mathrm{p,recoil}}=-\frac{k^3}{24\pi^2\varepsilon_0}\Im\left\{\mathbf{p}^*\times\mathbf{p}\right\}.
\end{aligned}
\end{equation}

\section{Derivation of the analytical expression of the optical torque on an induced electric quadrupole}\label{appendixE}
In this section, a detailed derivation for the analytical expression of the optical quadrupolar torque $\mathbf{\Gamma}_{\mathrm{Qe}}$ is provided. $\mathbf{\Gamma}_{\mathrm{Qe}}$ is derived in the same way as the dipolar torques using the total angular momentum flux, 
\begin{equation}
\begin{aligned}
\mathbf{\Gamma}=&\Re \int_0^{2\pi}\int_{0}^{\pi} (r\hat{\mathbf{n}})\times\left\{\frac{\varepsilon_0}{2}\mathbf{E}_{\mathrm{tot}}\left(\mathbf{E}_{\mathrm{tot}}^*\cdot\hat{\mathbf{n}}\right)+\frac{\mu_0}{2}\mathbf{H}_{\mathrm{tot}}\left(\mathbf{H}_{\mathrm{tot}}^*\cdot\hat{\mathbf{n}}\right)\right\}  r^2\sin\theta \mathrm{d}\theta \mathrm{d}\phi,
\end{aligned}
\end{equation}
where we only consider the torque contributed by the induced electric quadrupole and the incident field so that $\mathbf{E}_{\mathrm{tot}}(\mathbf{r}')=\mathbf{E}_{\mathrm{inc}}(\mathbf{r}')+\mathbf{E}_{\mathrm{Qe}}(\mathbf{r}')$ and $\mathbf{H}_{\mathrm{tot}}(\mathbf{r}')=\mathbf{H}_{\mathrm{inc}}(\mathbf{r}')+\mathbf{H}_{\mathrm{Qe}}(\mathbf{r}')$. 

Like the dipolar torque, $\mathbf{\Gamma}_{\mathrm{Qe}}$ attributed to the interaction between the induced electric quadrupole and incident field can be decomposed into different parts, 
\begin{equation}
\begin{aligned}
\mathbf{\Gamma}_{\mathrm{Qe}}=\mathbf{\Gamma}_{\mathrm{inc}}+\mathbf{\Gamma}_{\mathrm{Qe,mix}}+\mathbf{\Gamma}_{\mathrm{Qe,recoil}},
\end{aligned}
\end{equation}
As known in previous sections, the torque component purely dependent on the incident field does not contribute to the total quadrupolar torque, 
\begin{equation}
\begin{aligned}
\mathbf{\Gamma}_{\mathrm{inc}}&=0,
\end{aligned}
\end{equation}
while the extinction torque $\mathbf{\Gamma}_{\mathrm{Qe,mix}}$, as a result of the interference between incident and radiation fields, can be separated into different components as
\begin{equation}
\begin{aligned}
\mathbf{\Gamma}_{\mathrm{Qe,mix}}&=\frac{\varepsilon_0r^3}{2}\Re \int_0^{2\pi}\int_{0}^{\pi} \left(\hat{\mathbf{n}}\times\mathbf{E}_{\mathrm{inc}}\right)\left(\mathbf{E}_{\mathrm{Qe}}^*\cdot\hat{\mathbf{n}}\right) \sin\theta \mathrm{d}\theta \mathrm{d}\phi \\
&+\frac{\varepsilon_0r^3}{2}\Re \int_0^{2\pi}\int_{0}^{\pi} \left(\hat{\mathbf{n}}\times\mathbf{E}_{\mathrm{Qe}}\right)\left(\mathbf{E}_{\mathrm{inc}}^*\cdot\hat{\mathbf{n}}\right) \sin\theta \mathrm{d}\theta \mathrm{d}\phi\\
&+\frac{\mu_0r^3}{2}\Re \int_0^{2\pi}\int_{0}^{\pi} \left(\hat{\mathbf{n}}\times\mathbf{H}_{\mathrm{inc}}\right)\left(\mathbf{H}_{\mathrm{Qe}}^*\cdot\hat{\mathbf{n}}\right) \sin\theta \mathrm{d}\theta \mathrm{d}\phi \\
&+\frac{\mu_0r^3}{2}\Re \int_0^{2\pi}\int_{0}^{\pi} \left(\hat{\mathbf{n}}\times\mathbf{H}_{\mathrm{Qe}}\right)\left(\mathbf{H}_{\mathrm{inc}}^*\cdot\hat{\mathbf{n}}\right) \sin\theta \mathrm{d}\theta \mathrm{d}\phi\\
\end{aligned}
\end{equation}
and the recoil torque $\mathbf{\Gamma}_{\mathrm{Qe,recoil}}$ as a result of self-interaction of the induced electric quadrupole is expressed as
\begin{equation}
\begin{aligned}
\mathbf{\Gamma}_{\mathrm{Qe,recoil}}&=\frac{\varepsilon_0r^3}{2}\Re \int_0^{2\pi}\int_{0}^{\pi} \left(\hat{\mathbf{n}}\times\mathbf{E}_{\mathrm{Qe}}\right)\left(\mathbf{E}_{\mathrm{Qe}}^*\cdot\hat{\mathbf{n}}\right) \sin\theta \mathrm{d}\theta \mathrm{d}\phi \\
&+ \frac{\mu_0r^3}{2}\Re \int_0^{2\pi}\int_{0}^{\pi}\left(\hat{\mathbf{n}}\times\mathbf{H}_{\mathrm{Qe}}\right)\left(\mathbf{H}_{\mathrm{Qe}}^*\cdot\hat{\mathbf{n}}\right) \sin\theta \mathrm{d}\theta \mathrm{d}\phi. \\
\end{aligned}
\end{equation}

Due to the fact that the angular momenta of the incident, radiation and total fields are separately conserved quantities, the corresponding contributions to the dipolar torque can be calculated independent of the radius of the enclose surface,
\begin{equation}
\begin{aligned}
&\mathbf{\Gamma}_{\mathrm{Qe}}=\lim_{r\rightarrow \infty}\mathbf{\Gamma}_{\mathrm{Qe}}(r)=\lim_{r\rightarrow 0}\mathbf{\Gamma}_{\mathrm{Qe}}(r), \\
&\mathbf{\Gamma}_{\mathrm{Qe,recoil}}=\lim_{r\rightarrow \infty}\mathbf{\Gamma}_{\mathrm{Qe,recoil}}(r)=\lim_{r\rightarrow 0}\mathbf{\Gamma}_{\mathrm{Qe,recoil}}(r), \\
&\mathbf{\Gamma}_{\mathrm{Qe,mix}}=\lim_{r\rightarrow \infty}\mathbf{\Gamma}_{\mathrm{Qe,mix}}(r)=\lim_{r\rightarrow 0}\mathbf{\Gamma}_{\mathrm{Qe,mix}}(r). 
\end{aligned}
\end{equation}

It follows from Eq. \ref{eq:emquadrupoleradiation} and Eq. \ref{eq:emquadrupoleGreen} that
\begin{equation}
\begin{aligned}
\mathbf{H}_{\mathrm{Qe}}^*\cdot\hat{\mathbf{n}}=0,
\end{aligned}
\end{equation}

It is thus easy to prove that 
\begin{equation}
\begin{aligned}
&\frac{\mu_0r^3}{2}\Re \int_0^{2\pi}\int_{0}^{\pi} \left(\hat{\mathbf{n}}\times\mathbf{H}_{\mathrm{inc}}\right)\left(\mathbf{H}_{\mathrm{Qe}}^*\cdot\hat{\mathbf{n}}\right) \sin\theta \mathrm{d}\theta \mathrm{d}\phi=0,\\
&\frac{\mu_0r^3}{2}\Re \int_0^{2\pi}\int_{0}^{\pi} \left(\hat{\mathbf{n}}\times\mathbf{H}_{\mathrm{Qe}}\right)\left(\mathbf{H}_{\mathrm{Qe}}^*\cdot\hat{\mathbf{n}}\right) \sin\theta \mathrm{d}\theta \mathrm{d}\phi=0.
\end{aligned}
\end{equation}

The analytical expressions of the remaining non-zero terms of $\mathbf{\Gamma}_{\mathrm{Qe}}$ can be derived as follows. 
 
It follows from Eq. \ref{eq:emquadrupoleradiation} and Eq. \ref{eq:emquadrupoleGreen} that
\begin{equation}
\begin{aligned}
\mathbf{E}_{\mathrm{Qe}}^*\cdot\hat{\mathbf{n}}=&\frac{-i k^3 e^{-ikr}}{24\pi \varepsilon_0 r}\left[\frac{-3i}{kr}+\frac{9}{(kr)^2}+\frac{9i}{(kr)^3}\right]\left[\left(\overleftrightarrow{\mathbf{Q}}^{\mathrm{e}*}\cdot\hat{\mathbf{n}}\right)\cdot\hat{\mathbf{n}} \right],
\end{aligned}
\end{equation}

Using the aforementioned properties of $\hat{\mathbf{n}}$ and the approximation that $\mathbf{E}_{\mathrm{inc}}(\mathbf{r})\approx\mathbf{E}_{\mathrm{0}}+r(\hat{\mathbf{n}}\cdot\nabla)\mathbf{E}_{\mathrm{0}}$, we can derive that: 

\begin{equation}
\begin{aligned}\label{eq:AppEtQemix2}
&\lim_{r\rightarrow0}\frac{\varepsilon_0r^3}{2}\Re \int_0^{2\pi}\int_{0}^{\pi} \left(\hat{\mathbf{n}}\times\mathbf{E}_{\mathrm{inc}}\right)\left(\mathbf{E}_{\mathrm{Qe}}^*\cdot\hat{\mathbf{n}}\right) \sin\theta \mathrm{d}\theta \mathrm{d}\phi\\
&=\frac{\varepsilon_0r^3}{2}\Re \int_0^{2\pi}\int_{0}^{\pi} \left(\hat{\mathbf{n}}\times\mathbf{E}_{\mathrm{0}}\right) \times\frac{-i k^3}{24\pi \varepsilon_0 r}\left[\frac{9}{(kr)^2}\right]\left[\left(\overleftrightarrow{\mathbf{Q}}^{\mathrm{e}*}\cdot\hat{\mathbf{n}}\right)\cdot\hat{\mathbf{n}} \right]\sin\theta \mathrm{d}\theta \mathrm{d}\phi\\
&+\frac{\varepsilon_0r^3}{2}\Re \int_0^{2\pi}\int_{0}^{\pi} \left\{\hat{\mathbf{n}}\times\left[(r\hat{\mathbf{n}}\cdot\nabla)\mathbf{E}_{\mathrm{0}}\right]\right\}\frac{-i k^3 }{24\pi \varepsilon_0 r}\left[\frac{9i}{(kr)^3}\right]\left[\left(\overleftrightarrow{\mathbf{Q}}^{\mathrm{e}*}\cdot\hat{\mathbf{n}}\right)\cdot\hat{\mathbf{n}} \right] \sin\theta \mathrm{d}\theta \mathrm{d}\phi\\
&=\Re \frac{-3 i k}{16\pi}\int_0^{2\pi}\int_{0}^{\pi} \left(\hat{\mathbf{n}}\times\mathbf{E}_{\mathrm{0}}\right)\left[\left(\overleftrightarrow{\mathbf{Q}}^{\mathrm{e}*}\cdot\hat{\mathbf{n}}\right)\cdot\hat{\mathbf{n}} \right]\sin\theta \mathrm{d}\theta \mathrm{d}\phi+\Re \frac{3}{16\pi}\int_0^{2\pi}\int_{0}^{\pi} \left\{\hat{\mathbf{n}}\times\left[(\hat{\mathbf{n}}\cdot\nabla)\mathbf{E}_{\mathrm{0}}\right]\right\}\left[\left(\overleftrightarrow{\mathbf{Q}}^{\mathrm{e}*}\cdot\hat{\mathbf{n}}\right)\cdot\hat{\mathbf{n}} \right]\sin\theta \mathrm{d}\theta \mathrm{d}\phi\\
&=\Re \frac{3}{16\pi}\int_0^{2\pi}\int_{0}^{\pi} \left\{\hat{\mathbf{n}}\times\left[(\hat{\mathbf{n}}\cdot\nabla)\mathbf{E}_{\mathrm{0}}\right]\right\}\left[\left(\overleftrightarrow{\mathbf{Q}}^{\mathrm{e}*}\cdot\hat{\mathbf{n}}\right)\cdot\hat{\mathbf{n}} \right]\sin\theta \mathrm{d}\theta \mathrm{d}\phi\\
&=\Re\left\{\frac{ikZ_0}{10}\mathbf{H}_{\mathrm{0}}\left(\sum_{u=x,y,z}Q^{e*}_{uu}\right)\right\}-\Re\left\{\frac{ikZ_0}{20}\left(\overleftrightarrow{\mathbf{Q}}^{\mathrm{e}*}\cdot\mathbf{H}_{\mathrm{0}}\right)\right\}+\frac{1}{10}\Re\left\{\sum_{u=x,y,z}\mathbf{Q}^{\mathrm{e}*}_{u}\times\mathbf{D}^{\mathrm{e}}_{u}\right\},
\end{aligned}
\end{equation}
where we introduce the notation that $\mathbf{D}^{e}_{u}=\sum_{v=x,y,z}\mathbf{\hat{e}}_{v}\left[\overleftrightarrow{\mathbcal{D}^e}\right]_{uv}$ and  $\mathbf{Q}^{e}_{u}=\sum_{v=x,y,z}\mathbf{\hat{e}}_{v}\left[\overleftrightarrow{\mathbf{Q}^e}\right]_{uv}$

For an induced electric quadrupole moment in an isotropic Mie particle in which $\mathbf{Q}^{e}_{u}=\alpha_{\mathrm{Qe}}\mathbf{D}^{e}_{u}$, using the relations in Eq. \ref{eq:emquadrupolemoment} and Eq. \ref{eq:emquadrupolefield} so that $\sum_{u=x,y,z}Q^{e*}_{uu}=0$ and the fact that $\mathbf{D}^{\mathrm{e}*}_{u}\times\mathbf{D}^{\mathrm{e}}_{u}$ are purely imaginary, Eq. \ref{eq:AppEtQemix2} is reduced to 
\begin{equation}
\begin{aligned}
&\lim_{r\rightarrow0}\frac{\varepsilon_0r^3}{2}\Re \int_0^{2\pi}\int_{0}^{\pi} \left(\hat{\mathbf{n}}\times\mathbf{E}_{\mathrm{inc}}\right)\left(\mathbf{E}_{\mathrm{Qe}}^*\cdot\hat{\mathbf{n}}\right) \sin\theta \mathrm{d}\theta \mathrm{d}\phi\\
&=\Re\left\{\frac{-ikZ_0}{20}\left(\overleftrightarrow{\mathbf{Q}}^{\mathrm{e}*}\cdot\mathbf{H}_{\mathrm{0}}\right)\right\}+\frac{\Im\{\alpha_{\mathrm{Qe}}\}}{10}\sum_{u=x,y,z}\Im\left\{\mathbf{D}^{\mathrm{e}*}_{u}\times\mathbf{D}^{\mathrm{e}}_{u}\right\}, 
\end{aligned}
\end{equation}

It follows from Eq. \ref{eq:emquadrupoleradiation} and Eq. \ref{eq:emquadrupoleGreen} that
\begin{equation}
\begin{aligned}
\hat{\mathbf{n}}\times\mathbf{E}_{\mathrm{Qe}}&=\frac{i k^3 e^{ikr}}{24\pi \varepsilon_0 r}\left[-1-\frac{3i}{kr}+\frac{6}{(kr)^2}+\frac{6i}{(kr)^3}\right]\left[\hat{\mathbf{n}}\times\left(\overleftrightarrow{\mathbf{Q}}^{\mathrm{e}}\cdot\hat{\mathbf{n}}\right) \right],
\end{aligned}
\end{equation}
and using the aforementioned properties of $\hat{\mathbf{n}}$ and the approximation that $\mathbf{E}_{\mathrm{inc}}(\mathbf{r})\approx\mathbf{E}_{\mathrm{0}}+r(\hat{\mathbf{n}}\cdot\nabla)\mathbf{E}_{\mathrm{0}}$, we can derive that
\begin{equation}
\begin{aligned}
&\lim_{r\rightarrow0}\frac{\varepsilon_0r^3}{2}\Re \int_0^{2\pi}\int_{0}^{\pi} \left(\hat{\mathbf{n}}\times\mathbf{E}_{\mathrm{Qe}}\right)\left(\mathbf{E}_{\mathrm{inc}}^*\cdot\hat{\mathbf{n}}\right) \sin\theta \mathrm{d}\theta \mathrm{d}\phi\\
&=\frac{\varepsilon_0r^3}{2}\Re \int_0^{2\pi}\int_{0}^{\pi} \frac{i k^3 }{24\pi \varepsilon_0 r}\left[\frac{6}{(kr)^2}\right]\left[\hat{\mathbf{n}}\times\left(\overleftrightarrow{\mathbf{Q}}^{\mathrm{e}}\cdot\hat{\mathbf{n}}\right) \right]\left(\mathbf{E}_{\mathrm{0}}^*\cdot\hat{\mathbf{n}}\right) \sin\theta \mathrm{d}\theta \mathrm{d}\phi\\
&+\frac{\varepsilon_0r^3}{2}\Re \int_0^{2\pi}\int_{0}^{\pi}\frac{i k^3}{24\pi \varepsilon_0 r}\left[\frac{6i}{(kr)^3}\right]\left[\hat{\mathbf{n}}\times\left(\overleftrightarrow{\mathbf{Q}}^{\mathrm{e}}\cdot\hat{\mathbf{n}}\right)\right]\left\{\left[(r\hat{\mathbf{n}}\cdot\nabla)\mathbf{E}^*_{\mathrm{0}}\right]\cdot\hat{\mathbf{n}}\right\} \sin\theta \mathrm{d}\theta \mathrm{d}\phi \\
&=\Re \frac{-1}{8\pi}\int_0^{2\pi}\int_{0}^{\pi} \left[\hat{\mathbf{n}}\times\left(\overleftrightarrow{\mathbf{Q}}^{\mathrm{e}}\cdot\hat{\mathbf{n}}\right)\right]\left\{\left[(\hat{\mathbf{n}}\cdot\nabla)\mathbf{E}^*_{\mathrm{0}}\right]\cdot\hat{\mathbf{n}}\right\} \sin\theta \mathrm{d}\theta \mathrm{d}\phi \\
&=\frac{1}{15}\Re\left\{\sum_{u=x,y,z}\left(\overleftrightarrow{\mathbf{Q}}^{\mathrm{e}*}\cdot\hat{\mathbf{e}}_{u}\right)\times\left(\overleftrightarrow{\mathbf{D}}^{\mathrm{e}}\cdot\hat{\mathbf{e}}_{u}\right)\right\}\\
&=\frac{1}{15}\Re\left\{\sum_{u=x,y,z}\mathbf{Q}^{\mathrm{e}*}_{u}\times\mathbf{D}^{\mathrm{e}}_{u}\right\},
\end{aligned}
\end{equation}

For an induced electric quadrupole moment in an isotropic Mie particle, this is equivalent to 
\begin{equation}
\begin{aligned}
&\lim_{r\rightarrow0}\frac{\varepsilon_0r^3}{2}\Re \int_0^{2\pi}\int_{0}^{\pi} \left(\hat{\mathbf{n}}\times\mathbf{E}_{\mathrm{Qe}}\right)\left(\mathbf{E}_{\mathrm{inc}}^*\cdot\hat{\mathbf{n}}\right) \sin\theta \mathrm{d}\theta \mathrm{d}\phi\\
&=\frac{1}{15}\Re\left\{\sum_{u=x,y,z}\mathbf{Q}^{\mathrm{e}*}_{u}\times\mathbf{D}^{\mathrm{e}}_{u}\right\}\\
&=\frac{\Im\{\alpha_{\mathrm{Qe}}\}}{15}\sum_{u=x,y,z}\Im\left\{\mathbf{D}^{\mathrm{e}*}_{u}\times\mathbf{D}^{\mathrm{e}}_{u}\right\}\\
&=\frac{\Im\{\alpha_{\mathrm{Qe}}\}}{15}\Im\left\{\sum_j \hat{\mathbf{e}}_j \sum_{l}\sum_{u}\sum_{v}\epsilon_{jlv}\left[\overleftrightarrow{\mathbcal{D}^\mathrm{e}}\right]_{lu}^*\left[\overleftrightarrow{\mathbcal{D}^\mathrm{e}}\right]_{uv}\right\},
\end{aligned}
\end{equation}

It follows from Eq. \ref{eq:emquadrupoleradiation} and Eq. \ref{eq:emquadrupoleGreen} that
\begin{equation}
\begin{aligned}
\hat{\mathbf{n}}\times\mathbf{H}_{\mathrm{Qe}}&=\frac{-i k^3 c_0 e^{ikr}}{24\pi r}\left[1+\frac{3i}{kr}-\frac{3}{(kr)^2}\right]\left\{\hat{\mathbf{n}}\times\left[\hat{\mathbf{n}}\times\left(\overleftrightarrow{\mathbf{Q}}^{\mathrm{e}}\cdot\hat{\mathbf{n}}\right) \right]\right\},
\end{aligned}
\end{equation}
using the aforementioned properties of $\hat{\mathbf{n}}$ and the approximation that $\mathbf{H}^*_{\mathrm{inc}}(\mathbf{r})\approx\mathbf{H}^*_{\mathrm{0}}+r(\hat{\mathbf{n}}\cdot\nabla)\mathbf{H}^*_{\mathrm{0}}$, we can derive that in the small $r$ limit,
\begin{equation}
\begin{aligned}
&\lim_{r\rightarrow0}\frac{\mu_0r^4}{2}\Re \int_0^{2\pi}\int_{0}^{\pi} \left(\hat{\mathbf{n}}\times\mathbf{H}_{\mathrm{Qe}}\right)\left\{\left[(\hat{\mathbf{n}}\cdot\nabla)\mathbf{H}_{\mathrm{0}}^*\right]\cdot\hat{\mathbf{n}}\right\}\sin\theta \mathrm{d}\theta \mathrm{d}\phi=0,
\end{aligned}
\end{equation}

\begin{equation}
\begin{aligned}
&\lim_{r\rightarrow0}\frac{\mu_0r^3}{2}\Re \int_0^{2\pi}\int_{0}^{\pi} \left(\hat{\mathbf{n}}\times\mathbf{H}_{\mathrm{Qe}}\right)\left(\mathbf{H}_{\mathrm{inc}}^*\cdot\hat{\mathbf{n}}\right) \sin\theta \mathrm{d}\theta \mathrm{d}\phi\\
&=\Re  \frac{i k c_0\mu_0}{16\pi}\int_0^{2\pi}\int_{0}^{\pi}\left\{\hat{\mathbf{n}}\times\left[\hat{\mathbf{n}}\times\left(\overleftrightarrow{\mathbf{Q}}^{\mathrm{e}}\cdot\hat{\mathbf{n}}\right) \right]\right\}\left(\mathbf{H}_{\mathrm{0}}^*\cdot\hat{\mathbf{n}}\right) \sin\theta \mathrm{d}\theta \mathrm{d}\phi\\
&=\Re\left\{\frac{ikZ_0}{60}\mathbf{H}^*_{\mathrm{0}}\left(\sum_{u=x,y,z}Q^{\mathrm{e}}_{uu}\right)\right\}-\Re\left\{\frac{ikZ_0}{20}\left(\overleftrightarrow{\mathbf{Q}}^{\mathrm{e}}\cdot\mathbf{H}_{\mathrm{0}}^*\right)\right\}\\
&=\Re\left\{\frac{-ikZ_0}{60}\mathbf{H}_{\mathrm{0}}\left(\sum_{u=x,y,z}Q^{\mathrm{e}*}_{uu}\right)\right\}+\Re\left\{\frac{ikZ_0}{20}\left(\overleftrightarrow{\mathbf{Q}}^{\mathrm{e}*}\cdot\mathbf{H}_{\mathrm{0}}\right)\right\},
\end{aligned}
\end{equation}

For an induced electric quadrupole moment in an isotropic Mie particle, this reduces to 
\begin{equation}
\begin{aligned}
&\lim_{r\rightarrow0}\frac{\mu_0r^3}{2}\Re \int_0^{2\pi}\int_{0}^{\pi} \left(\hat{\mathbf{n}}\times\mathbf{H}_{\mathrm{Qe}}\right)\left(\mathbf{H}_{\mathrm{inc}}^*\cdot\hat{\mathbf{n}}\right) \sin\theta \mathrm{d}\theta \mathrm{d}\phi\\
&=\Re\left\{\frac{ikZ_0}{20}\left(\overleftrightarrow{\mathbf{Q}}^{\mathrm{e}*}\cdot\mathbf{H}_{\mathrm{0}}\right)\right\}
\end{aligned}
\end{equation}

To summarise the above results, the analytical expression for $\mathbf{\Gamma}_{\mathrm{Qe,mix}}$ is
\begin{equation}
\begin{aligned}
\mathbf{\Gamma}_{\mathrm{Qe,mix}}=&\lim_{r\rightarrow0}\frac{\varepsilon_0r^3}{2}\Re \int_0^{2\pi}\int_{0}^{\pi} \left(\hat{\mathbf{n}}\times\mathbf{E}_{\mathrm{inc}}\right)\left(\mathbf{E}_{\mathrm{Qe}}^*\cdot\hat{\mathbf{n}}\right) \sin\theta \mathrm{d}\theta \mathrm{d}\phi\\
&+\lim_{r\rightarrow0}\frac{\varepsilon_0r^3}{2}\Re \int_0^{2\pi}\int_{0}^{\pi} \left(\hat{\mathbf{n}}\times\mathbf{E}_{\mathrm{Qe}}\right)\left(\mathbf{E}_{\mathrm{inc}}^*\cdot\hat{\mathbf{n}}\right) \sin\theta \mathrm{d}\theta \mathrm{d}\phi\\
&+\lim_{r\rightarrow0}\frac{\mu_0r^3}{2}\Re \int_0^{2\pi}\int_{0}^{\pi} \left(\hat{\mathbf{n}}\times\mathbf{H}_{\mathrm{Qe}}\right)\left(\mathbf{H}_{\mathrm{inc}}^*\cdot\hat{\mathbf{n}}\right) \sin\theta \mathrm{d}\theta \mathrm{d}\phi\\
=&\Re\left\{\frac{ikZ_0}{12}\mathbf{H}_{\mathrm{0}}\left(\sum_{u=x,y,z}Q^{\mathrm{e}*}_{uu}\right)\right\}+\frac{1}{6}\Re\left\{\sum_{u=x,y,z}\mathbf{Q}^{\mathrm{e}*}_{u}\times\mathbf{D}^{\mathrm{e}}_{u}\right\}\\
\end{aligned}
\end{equation}

For an induced electric quadrupole moment in an isotropic Mie particle, this is equivalent to 
\begin{equation}
\begin{aligned}
\mathbf{\Gamma}_{\mathrm{Qe,mix}}=&\frac{\Im\{\alpha_{\mathrm{Qe}}\}}{6}\Im\left\{\sum_{u=x,y,z}\mathbf{D}^{\mathrm{e}*}_{u}\times\mathbf{D}^{\mathrm{e}}_{u}\right\}=\frac{120\pi\varepsilon_0}{k^5}\Re(a_2)\mathbf{s}^{\mathrm{Qe}},\\
\mathbf{s}^{\mathrm{Qe}}=&\frac{\varepsilon_0}{6}\Im\Bigg\{\sum_{u=x,y,z}\mathbf{D}^{\mathrm{e}*}_{u}\times\mathbf{D}^{\mathrm{e}}_{u}\Bigg\},\\
\end{aligned}
\end{equation}

The last nonzero term contributing to the electric quadrupolar torque is the `recoil' torque $\mathbf{\Gamma}_{\mathrm{p,recoil}}$, which can be derived using the dipolar radiation field properties in Eq. \ref{eq:emquadrupoleradiation}, Eq. \ref{eq:emquadrupoleGreen}, Eq. \ref{eq:emquadrupolefield} and the aforementioned integration properties of the unit vector $\hat{\mathbf{n}}$ in the large $r$ limit as $r\rightarrow\infty$,
\begin{equation}
\begin{aligned}
\mathbf{\Gamma}_{\mathrm{Qe,recoil}}=&\lim_{r\to\infty}\frac{\varepsilon_0r^3}{2}\Re \int_0^{2\pi}\int_{0}^{\pi} \left(\hat{\mathbf{n}}\times\mathbf{E}_{\mathrm{Qe}}\right)\left(\mathbf{E}_{\mathrm{Qe}}^*\cdot\hat{\mathbf{n}}\right) \sin\theta \mathrm{d}\theta \mathrm{d}\phi\\
&=\lim_{r\to\infty}\Re \frac{3i k^5}{2\varepsilon_0\cdot(24\pi)^2}\left[1+\frac{3i}{(kr)^3}+\frac{18i}{(kr)^5}\right]\int_0^{2\pi}\int_{0}^{\pi} \left[\hat{\mathbf{n}}\times\left(\overleftrightarrow{\mathbf{Q}}^{\mathrm{e}}\cdot\hat{\mathbf{n}}\right) \right] \left[\left(\overleftrightarrow{\mathbf{Q}}^{\mathrm{e}*}\cdot\hat{\mathbf{n}}\right)\cdot\hat{\mathbf{n}} \right] \sin\theta \mathrm{d}\theta \mathrm{d}\phi\\
&=\Re \frac{3i k^5}{2\varepsilon_0\cdot(24\pi)^2}\left[1\right]\int_0^{2\pi}\int_{0}^{\pi} \left[\hat{\mathbf{n}}\times\left(\overleftrightarrow{\mathbf{Q}}^{\mathrm{e}}\cdot\hat{\mathbf{n}}\right) \right] \left[\left(\overleftrightarrow{\mathbf{Q}}^{\mathrm{e}*}\cdot\hat{\mathbf{n}}\right)\cdot\hat{\mathbf{n}} \right] \sin\theta \mathrm{d}\theta \mathrm{d}\phi\\
&= \frac{3i k^5}{2\varepsilon_0\cdot(24\pi)^2}\left[\frac{-i8\pi}{15}\right]\sum_j\hat{\mathbf{e}}_j\sum_l\sum_u\sum_v \Im\left\{\epsilon_{lvj}Q^{\mathrm{e}}_{lu}Q^{\mathrm{e}*}_{uv}\right\}\\
&=-\frac{k^5}{720\pi\varepsilon_0}\Im\left\{\sum_{u=x,y,z}\mathbf{Q}^{\mathrm{e}*}_{u}\times\mathbf{Q}^{\mathrm{e}}_{u}\right\},
\end{aligned}
\end{equation}

For an induced electric quadrupole moment in an isotropic Mie particle, this is equivalent to 
\begin{equation}
\begin{aligned}
\mathbf{\Gamma}_{\mathrm{Qe,recoil}}=&=-\frac{k^5}{720\pi\varepsilon_0}\Im\left\{\sum_{u=x,y,z}\mathbf{Q}^{\mathrm{e}*}_{u}\times\mathbf{Q}^{\mathrm{e}}_{u}\right\}\\
&=-\frac{ k^5}{720\pi\varepsilon_0}\Im\left\{|\alpha_{\mathrm{Qe}}|^2\sum_{u=x,y,z}\mathbf{D}^{\mathrm{e}*}_{u}\times\mathbf{D}^{\mathrm{e}}_{u}\right\}\\
&=-\frac{120\pi\varepsilon_0}{k^5}|a_2|^2\mathbf{s}^{\mathrm{Qe}}.
\end{aligned}
\end{equation}

To summarise above results, the torque acting on the isotropic Mie particle that is attributed to the induced electric quadrupole can be written analytically as, 
\begin{equation}
\begin{aligned}
\mathbf{\Gamma}_{\mathrm{Qe}}=\mathbf{\Gamma}_{\mathrm{Qe,mix}}+\mathbf{\Gamma}_{\mathrm{Qe,recoil}}=\frac{120\pi\varepsilon_0}{k^5}\left[\Re(a_2)-|a_2|^2\right]\mathbf{s}^{\mathrm{Qe}}
\end{aligned}
\end{equation}


\begin{thebibliography}{0}%
\makeatletter
\providecommand \@ifxundefined [1]{%
 \@ifx{#1\undefined}
}%
\providecommand \@ifnum [1]{%
 \ifnum #1\expandafter \@firstoftwo
 \else \expandafter \@secondoftwo
 \fi
}%
\providecommand \@ifx [1]{%
 \ifx #1\expandafter \@firstoftwo
 \else \expandafter \@secondoftwo
 \fi
}%
\providecommand \natexlab [1]{#1}%
\providecommand \enquote  [1]{``#1''}%
\providecommand \bibnamefont  [1]{#1}%
\providecommand \bibfnamefont [1]{#1}%
\providecommand \citenamefont [1]{#1}%
\providecommand \href@noop [0]{\@secondoftwo}%
\providecommand \href [0]{\begingroup \@sanitize@url \@href}%
\providecommand \@href[1]{\@@startlink{#1}\@@href}%
\providecommand \@@href[1]{\endgroup#1\@@endlink}%
\providecommand \@sanitize@url [0]{\catcode `\\12\catcode `\$12\catcode
  `\&12\catcode `\#12\catcode `\^12\catcode `\_12\catcode `\%12\relax}%
\providecommand \@@startlink[1]{}%
\providecommand \@@endlink[0]{}%
\providecommand \url  [0]{\begingroup\@sanitize@url \@url }%
\providecommand \@url [1]{\endgroup\@href {#1}{\urlprefix }}%
\providecommand \urlprefix  [0]{URL }%
\providecommand \Eprint [0]{\href }%
\providecommand \doibase [0]{https://doi.org/}%
\providecommand \selectlanguage [0]{\@gobble}%
\providecommand \bibinfo  [0]{\@secondoftwo}%
\providecommand \bibfield  [0]{\@secondoftwo}%
\providecommand \translation [1]{[#1]}%
\providecommand \BibitemOpen [0]{}%
\providecommand \bibitemStop [0]{}%
\providecommand \bibitemNoStop [0]{.\EOS\space}%
\providecommand \EOS [0]{\spacefactor3000\relax}%
\providecommand \BibitemShut  [1]{\csname bibitem#1\endcsname}%
\let\auto@bib@innerbib\@empty
\end{thebibliography}%


\begin{thebibliography}{22}%
\makeatletter
\providecommand \@ifxundefined [1]{%
 \@ifx{#1\undefined}
}%
\providecommand \@ifnum [1]{%
 \ifnum #1\expandafter \@firstoftwo
 \else \expandafter \@secondoftwo
 \fi
}%
\providecommand \@ifx [1]{%
 \ifx #1\expandafter \@firstoftwo
 \else \expandafter \@secondoftwo
 \fi
}%
\providecommand \natexlab [1]{#1}%
\providecommand \enquote  [1]{``#1''}%
\providecommand \bibnamefont  [1]{#1}%
\providecommand \bibfnamefont [1]{#1}%
\providecommand \citenamefont [1]{#1}%
\providecommand \href@noop [0]{\@secondoftwo}%
\providecommand \href [0]{\begingroup \@sanitize@url \@href}%
\providecommand \@href[1]{\@@startlink{#1}\@@href}%
\providecommand \@@href[1]{\endgroup#1\@@endlink}%
\providecommand \@sanitize@url [0]{\catcode `\\12\catcode `\$12\catcode
  `\&12\catcode `\#12\catcode `\^12\catcode `\_12\catcode `\%12\relax}%
\providecommand \@@startlink[1]{}%
\providecommand \@@endlink[0]{}%
\providecommand \url  [0]{\begingroup\@sanitize@url \@url }%
\providecommand \@url [1]{\endgroup\@href {#1}{\urlprefix }}%
\providecommand \urlprefix  [0]{URL }%
\providecommand \Eprint [0]{\href }%
\providecommand \doibase [0]{https://doi.org/}%
\providecommand \selectlanguage [0]{\@gobble}%
\providecommand \bibinfo  [0]{\@secondoftwo}%
\providecommand \bibfield  [0]{\@secondoftwo}%
\providecommand \translation [1]{[#1]}%
\providecommand \BibitemOpen [0]{}%
\providecommand \bibitemStop [0]{}%
\providecommand \bibitemNoStop [0]{.\EOS\space}%
\providecommand \EOS [0]{\spacefactor3000\relax}%
\providecommand \BibitemShut  [1]{\csname bibitem#1\endcsname}%
\let\auto@bib@innerbib\@empty
\bibitem [{\citenamefont {Barnett}\ \emph {et~al.}(2016)\citenamefont
  {Barnett}, \citenamefont {Allen}, \citenamefont {Cameron}, \citenamefont
  {Gilson}, \citenamefont {Padgett}, \citenamefont {Speirits},\ and\
  \citenamefont {Yao}}]{SMBarnettJO2016}%
  \BibitemOpen
  \bibfield  {author} {\bibinfo {author} {\bibfnamefont {S.~M.}\ \bibnamefont
  {Barnett}}, \bibinfo {author} {\bibfnamefont {L.}~\bibnamefont {Allen}},
  \bibinfo {author} {\bibfnamefont {R.~P.}\ \bibnamefont {Cameron}}, \bibinfo
  {author} {\bibfnamefont {C.~R.}\ \bibnamefont {Gilson}}, \bibinfo {author}
  {\bibfnamefont {M.~J.}\ \bibnamefont {Padgett}}, \bibinfo {author}
  {\bibfnamefont {F.~C.}\ \bibnamefont {Speirits}},\ and\ \bibinfo {author}
  {\bibfnamefont {A.~M.}\ \bibnamefont {Yao}},\ }\bibfield  {title} {\bibinfo
  {title} {{On the natures of the spin and orbital parts of optical angular
  momentum}},\ }\href@noop {} {\bibfield  {journal} {\bibinfo  {journal}
  {Journal of Optics}\ }\textbf {\bibinfo {volume} {18}},\ \bibinfo {pages}
  {064004} (\bibinfo {year} {2016})}\BibitemShut {NoStop}%
\bibitem [{\citenamefont {Bliokh}\ \emph {et~al.}(2014)\citenamefont {Bliokh},
  \citenamefont {Dressel},\ and\ \citenamefont {Nori}}]{KYBliokhNJP2014}%
  \BibitemOpen
  \bibfield  {author} {\bibinfo {author} {\bibfnamefont {K.~Y.}\ \bibnamefont
  {Bliokh}}, \bibinfo {author} {\bibfnamefont {J.}~\bibnamefont {Dressel}},\
  and\ \bibinfo {author} {\bibfnamefont {F.}~\bibnamefont {Nori}},\ }\bibfield
  {title} {\bibinfo {title} {{Conservation of the spin and orbital angular
  momenta in electromagnetism}},\ }\href@noop {} {\bibfield  {journal}
  {\bibinfo  {journal} {New Journal of Physics}\ }\textbf {\bibinfo {volume}
  {16}},\ \bibinfo {pages} {093037} (\bibinfo {year} {2014})}\BibitemShut
  {NoStop}%
\bibitem [{\citenamefont {Cameron}\ \emph {et~al.}(2012)\citenamefont
  {Cameron}, \citenamefont {Barnett},\ and\ \citenamefont
  {Yao}}]{RPCameronNJP2012}%
  \BibitemOpen
  \bibfield  {author} {\bibinfo {author} {\bibfnamefont {R.~P.}\ \bibnamefont
  {Cameron}}, \bibinfo {author} {\bibfnamefont {S.~M.}\ \bibnamefont
  {Barnett}},\ and\ \bibinfo {author} {\bibfnamefont {A.~M.}\ \bibnamefont
  {Yao}},\ }\bibfield  {title} {\bibinfo {title} {{Optical helicity, optical
  spin and related quantities in electromagnetic theory }},\ }\href@noop {}
  {\bibfield  {journal} {\bibinfo  {journal} {New Journal of Physics}\ }\textbf
  {\bibinfo {volume} {14}},\ \bibinfo {pages} {053050} (\bibinfo {year}
  {2012})}\BibitemShut {NoStop}%
\bibitem [{\citenamefont
  {Nieto-Vesperinas}(2015{\natexlab{a}})}]{MNVesperinasPRA2015}%
  \BibitemOpen
  \bibfield  {author} {\bibinfo {author} {\bibfnamefont {M.}~\bibnamefont
  {Nieto-Vesperinas}},\ }\bibfield  {title} {\bibinfo {title} {{Optical theorem
  for the conservation of electromagnetic helicity: Significance for molecular
  energy transfer and enantiomeric discrimination by circular dichroism}},\
  }\href@noop {} {\bibfield  {journal} {\bibinfo  {journal} {Physical Review
  A}\ }\textbf {\bibinfo {volume} {92}},\ \bibinfo {pages} {023813} (\bibinfo
  {year} {2015}{\natexlab{a}})}\BibitemShut {NoStop}%
\bibitem [{\citenamefont
  {Nieto-Vesperinas}(2015{\natexlab{b}})}]{MNVesperinasPRA2015j2}%
  \BibitemOpen
  \bibfield  {author} {\bibinfo {author} {\bibfnamefont {M.}~\bibnamefont
  {Nieto-Vesperinas}},\ }\bibfield  {title} {\bibinfo {title} {{Optical torque:
  Electromagnetic spin and orbital-angular-momentum conservation laws and their
  significance}},\ }\href@noop {} {\bibfield  {journal} {\bibinfo  {journal}
  {Physical Review A}\ }\textbf {\bibinfo {volume} {92}},\ \bibinfo {pages}
  {043843} (\bibinfo {year} {2015}{\natexlab{b}})}\BibitemShut {NoStop}%
\bibitem [{\citenamefont {Lee}\ \emph {et~al.}(2014)\citenamefont {Lee},
  \citenamefont {Fung}, \citenamefont {Jin},\ and\ \citenamefont
  {Fang}}]{YELeeNanophoton2014}%
  \BibitemOpen
  \bibfield  {author} {\bibinfo {author} {\bibfnamefont {Y.~E.}\ \bibnamefont
  {Lee}}, \bibinfo {author} {\bibfnamefont {K.~H.}\ \bibnamefont {Fung}},
  \bibinfo {author} {\bibfnamefont {D.}~\bibnamefont {Jin}},\ and\ \bibinfo
  {author} {\bibfnamefont {N.~X.}\ \bibnamefont {Fang}},\ }\bibfield  {title}
  {\bibinfo {title} {{Optical torque from enhanced scattering by multipolar
  plasmonic resonance}},\ }\href@noop {} {\bibfield  {journal} {\bibinfo
  {journal} {Nanophotonics}\ }\textbf {\bibinfo {volume} {3}},\ \bibinfo
  {pages} {343} (\bibinfo {year} {2014})}\BibitemShut {NoStop}%
\bibitem [{\citenamefont {Wu}\ \emph {et~al.}(2020)\citenamefont {Wu},
  \citenamefont {Tanaka}, \citenamefont {Fukuhara},\ and\ \citenamefont
  {Shimura}}]{AAWuPRA2020}%
  \BibitemOpen
  \bibfield  {author} {\bibinfo {author} {\bibfnamefont {A.~A.}\ \bibnamefont
  {Wu}}, \bibinfo {author} {\bibfnamefont {Y.~Y.}\ \bibnamefont {Tanaka}},
  \bibinfo {author} {\bibfnamefont {R.}~\bibnamefont {Fukuhara}},\ and\
  \bibinfo {author} {\bibfnamefont {T.}~\bibnamefont {Shimura}},\ }\bibfield
  {title} {\bibinfo {title} {{Continuity equation for spin angular momentum in
  relation to optical chirality}},\ }\href@noop {} {\bibfield  {journal}
  {\bibinfo  {journal} {Physical Review A}\ }\textbf {\bibinfo {volume}
  {102}},\ \bibinfo {pages} {023531} (\bibinfo {year} {2020})}\BibitemShut
  {NoStop}%
\bibitem [{\citenamefont {Ashkin}(1970)}]{AAshkinPRL1970}%
  \BibitemOpen
  \bibfield  {author} {\bibinfo {author} {\bibfnamefont {A.}~\bibnamefont
  {Ashkin}},\ }\bibfield  {title} {\bibinfo {title} {{Acceleration and trapping
  of particles by radiation pressure}},\ }\href@noop {} {\bibfield  {journal}
  {\bibinfo  {journal} {Physical Review Letters}\ }\textbf {\bibinfo {volume}
  {24}},\ \bibinfo {pages} {156} (\bibinfo {year} {1970})}\BibitemShut
  {NoStop}%
\bibitem [{\citenamefont {Ashkin}\ \emph {et~al.}(1986)\citenamefont {Ashkin},
  \citenamefont {Dziedzic}, \citenamefont {Bjorkhom},\ and\ \citenamefont
  {Chu}}]{AAshkinOL1986}%
  \BibitemOpen
  \bibfield  {author} {\bibinfo {author} {\bibfnamefont {A.}~\bibnamefont
  {Ashkin}}, \bibinfo {author} {\bibfnamefont {J.~M.}\ \bibnamefont
  {Dziedzic}}, \bibinfo {author} {\bibfnamefont {J.~E.}\ \bibnamefont
  {Bjorkhom}},\ and\ \bibinfo {author} {\bibfnamefont {S.}~\bibnamefont
  {Chu}},\ }\bibfield  {title} {\bibinfo {title} {{Observation of a single-beam
  gradient force optical trap for dielectric particles}},\ }\href@noop {}
  {\bibfield  {journal} {\bibinfo  {journal} {Optics Letters}\ }\textbf
  {\bibinfo {volume} {11}},\ \bibinfo {pages} {288} (\bibinfo {year}
  {1986})}\BibitemShut {NoStop}%
\bibitem [{\citenamefont {Marston}\ and\ \citenamefont
  {Crichton}(1984)}]{PLMarstonPRA1984}%
  \BibitemOpen
  \bibfield  {author} {\bibinfo {author} {\bibfnamefont {P.~L.}\ \bibnamefont
  {Marston}}\ and\ \bibinfo {author} {\bibfnamefont {J.~H.}\ \bibnamefont
  {Crichton}},\ }\bibfield  {title} {\bibinfo {title} {{Radiation torque on a
  sphere caused by a circularly-polarized electromagnetic wave}},\ }\href@noop
  {} {\bibfield  {journal} {\bibinfo  {journal} {Physical Review A}\ }\textbf
  {\bibinfo {volume} {30}},\ \bibinfo {pages} {2508} (\bibinfo {year}
  {1984})}\BibitemShut {NoStop}%
\bibitem [{\citenamefont {Chang}\ and\ \citenamefont
  {Lee}(1985)}]{SChangJOSAB1985}%
  \BibitemOpen
  \bibfield  {author} {\bibinfo {author} {\bibfnamefont {S.}~\bibnamefont
  {Chang}}\ and\ \bibinfo {author} {\bibfnamefont {S.~S.}\ \bibnamefont
  {Lee}},\ }\bibfield  {title} {\bibinfo {title} {{Optical torque exerted on a
  homogeneous sphere levitated in the circularly polarized fundamental-mode
  laser beam}},\ }\href@noop {} {\bibfield  {journal} {\bibinfo  {journal} {J.
  Opt. Soc. Am. B}\ }\textbf {\bibinfo {volume} {2}},\ \bibinfo {pages} {1853}
  (\bibinfo {year} {1985})}\BibitemShut {NoStop}%
\bibitem [{\citenamefont {Barton}\ \emph {et~al.}(1989)\citenamefont {Barton},
  \citenamefont {Alexander},\ and\ \citenamefont {Schaub}}]{JPBartonJAP1989}%
  \BibitemOpen
  \bibfield  {author} {\bibinfo {author} {\bibfnamefont {J.~P.}\ \bibnamefont
  {Barton}}, \bibinfo {author} {\bibfnamefont {D.~R.}\ \bibnamefont
  {Alexander}},\ and\ \bibinfo {author} {\bibfnamefont {S.~A.}\ \bibnamefont
  {Schaub}},\ }\bibfield  {title} {\bibinfo {title} {{Theoretical determination
  of net radiation force and torque for a spherical particle illuminated by a
  focused laser beam}},\ }\href@noop {} {\bibfield  {journal} {\bibinfo
  {journal} {J. Opt. Soc. Am. B}\ }\textbf {\bibinfo {volume} {66}},\ \bibinfo
  {pages} {4594} (\bibinfo {year} {1989})}\BibitemShut {NoStop}%
\bibitem [{\citenamefont {Chang}\ and\ \citenamefont
  {Lee}(1998)}]{SChangOC1998}%
  \BibitemOpen
  \bibfield  {author} {\bibinfo {author} {\bibfnamefont {S.}~\bibnamefont
  {Chang}}\ and\ \bibinfo {author} {\bibfnamefont {S.~S.}\ \bibnamefont
  {Lee}},\ }\bibfield  {title} {\bibinfo {title} {{Optical torque exerted on a
  sphere in the evanescent field of a circularly-polarized Gaussian laser
  beam}},\ }\href@noop {} {\bibfield  {journal} {\bibinfo  {journal} {Optics
  Communications}\ }\textbf {\bibinfo {volume} {151}},\ \bibinfo {pages} {286}
  (\bibinfo {year} {1998})}\BibitemShut {NoStop}%
\bibitem [{\citenamefont {Canaguier-Durand}\ and\ \citenamefont
  {Genet}(2014)}]{ACanaguierDurandPRA2014}%
  \BibitemOpen
  \bibfield  {author} {\bibinfo {author} {\bibfnamefont {A.}~\bibnamefont
  {Canaguier-Durand}}\ and\ \bibinfo {author} {\bibfnamefont {C.}~\bibnamefont
  {Genet}},\ }\bibfield  {title} {\bibinfo {title} {{Transverse spinning of a
  sphere in a plasmonic field}},\ }\href@noop {} {\bibfield  {journal}
  {\bibinfo  {journal} {Physical Review A}\ }\textbf {\bibinfo {volume} {89}},\
  \bibinfo {pages} {033841} (\bibinfo {year} {2014})}\BibitemShut {NoStop}%
\bibitem [{\citenamefont
  {Nieto-Vesperinas}(2015{\natexlab{c}})}]{MNVesperinasOL2015}%
  \BibitemOpen
  \bibfield  {author} {\bibinfo {author} {\bibfnamefont {M.}~\bibnamefont
  {Nieto-Vesperinas}},\ }\bibfield  {title} {\bibinfo {title} {{Optical torque
  on small bi-isotropic particles}},\ }\href@noop {} {\bibfield  {journal}
  {\bibinfo  {journal} {Optics Letters}\ }\textbf {\bibinfo {volume} {40}},\
  \bibinfo {pages} {3021} (\bibinfo {year} {2015}{\natexlab{c}})}\BibitemShut
  {NoStop}%
\bibitem [{\citenamefont {Jiang}\ \emph {et~al.}(2015)\citenamefont {Jiang},
  \citenamefont {Chen}, \citenamefont {Chen}, \citenamefont {Ng},\ and\
  \citenamefont {Lin}}]{YJiangArxiv2015}%
  \BibitemOpen
  \bibfield  {author} {\bibinfo {author} {\bibfnamefont {Y.}~\bibnamefont
  {Jiang}}, \bibinfo {author} {\bibfnamefont {H.}~\bibnamefont {Chen}},
  \bibinfo {author} {\bibfnamefont {J.}~\bibnamefont {Chen}}, \bibinfo {author}
  {\bibfnamefont {J.}~\bibnamefont {Ng}},\ and\ \bibinfo {author}
  {\bibfnamefont {Z.}~\bibnamefont {Lin}},\ }\bibfield  {title} {\bibinfo
  {title} {{Universal relationships between optical force/torque and orbital
  versus spin momentum/angular momentum of light}},\ }\href@noop {} {\bibfield
  {journal} {\bibinfo  {journal} {arXiv:1511.08546v3}\ } (\bibinfo {year}
  {2015})}\BibitemShut {NoStop}%
\bibitem [{\citenamefont {Chen}\ \emph {et~al.}(2014)\citenamefont {Chen},
  \citenamefont {Ng}, \citenamefont {Ding}, \citenamefont {Fung}, \citenamefont
  {Lin},\ and\ \citenamefont {Chan}}]{JChenSC2014}%
  \BibitemOpen
  \bibfield  {author} {\bibinfo {author} {\bibfnamefont {J.}~\bibnamefont
  {Chen}}, \bibinfo {author} {\bibfnamefont {J.}~\bibnamefont {Ng}}, \bibinfo
  {author} {\bibfnamefont {K.}~\bibnamefont {Ding}}, \bibinfo {author}
  {\bibfnamefont {K.~H.}\ \bibnamefont {Fung}}, \bibinfo {author}
  {\bibfnamefont {Z.}~\bibnamefont {Lin}},\ and\ \bibinfo {author}
  {\bibfnamefont {C.~T.}\ \bibnamefont {Chan}},\ }\bibfield  {title} {\bibinfo
  {title} {{Negative optical torque}},\ }\href@noop {} {\bibfield  {journal}
  {\bibinfo  {journal} {Scientific Reports}\ }\textbf {\bibinfo {volume} {4}},\
  \bibinfo {pages} {1} (\bibinfo {year} {2014})}\BibitemShut {NoStop}%
\bibitem [{\citenamefont {Tkachenko}\ \emph {et~al.}(2020)\citenamefont
  {Tkachenko}, \citenamefont {Toftul}, \citenamefont {Esporlas}, \citenamefont
  {Maimaiti}, \citenamefont {Le~Kien}, \citenamefont {Truong},\ and\
  \citenamefont {Chormaic}}]{GTkachenkoOptica2020}%
  \BibitemOpen
  \bibfield  {author} {\bibinfo {author} {\bibfnamefont {G.}~\bibnamefont
  {Tkachenko}}, \bibinfo {author} {\bibfnamefont {I.}~\bibnamefont {Toftul}},
  \bibinfo {author} {\bibfnamefont {C.}~\bibnamefont {Esporlas}}, \bibinfo
  {author} {\bibfnamefont {A.}~\bibnamefont {Maimaiti}}, \bibinfo {author}
  {\bibfnamefont {F.}~\bibnamefont {Le~Kien}}, \bibinfo {author} {\bibfnamefont
  {V.~G.}\ \bibnamefont {Truong}},\ and\ \bibinfo {author} {\bibfnamefont
  {S.~N.}\ \bibnamefont {Chormaic}},\ }\bibfield  {title} {\bibinfo {title}
  {{Light-induced rotation of dielectric microparticles around an optical
  nanofiber}},\ }\href@noop {} {\bibfield  {journal} {\bibinfo  {journal}
  {Optica}\ }\textbf {\bibinfo {volume} {7}},\ \bibinfo {pages} {59} (\bibinfo
  {year} {2020})}\BibitemShut {NoStop}%
\bibitem [{\citenamefont {Han}\ \emph {et~al.}(2009)\citenamefont {Han},
  \citenamefont {Lai}, \citenamefont {Fung}, \citenamefont {Zhang},\ and\
  \citenamefont {Chan}}]{DZHanPRB2009}%
  \BibitemOpen
  \bibfield  {author} {\bibinfo {author} {\bibfnamefont {D.}~\bibnamefont
  {Han}}, \bibinfo {author} {\bibfnamefont {Y.}~\bibnamefont {Lai}}, \bibinfo
  {author} {\bibfnamefont {K.~H.}\ \bibnamefont {Fung}}, \bibinfo {author}
  {\bibfnamefont {Z.~Q.}\ \bibnamefont {Zhang}},\ and\ \bibinfo {author}
  {\bibfnamefont {C.~T.}\ \bibnamefont {Chan}},\ }\bibfield  {title} {\bibinfo
  {title} {{Negative group velocity from quadrupole resonances of plasmonic
  spheres}},\ }\href@noop {} {\bibfield  {journal} {\bibinfo  {journal}
  {Physical Review B}\ }\textbf {\bibinfo {volume} {79}},\ \bibinfo {pages}
  {195444} (\bibinfo {year} {2009})}\BibitemShut {NoStop}%
\bibitem [{\citenamefont {Arango}\ \emph {et~al.}(2014)\citenamefont {Arango},
  \citenamefont {Coenen},\ and\ \citenamefont
  {Koenderink}}]{FArangoACSPhoton2011}%
  \BibitemOpen
  \bibfield  {author} {\bibinfo {author} {\bibfnamefont {F.~B.}\ \bibnamefont
  {Arango}}, \bibinfo {author} {\bibfnamefont {T.}~\bibnamefont {Coenen}},\
  and\ \bibinfo {author} {\bibfnamefont {A.}~\bibnamefont {Koenderink}},\
  }\bibfield  {title} {\bibinfo {title} {{Underpinning Hybridization Intuition
  for Complex Nanoantennas by Magnetoelectric Quadrupolar Polarizability
  Retrieval}},\ }\href@noop {} {\bibfield  {journal} {\bibinfo  {journal} {ACS
  Photonics}\ }\textbf {\bibinfo {volume} {1}},\ \bibinfo {pages} {444}
  (\bibinfo {year} {2014})}\BibitemShut {NoStop}%
\bibitem [{\citenamefont {Das}\ \emph {et~al.}(2015)\citenamefont {Das},
  \citenamefont {Iyer}, \citenamefont {DeCrescent},\ and\ \citenamefont
  {Schuller}}]{TDasPRB2015}%
  \BibitemOpen
  \bibfield  {author} {\bibinfo {author} {\bibfnamefont {T.}~\bibnamefont
  {Das}}, \bibinfo {author} {\bibfnamefont {P.~P.}\ \bibnamefont {Iyer}},
  \bibinfo {author} {\bibfnamefont {R.~A.}\ \bibnamefont {DeCrescent}},\ and\
  \bibinfo {author} {\bibfnamefont {J.~A.}\ \bibnamefont {Schuller}},\
  }\bibfield  {title} {\bibinfo {title} {{Beam engineering for selective and
  enhanced coupling to multipolar resonances}},\ }\href@noop {} {\bibfield
  {journal} {\bibinfo  {journal} {Physical Review B}\ }\textbf {\bibinfo
  {volume} {92}},\ \bibinfo {pages} {241110(R)} (\bibinfo {year}
  {2015})}\BibitemShut {NoStop}%
\bibitem [{\citenamefont {Babicheva}\ and\ \citenamefont
  {Evlyukhin}(2019)}]{VEBabichevaPRB2019}%
  \BibitemOpen
  \bibfield  {author} {\bibinfo {author} {\bibfnamefont {V.~E.}\ \bibnamefont
  {Babicheva}}\ and\ \bibinfo {author} {\bibfnamefont {A.~B.}\ \bibnamefont
  {Evlyukhin}},\ }\bibfield  {title} {\bibinfo {title} {{Analytical model of
  resonant electromagnetic dipole-quadrupole coupling in nanoparticle
  arrays}},\ }\href@noop {} {\bibfield  {journal} {\bibinfo  {journal}
  {Physical Review B}\ }\textbf {\bibinfo {volume} {99}},\ \bibinfo {pages}
  {195444} (\bibinfo {year} {2019})}\BibitemShut {NoStop}%
\end{thebibliography}
\end{document}